\documentclass[twocolumn]{aastex631}
\usepackage{comment}

\shorttitle{Reconstructing the XUV Spectra of Active Sun-like Stars}
\shortauthors{Namekata et al.}
\graphicspath{{./}}

\begin{document}

\title{Reconstructing the XUV Spectra of Active Sun-like Stars Using Solar Scaling Relations with Magnetic Flux}

\author[0000-0002-1297-9485]{Kosuke Namekata}
\affiliation{ALMA Project, National Astronomical Observatory of Japan, NINS, Osawa, Mitaka, Tokyo, 181-8588, Japan}
\email{kosuke.namekata@astro.nao.ac.jp, namekata@kusastro.kyoto-u.ac.jp}
\author[0000-0002-1276-2403]{Shin Toriumi}
\affiliation{Institute of Space and Astronautical Science, Japan Aerospace Exploration Agency, 3-1-1 Yoshinodai, Chuo-ku, Sagamihara, Kanagawa 252-5210, Japan}
\author[0000-0003-4452-0588]{Vladimir S. Airapetian}
\affiliation{Sellers Exoplanetary Environments Collaboration, NASA Goddard Space Flight Center, Greenbelt, MD, USA}
\affiliation{Department of Physics, American University, Washington, DC, USA}
\author[0000-0002-7136-8190]{Munehito Shoda}
\affiliation{Department of Earth and Planetary Science, School of Science, The University of Tokyo, 7-3-1, Hongo, Bunkyo, Tokyo, 113-0033, Japan}
\author[0000-0003-0321-7881]{Kyoko Watanabe}
\affiliation{National Defense Academy of Japan, 1-10-20 Hashirimizu, Yokosuka 239-8686, Japan}
\author[0000-0002-0412-0849]{Yuta Notsu}
\affiliation{Laboratory for Atmospheric and Space Physics, University of Colorado Boulder, 3665 Discovery Drive, Boulder, CO 80303, USA}
\affiliation{National Solar Observatory, 3665 Discovery Drive, Boulder, CO 80303, USA}
\affiliation{Department of Earth and Planetary Sciences, Tokyo Institute of Technology, 2-12-1 Ookayama, Meguro-ku, Tokyo 152-8551, Japan}



\begin{abstract}

Kepler Space Telescope and Transiting Exoplanet Survey Satellite unveiled that Sun-like stars frequently host exoplanets. 
These exoplanets are subject to fluxes of ionizing radiation in the form of X-ray and extreme-ultraviolet (EUV) radiation that may cause changes in their atmospheric dynamics and chemistry. 
While X-ray fluxes can be observed directly, EUV fluxes cannot be observed because of severe interstellar medium absorption.
Here, we present a new empirical method to estimate the whole stellar XUV (X-ray plus EUV) and FUV spectra as a function of total unsigned magnetic fluxes of stars.
The response of the solar XUV and FUV spectrum (0.1--180 nm) to the solar total unsigned magnetic flux is investigated by using the long-term Sun-as-a-star dataset over 10 yrs, and the power-law relation is obtained for each wavelength with a spectral resolution of 0.1--1 nm.
We applied the scaling relations to active young Sun-like stars (G-dwarfs), EK Dra (G1.5V), $\pi^1$ Uma (G1.5V) and $\kappa^1$ Ceti (G5V), and found that the observed spectra (except for the unobservable longward EUV wavelength) are roughly consistent with the extension of the derived power-law relations with errors of an order of magnitude.
This suggests that our model is a valuable method to derive the XUV/FUV fluxes of Sun-like stars including the EUV band mostly absorbed at wavelengths longward of 36 nm.
We also discuss differences between the solar extensions and stellar observations at the wavelength in the 2--30 nm band and concluded that simultaneous observations of magnetic and XUV/FUV fluxes are necessary for further validations. 





\end{abstract}



\section{Introduction}\label{sec:1}


The solar atmosphere spans from the photosphere ($\sim$6000 K), chromosphere ($\sim$10$^{4}$ K), transition region (TR), and corona ($\sim$10$^6$ K) \citep{1943ZA.....22...30E,1971SoPh...18..347G,1981ApJS...45..635V}.
The corona and TR are the source of the X-ray and extreme ultraviolet (EUV) radiations (hereafter, XUV is referred to as X-ray plus EUV bands (at wavelength between 1 to 92 nm).
The large magnetic fluxes are usually associated with solar coronal active regions that produce strong quiescent XUV emission \citep{2003ApJ...598.1387P,2019LRSP...16....3T}. Solar active regions are also the sources of the transient XUV emission triggered by solar flares \citep{2011LRSP....8....6S}.
The quiescent and transient coronal activity have been universally observed not only on the Sun, but also on the F, G, K, and M-type main-sequence stars in the XUV bands. \citep{1997ApJ...483..947G,2005ApJ...622..680R,2007LRSP....4....3G}.


Recent Kepler space telescope and Transiting Exoplanet Survey Satellite (TESS) observations unveiled that F, G, K, and M dwarfs frequently host close-in exoplanets.
Moreover, some exoplanet-hosting stars are much younger and have much higher magnetic activity than the Sun.
The exoplanets around active stars are subject to high-energy XUV and Far UV (FUV) radiation \citep[e.g.,][]{2011ApJ...743...48W,2020ApJ...901...70T,2021A&A...649A..96J}, stellar winds \citep[e.g.,][]{2013PASJ...65...98S,2020ApJ...896..123S,2021ApJ...915...37W}, and eruptive phenomena \citep[e.g.,][]{2016A&A...590A..11V,2020PASJ..tmp..253M,2021NatAs...5..697V,2022NatAs...6..241N,2022ApJ...936..170L} that may cause significant changes in their atmospheric dynamics and chemistry. 
Stellar XUV radiation causes expansion and thermal escape of planetary atmospheres \citep[e.g.,][]{2016NatGe...9..452A,2017ApJ...836L...3A,2019LNP...955.....L,2020IJAsB..19..136A}, and therefore is critical for understanding planetary habitability.
Stellar XUV fluxes usually consist of the quasi-steady quiescent components that come from quiescent coronae and transient flare events.
Stellar flares have been widely investigated from X-ray to radio on the cool M/K dwarfs \citep[][]{2013ApJS..207...15K,2014ApJ...797..121H,2018ApJ...855L...2M,2020PASJ...72...68N} and on G type dwarfs \citep{2012Natur.485..478M,2015EP&S...67...59M,2015AJ....150....7A,2019ApJ...876...58N,2019A&A...622A.210G,2022NatAs...6..241N,2022ApJ...926L...5N}.
The flare XUV emission significantly enhances the stellar irradiances by a factor of 10-100, and if these events are frequent, then its radiation can produce the dramatic changes in the exoplanetary atmospheres on a short timescale, typically from tens of minutes to hours \citep[e.g.,][]{2010AsBio..10..751S,2022MNRAS.509.5858H}.
On the other hand, from the exoplanetary perspective, the quasi-steady XUV component is associated with coronal active regions that can last for a few months \citep[e.g.,][]{2019ApJ...871..187N,2020ApJ...891..103N}, and thus provide much greater overall impact on exoplanetary atmospheric dynamics.
Theoretical models of exoplanetary atmospheric escape require the knowledge of distribution of XUV flux from the wavelengths between 1 - 92 nm as the major source of upper atmospheric heating, photodissociation and photoionization \citep[][]{2003ApJ...598L.121L,2010A&A...511L...8S,2017ApJ...836L...3A,2019A&A...624L..10J}. Thus, XUV flux evolution on time scales of the lifetime of stellar active regions are required for assessment of habitability of exoplanetary environments.

While characterization of stellar XUV fluxes are critically required from the exoplanetary studies, it is nearly impossible to detect the emission from the full EUV band (10--92 nm), even for the closest stars to the Sun, such as $\alpha$ Centauri.
This is because the interstellar medium (ISM) absorption by hydrogen, helium and other resonance lines of many other important metal species contribute to the absorption of the longward EUV emission ($>$36 nm), while the short XUV bands ($<$36 nm) are relatively unaffected \citep[e.g.,][]{1974ApJ...187..497C}, and thus can be detected from nearby stars by the Extreme Ultraviolet Explorer (EUVE), ROentgen SATellite (ROSAT) and Chandra X-ray Observatory  \citep[e.g.,][]{2005ApJ...622..680R,2021A&A...649A..96J}.
Since the EUV spectrum in 36--91.2 nm can be obtained only for the Sun, it is essential to develop methods for estimating these elusive stellar EUV fluxes from stellar observations.


There have been suggested several reconstruction methods to derive solar and stellar quiescent XUV spectrum (usually in the non-flaring phase). 
An early approach by \cite{2005ApJ...622..680R} estimated the stellar EUV fluxes by assuming power-law relations between the EUV flux and stellar age, based on the X-ray/FUV observations \citep[][]{1997JGR...102.1641A}. 
Later, \cite{2012ApJ...757...95C} extended this method to derive the whole stellar XUV spectrum by multiplying the scaling relation as a function of age and whole ``solar" XUV spectrum.
Their method largely depends on the assumption of power-law index. 
Another approach introduced by \cite{2007A&A...461.1185L} reconstructed the whole EUV band flux (10--91.2 nm) by scaling ROSAT flux (11--20 nm) with the solar flux ratio of 10--91.2 nm to 11--20 nm flux. Later, a similar scaling approach is also developed by using EUVE fluxes in 10--36 nm \citep{2021A&A...649A..96J}.
These early approaches could be valid for inactive Sun-like stars, but they would not hold when it is applied to active Sun-like stars and M-dwarfs having hotter coronae and TR than the Sun.

In addition, the following two methodologies have been developed recently: One is based on the inversion methods from the differential emission measure (DEM) obtained from X-ray and/or UV spectral lines \citep{2011A&A...532A...6S}.
This inversion method is recently used for the purpose of the EUV reconstruction of exoplanet hosting stars \citep[e.g.,][]{2015Icar..250..357C,2017MNRAS.464.2396L,2021ApJ...913...40D}.
The other approach represents the flux-flux methodology where EUV flux/spectrum is estimated from the fluxes of magnetically-sensitive emission lines with (semi-)empirical power-law scaling relations.
\cite{2014ApJ...780...61L} developed semi-empirical power-law scaling laws to estimate the EUV spectrum from the Ly$\alpha$ flux with 10 nm spectral resolution.
The estimated EUV fluxes are consistent with those obtained by the differential emission measure methods, and therefore the above two methods are well used to complement each other \citep[e.g.,][]{2016ApJ...820...89F,2016ApJ...824..101Y,2017ApJ...843...31Y,2016ApJ...824..102L,2018ApJ...867...71L,2019ApJ...871L..26F,2021ApJ...911...18W}.
Here, we note that the direct measurements of Ly$\alpha$ emission would have uncertainties due to the ISM absorption \citep{2019LNP...955.....L}.
Thus, the application of the Ly$\alpha$ flux to predict the EUV flux is useful only when there is no access to X-rays.
More recently, \cite{2018ApJS..239...16F} present a similar relationship to estimate the intrinsic stellar EUV flux in the 9--36 nm band using a FUV spectral line (N V or Si IV spectra) and stellar bolometric flux.
\cite{2017ApJ...843...31Y} proposed a flux-flux methodology for Ca II K (later, \cite{2020A&A...644A..67S} extended this to Ca II H\&K). 
The estimation of the EUV flux based on Ca II H\&K could be useful because these lines can be observable from the ground, while other methods requires space-based observations.


What other observables could be good indicators of EUV emission from stars?
In light of the fact that stellar coronae are heated by magnetic field \citep[e.g.,][]{1972ApJ...174..499P,1988ApJ...330..474P,2005ApJ...618.1020G}, the stellar magnetic field is a major candidate of such an observable.
Indeed, the stellar magnetic flux is calculated from observations taken with the ground-based telescopes \citep[e.g.,][]{2001ASPC..223..292S,2009ApJ...692..538R,2014MNRAS.441.2361V,2020A&A...635A.142K,2022A&A...662A..41R}, and therefore it could be a good tool to reconstruct the whole XUV spectra. 
Previous studies show a common power-law relation between X-ray radiation and magnetic fluxes for the Sun and F, G, K, and M dwarfs, with a common power-law relation, $F\propto\Phi^{\alpha}$ ($\alpha$=1.1--1.8) \citep[e.g.,][]{2003ApJ...598.1387P,2014MNRAS.441.2361V}.
More recently, \cite{2022ApJ...927..179T} derived power-law relations between the total unsigned magnetic flux and irradiance of five XUV/FUV emission lines/bands by using the ``Sun-as-a-star" approach and found that the solar empirical relations can be extended to active Sun-like stars (G and K dwarfs).
They also found that the power-law index decreases from above- to sub-unity as the formation temperature decreases from the corona to the chromosphere and proposed that the decreasing trend can be understood from the first principles physics-based models \cite[also see,][]{2022ApJS..262...46T}.
These studies provide an observational evidence that the mechanism of atmospheric heating by the magnetic field is universal among the Sun and Sun-like stars, regardless of activity levels.
At the same time, these results open up an opportunity to estimate the whole XUV/FUV spectrum of stars from the magnetic flux measurements.

In this paper, we developed a new method to reconstruct stellar XUV/FUV spectra (0.1--180 nm) with a spectral resolution of 0.1--1 nm from the observionally derived total unsigned magnetic fluxes.
We have extended the flux-flux methodology employed in  \cite{2022ApJ...927..179T} and \cite{2022ApJS..262...46T} to each wavelength of the XUV/FUV spectrum, regardless of whether it is emission line or continuum. Here, we use the Sun-as-a-star multi-mission data (spatially integrated data over the whole solar disk) over more than 10 years to derive the power-law relations for XUV/FUV spectral bands.
The Sun-as-a-star approach is a very powerful tool in understanding phenomena on stars that cannot be spatially resolved and has been widely used \citep[e.g.,][]{2011A&A...530A..84K,2019A&A...627A.118M,2022NatAs...6..241N,2022ApJ...933..209N,2020ApJ...902...36T,2022ApJS..262...46T,2022ApJ...927..179T,2022ApJ...939...98O}.
Section \ref{sec:2} describes the Sun-as-a-star dataset used in our analysis.
In Section \ref{sec:3}, we derived a new power-law method to estimate the stellar XUV/FUV spectrum by using Sun-as-a-star observation data.
In Section \ref{sec:4}, we compared the reconstructed stellar XUV/FUV spectra/fluxes from the scaling laws and observed spectra/fluxes for young Sun-like stars.
Finally, Section \ref{sec:dis} represents the discussion and conclusion.

\begin{deluxetable*}{lccccccc}
\tabletypesize{\footnotesize}
\tablecaption{Summary of datasets}
\tablewidth{0pt}
\tablehead{
\colhead{Satellite/Instr.} & \colhead{Wavelength} & \colhead{Resl./Samp.$^\S$} & \colhead{Obs. Period} & \colhead{Used Period} & \colhead{Basal Period$^{\#}$} & \colhead{Ver.$^\ddag$} & \colhead{Lev.$^\ddag$} \\
\colhead{} & \colhead{[nm]} & \colhead{[{nm}]} & \colhead{[yr]} & \colhead{[month.yr]} & \colhead{[month.yr]}  & &
}
\startdata
\hline
\textsf{(1) X-ray \& short EUV}\\
SORCE/XPS & 0.1--40 & 0.1 & 2003--2020 & 5.2010--2.2020 & 3.2019--2.2020 & 18 & 4 \\
(TIMED/SEE)$^\dagger$ & 0.5--180 & 1 & 2002-- & 1.2002--12.2016  & 1.2009--12.2009 & 12  & 3 \\
\hline
\textsf{(2) EUV}\\
SDO/EVE & 33.3--106.6 & 0.1/0.02 & 2010-- & 5.2010--2.2020 & 3.2019--2.2020 & 7 & 3  \\
(TIMED/SEE)$^\dagger$ & 0.5--180 & 0.4/1$^{\ast}$ & 2002-- & 1.2002--12.2016  & 1.2009--12.2009 & 12 & 3 \\
\hline
\textsf{(3) FUV}\\
TIMED/SEE & 0.5--180 &  0.4/1$^{\ast}$ & 2002-- & 1.2002--12.2016  & 1.2009--12.2009 & 12 & 3 \\
SORCE/SOLSTICE$^\dagger$ & 115--310 & 0.1/0.025 & 2003--2020 & 5.2010--2.2020 & 3.2019--2.2020 & 18 & 3 \\
\hline
\textsf{(4) Magnetic field}\\
SoHO/MDI & -- & -- & 1996--2011 & 1.2002--4.2010  & -- &  -- &  --\\
SDO/HMI & -- & -- & 2010-- & 5.2010--2.2020  &  3.2019--2.2020   & -- &  --\\
\enddata
\tablecomments{$^\S$Spectral resolutions and data samplings used in our analysis.
$^{\ast}$See the following url: \url{https://lasp.colorado.edu/home/see/overview/instrument-overview/}.
$^\dagger$Data indicated with blacket were not used in the main analysis but presented in the Appendix.
$^\ddag$Data versions and data levels.
$^{\#}$We note that the different activity minima may have different flux values, although there are not consistent observations that can show the differences \citep[see for example,][]{2021JSWSC..11....2C}.
}
\label{tab:1}
\end{deluxetable*}

\begin{figure*}
\epsscale{0.7}
\plotone{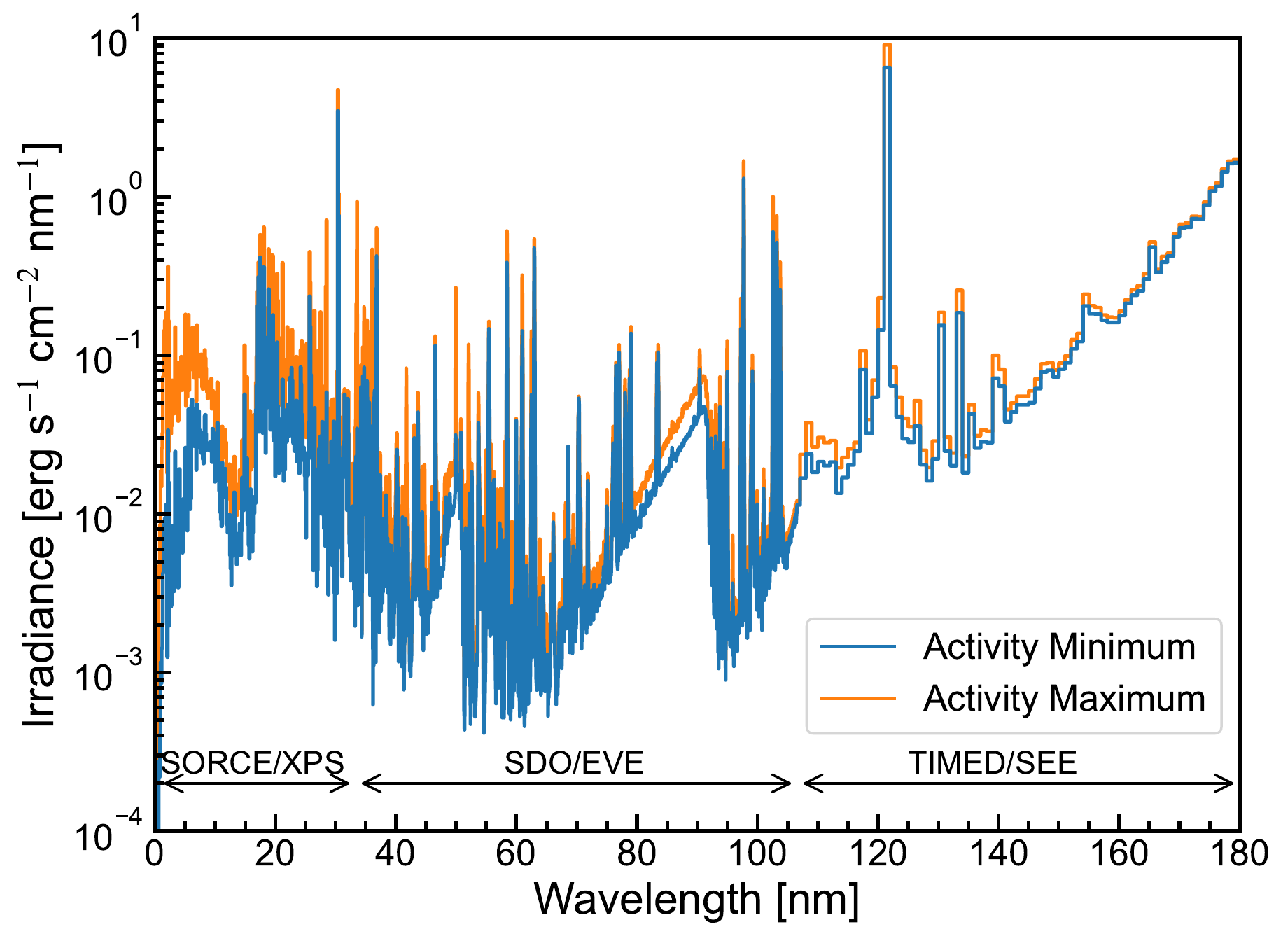}
\caption{X-ray, EUV, and FUV spectra of the Sun with SORCE/XPS (0.1--33.3 nm), SDO/EVE (33.3-106.6 nm), and TIMED/SEE (107--180 nm). The orange and blue lines indicate the solar maximum and minimum, respectively. The spectrum at solar minimum is the basal irradiance obtained in our analysis. The spectrum at solar maximum is a median value around days satisfying the following conditions: (1) the solar maximum from January 2014 to December 2015; (2) when the total unsigned magnetic flux is more than the 95th percentile for the all data used in the analysis. The fluxes are plotted in the unit of erg s$^{-1}$ cm$^{-2}$ nm$^{-1}$ at 1 AU.
}
\label{fig:1}
\end{figure*}

\section{Dataset}\label{sec:2}

\subsection{SORCE/XPS for 0.1-33.3 nm}\label{data:xps}

We used the daily averaged Sun-as-a-star spectra from X-ray to shortward EUV band (0.5-33.3 nm) obtained by the XUV Photometer System \citep[XPS;][]{2005SoPh..230..375W,2008SoPh..250..235W} onboard the Solar Radiation and Climate Experiment \citep[SORCE;][]{2005SoPh..230....7R} from May 2010 through February 2020 (see Figure \ref{fig:1} and Table \ref{tab:1}). 
The data were downloaded from the SORCE data archive\footnote{\url{https://lasp.colorado.edu/home/sorce/data/}}.
The XPS level-4 spectrum was used in this study spans from over 0.1 to 40 nm with a spectral resolution of 0.1 nm.
Note that the XPS level-4 spectrum is not directly derived from raw observations, but is based on the solar CHIANTI spectral model scaled to match the XPS photometer \citep{2008SoPh..250..235W}.
Model spectra representative of quiet Sun and active region are combined to match the signals from the XPS.
Because of this, we used SORCE/XPS data up to 33.3 nm in this study and we used a purely observational data obtained by SDO/EVE for 33.3 -- 106.6 nm range (see Figure \ref{fig:1} and Section \ref{data:eve}).
No data is available between 2013 July 13 and 2014 February 25 because all instruments on SORCE were powered down during a serious battery anomaly, but this is relatively short compared to the total period we analyzed.
The Solar Extreme Ultraviolet Experiment \citep[SEE;][]{2005JGRA..110.1312W,2018SoPh..293...76W} onboard Thermosphere Ionosphere Mesosphere Energetics Dynamics (TIMED) also covers this XUV spectral range with the spectral resolution of 1 nm (see Section \ref{data:see}), but SORCE/XPS data is much improved from the TIMED/SEE model because the SORCE/XPS model used SDO/EVE measurements to define reference spectra for the 6-40 nm range \citep{2022SoPh..297...64W}.

 
\subsection{SDO/EVE for 33.3-106.6 nm}\label{data:eve}

We analyzed the daily averaged spectra of EUV bands (33.3-106.6 nm) obtained by the Extreme ultraviolet Variability Experiment \citep[EVE;][]{2012SoPh..275..115W} on-board the Solar Dynamics Observatory \citep[SDO;][]{2012SoPh..275....3P} (see Figure \ref{fig:1} and Table \ref{tab:1}). 
The SDO/EVE instrument has performed Sun-as-a-star observations of the EUV solar spectrum from 6 nm to 106.6 nm since May 2010.
We used level-3 daily averaged spectrum of version 7 which was downloaded from the archive database\footnote{\url{https://lasp.colorado.edu/home/eve/data/}}. 
Due to the MEGS-A anomaly on May 2014, shortward EUV bands ($<$33 nm) have become not available.
As discussed in Section \ref{sec:power-law}, the data availability around activity minimum is essential because a basal-level subtraction is required for our analysis \citep[cf.][]{2022ApJ...927..179T,2022ApJS..262...46T}.
Then, only the data of longward EUV wavebands (33.3 nm to 106.6 nm) was used for our analysis, and other bands are covered by SORCE/XPS and TIMED/SEE as in Sections \ref{data:xps} and \ref{data:see}. 
We analyzed the SDO/EVE data as our primary source (see Figure \ref{fig:1}) because SDO/EVE is an updated and better calibrated system than other EUV spectroscopic instruments, such as TIMED/SEE \citep[Tom Woods, private communication;][]{2012SoPh..275..145H}, and it provides the highest wavelength resolution currently available (0.1 nm). 
In Appendix \ref{app:b}, a careful analysis of TIMED/SEE data for the XUV range is shown. 


\subsection{TIMED/SEE and SORCE/SOLSTICE for 106.6-180 nm}\label{data:see}

We analyzed the daily spectrum of FUV (106.6--180 nm) obtained by TIMED/SEE \citep{2005JGRA..110.1312W,2018SoPh..293...76W} (see Figure \ref{fig:1} and Table \ref{tab:1}).
The TIMED/SEE provides Sun-as-a-star spectra at 0.5--190 nm since January 2002 (for the shortward of 27 nm, a solar model is scaled to match the XPS broadband flux).
We used TIMED/SEE level 3 daily data with a spectral resolution of 1 nm. 
The level 3 data is the data averaged over one day, after applying corrections for atmospheric absorption, degradation, and removal of flares.
The data were downloaded from the archival database\footnote{\url{https://lasp.colorado.edu/home/see/data/} or \url{http://lasp.colorado.edu/data/timed_see/level3/}}.
TIMED/SEE continues to provide data to date, but data after 2017 are still not well calibrated. Because of this, we decided to use only data before 2016. 
As a result, the data around activity minimum during 2019-2020 is not available for TIMED/SEE, so we defined the fluxes around 2009 as basal flux for the TIMED/SEE data (see Section \ref{sec:power-law}).

The SORCE/SOLSTICE also has provided FUV data with a higher spectral resolution ($>$115 nm with a resolution of 0.1 nm) than TIMED/SEE, but there are some calibration issues in the longer FUV range, specifically in the $>$150--160 nm band.
Thus, we used the TIMED/SEE data as our primary source.
However, the higher spectral resolution of SORCE/SOLSTICE would be helpful to resolve the spectral lines in FUV range (for FUV $<$150 nm including important emission lines such as Ly$\alpha$). 
We present a careful analysis of SORCE/SOLSTICE data for FUV range in Appendix \ref{app:a}.

\subsection{Total Unsigned Magnetic Flux}

We used the full-disk magnetograms obtained by the Helioseismic and Magnetic Imager \citep[HMI;][]{2012SoPh..275..207S} onboard SDO and the Michelson Doppler Interferometer \citep[MDI;][]{1995SoPh..162..129S} onboard the Solar and Heliospheric Observatory (SoHO) to calculate the total unsigned magnetic flux in the visible solar hemisphere. The SDO/HMI has started the nominal observations in May 2010, while the observation program of SoHO/MDI was terminated on 2011 April 12.

Based on the HMI magnetograms, the daily total radial unsigned magnetic flux was calculated by following the method described in \cite{2022ApJ...927..179T} and \cite{2022ApJS..262...46T}. The line-of-sight magnetograms of 720 s cadence were downloaded from JSOC\footnote{\url{http://jsoc.stanford.edu}} for 0, 6, 12, and 18 UT for each day, and after binning the original 4096$\times$4096 pixels to 1024$\times$1024 pixels, the radial magnetic field component was derived by correcting the magnetic field strength by the heliocentric angle. The absolute radial field strength was integrated over the whole solar disk to obtain the total radial unsigned magnetic flux. The daily unsigned flux was calculated by averaging the four fluxes derived from the four magnetograms per day.

The SOHO/MDI derived full-disk magnetograms every 96 min and, in this study, the Level 1.8.2 magnetograms on JSOC were used. MDI has two observing modes: the first one is designed to use one set of observations to derive a single magnetogram (the so-called ``1-min magnetograms"); The other is the magnetogram derived by averaging the five 1-min magnetograms (the so-called ``5-min magnetograms"). The 5-min magnetograms have a lower noise level \citep{2007SoPh..241..185L,2012SoPh..279..295L} and, thus, only the 5-min magnetograms were used for the analysis. The radial component was calculated by correcting for the heliocentric angle as we did for the HMI data. The total flux was obtained by the full-disk integration, and the daily data was obtained by taking the median of 15 observations per day.

There is an overlap of HMI and MDI observations for about one year between May 2010 and April 2011. We compared the unsigned fluxes from the two observatories for this period and performed a linear fit to match the MDI flux with the HMI flux 
to smoothly connect the two observations. The conversion of MDI flux, $\Phi_{\rm MDI}$, into the HMI flux, $\Phi_{\rm HMI}$, is given as
\begin{eqnarray}
  \Phi_{\rm HMI}=(\Phi_{\rm MDI}-1.51\times 10^{23}\ {\rm Mx})/1.21.
\end{eqnarray}
Note that there is a center-to-limb variation in the conversion factors of the magnetic field strength between MDI and HMI \citep{2012SoPh..279..295L}. However, for simplicity, the comparison method described above was adopted in this study.

\section{XUV fluxes from the Sun as a Star}\label{sec:3}

\subsection{Light Curves}

\begin{figure*}
\epsscale{0.55}
\plotone{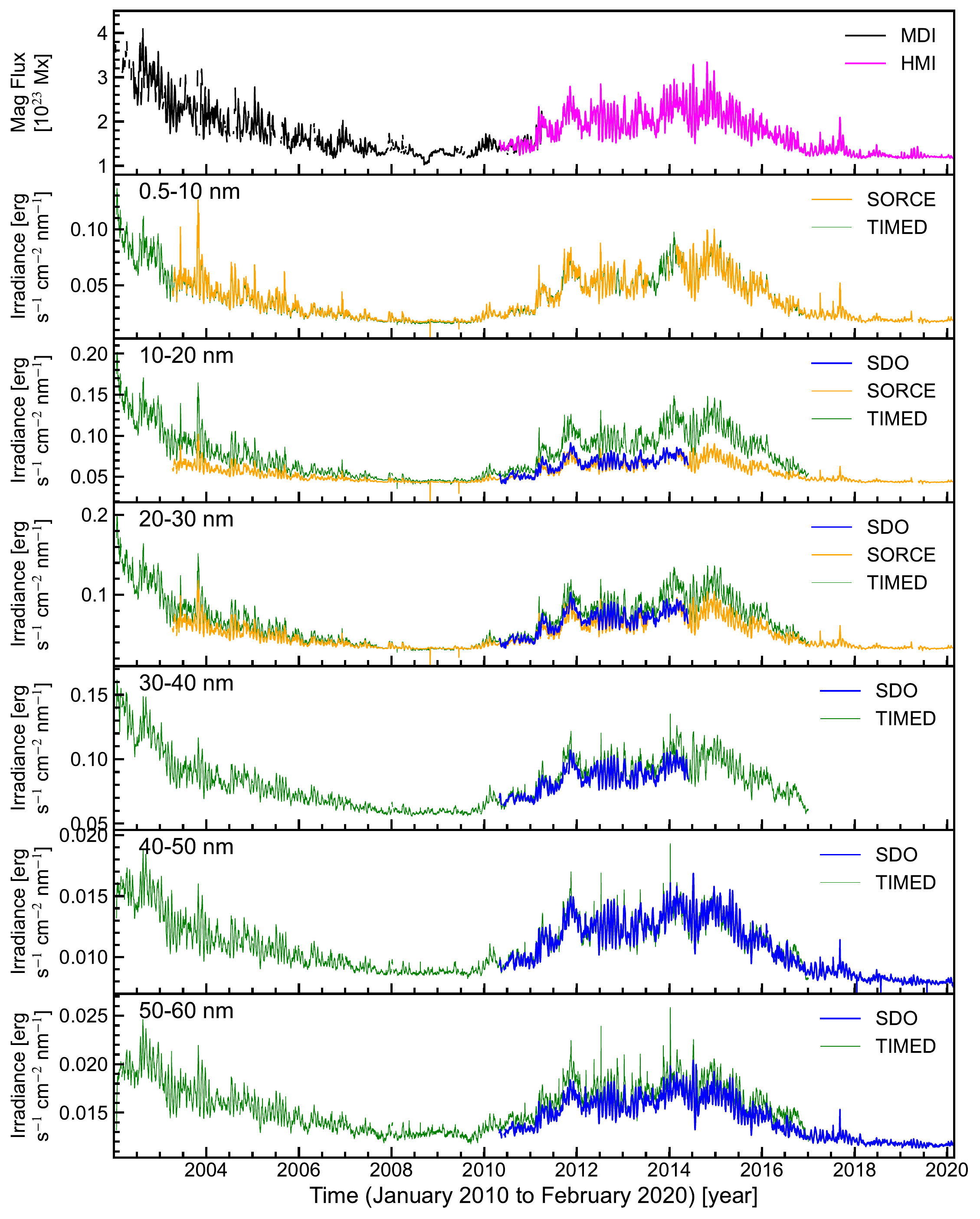}
\plotone{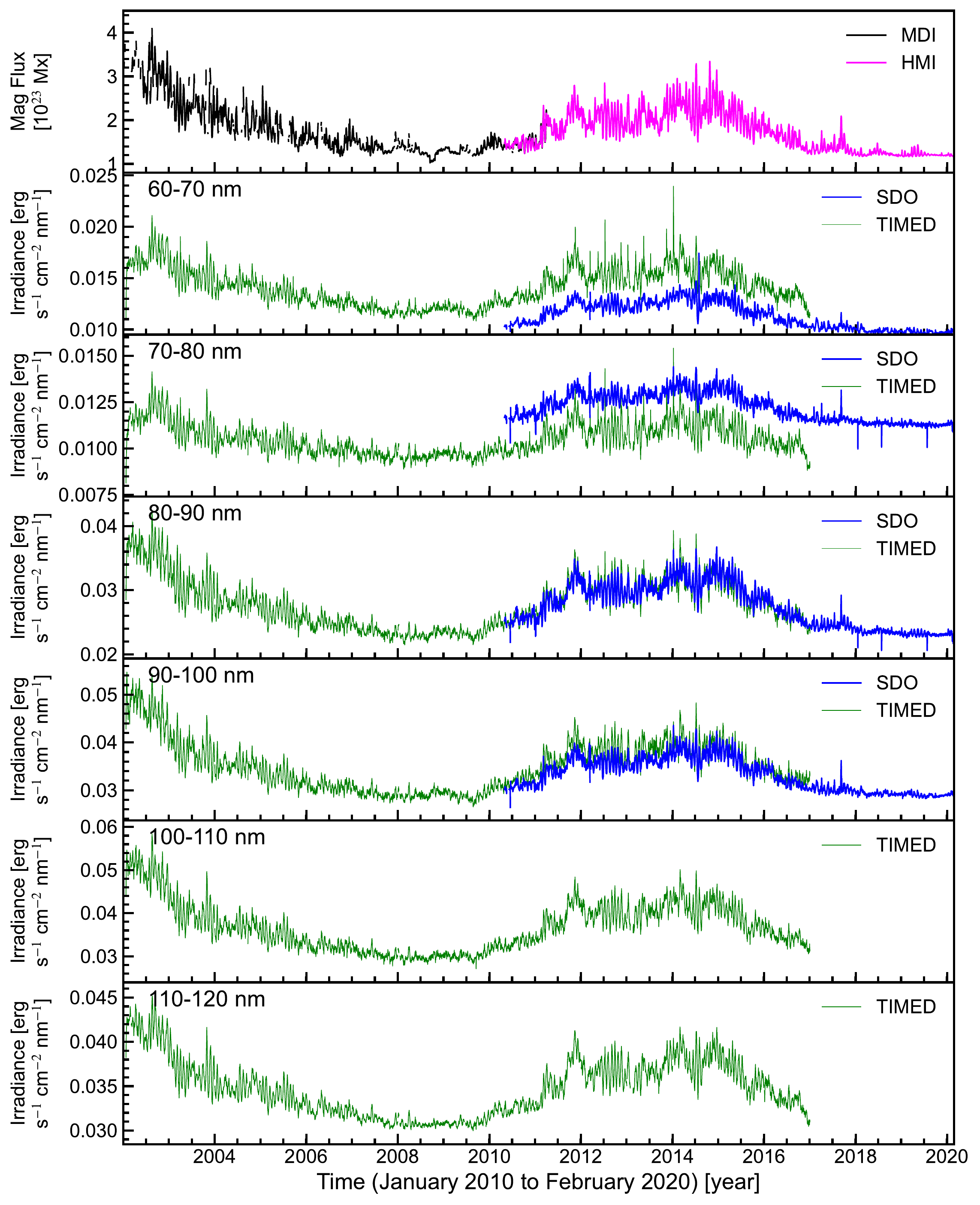}
\plotone{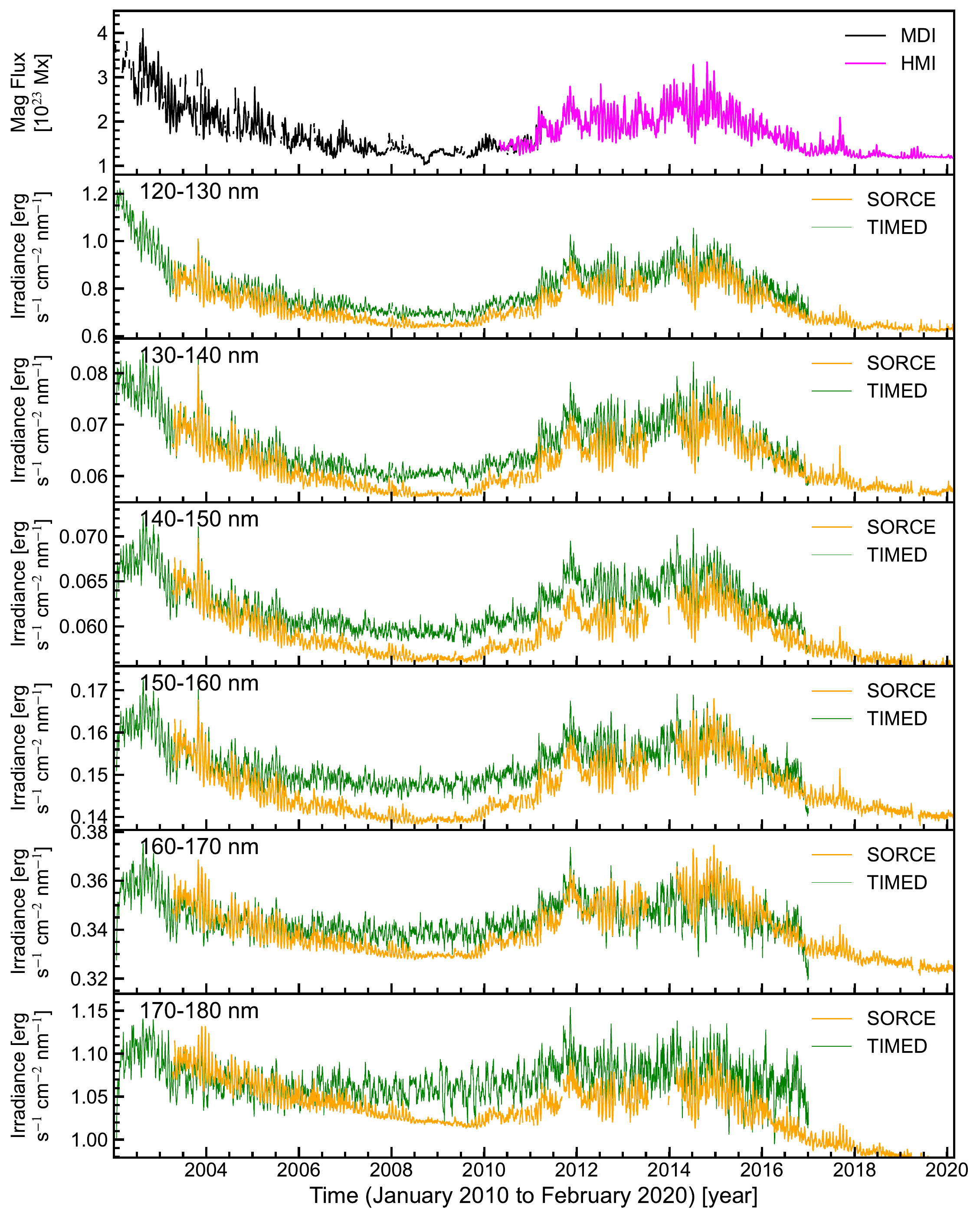}
\caption{Time series of total unsigned magnetic flux (the first column of each panel) and Sun-as-a-star light curve of X-ray, EUV, and FUV fluxes (in the unit of erg s$^{-1}$ cm$^{-2}$ nm$^{-1}$) averaged over each 10 nm bin (from the second to bottom columns). Each wavelength bin is described at top left of each panel. Different instruments are plotted in different colors.
}
\label{fig:2}
\end{figure*}

Figure \ref{fig:2} shows the time series of the total unsigned magnetic flux of the Sun and the light curves of daily Sun-as-a-star X-ray, EUV, and FUV irradiance in each 10 nm bin.
The total unsigned magnetic fluxes observed by SoHO/MDI and SDO/HMI data match well each other during the overlapping period between May 2010 and April 2011.
For the 10--30 nm band, SDO/EVE MEGS-A data is available before its anomaly on May 2014, and during this period SORCE/XPS model flux is well calibrated and match the SDO/EVE data well.
This is one reason why we use SORCE/XPS for X-ray and shortward EUV range ($<33$ nm) as mentioned in Section \ref{sec:2}.
For 40--100 nm band, SDO/EVE and TIMED/SEE are available, but there are small gaps between each dataset ($\sim$20--30 \%), in particular in 60-80 nm range.
SDO/EVE data was used in our analysis for the EUV range (40--100 nm) because it is new and well-calibrated compared to TIMED/SEE.
For 120--180 nm band, TIMED/SEE and SORCE/XPS data are available. However, SORCE/XPS data shows decreasing trends, and values in activity minimum during 2009 do not match those around 2019--2020, especially for 150--180 nm band.
This indicates that calibrations are not well done even in the latest version \citep{2022SoPh..297...55S}.
That is why we decided to use TIMED/SEE data for the FUV range, although we need to note that TIMED/SEE has a relatively large scatter in data for long FUV ranges.

\subsection{Power-law Relation}\label{sec:power-law}

\begin{figure*}
\epsscale{0.35}
\plotone{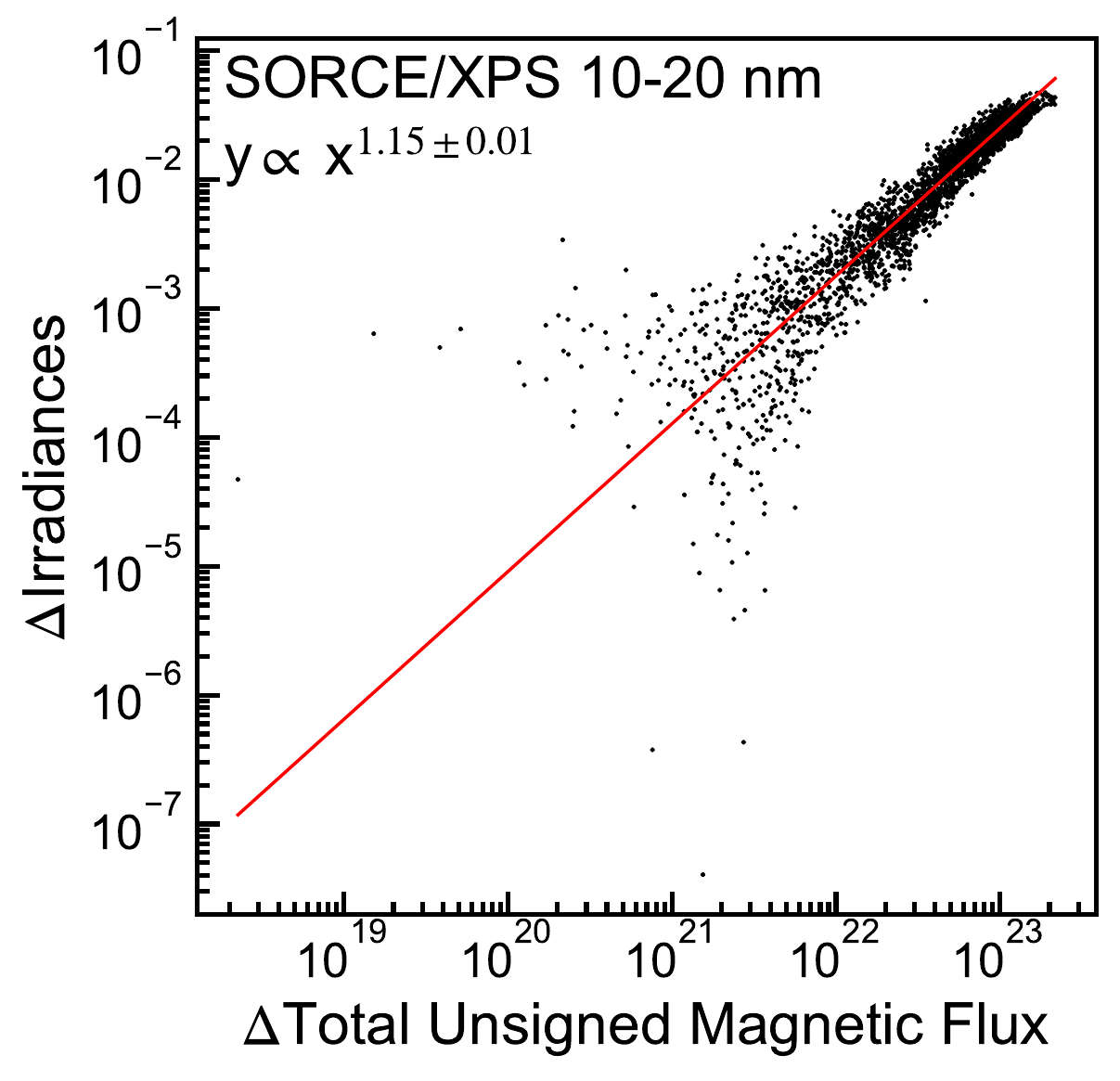}
\plotone{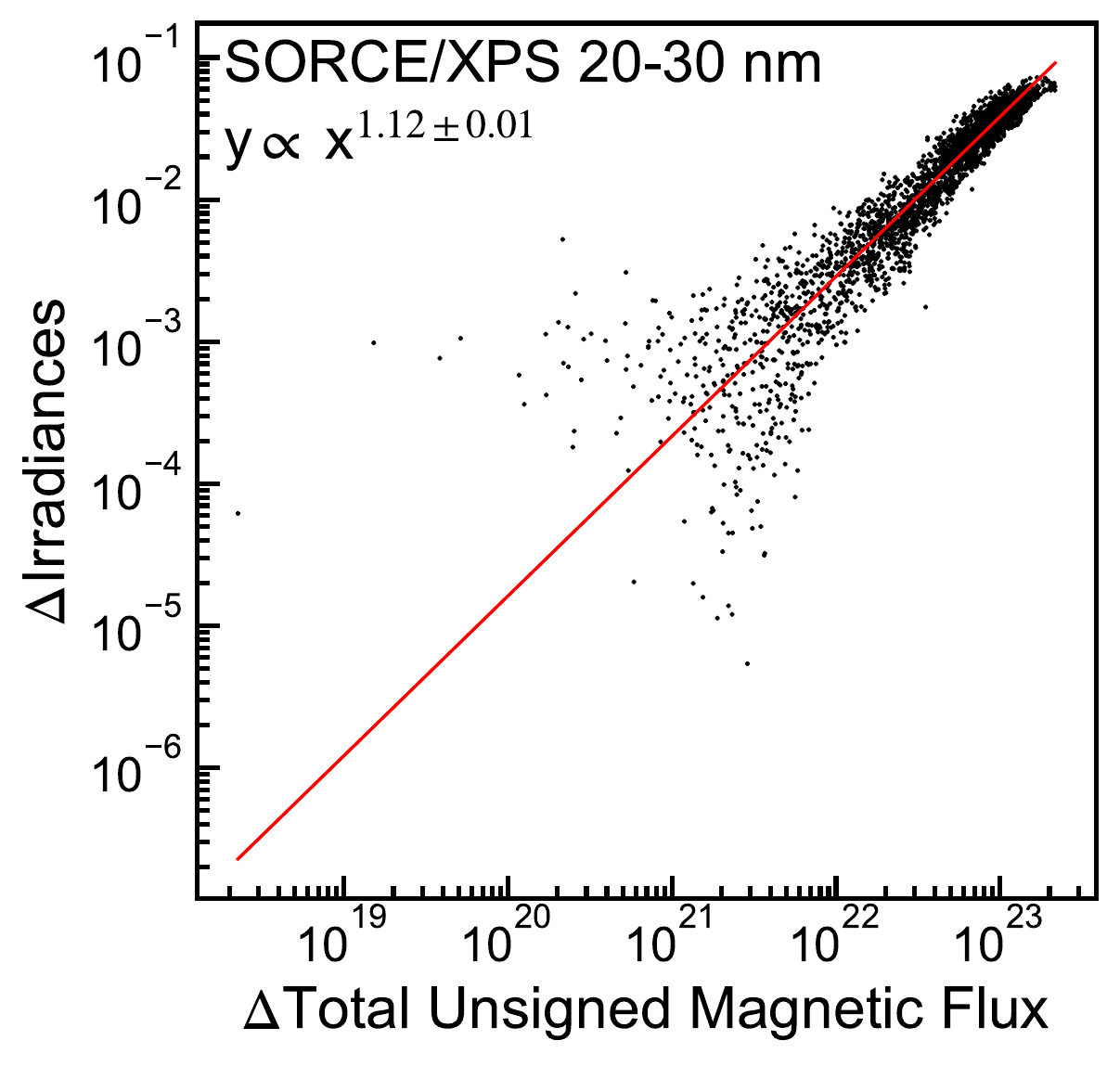}
\plotone{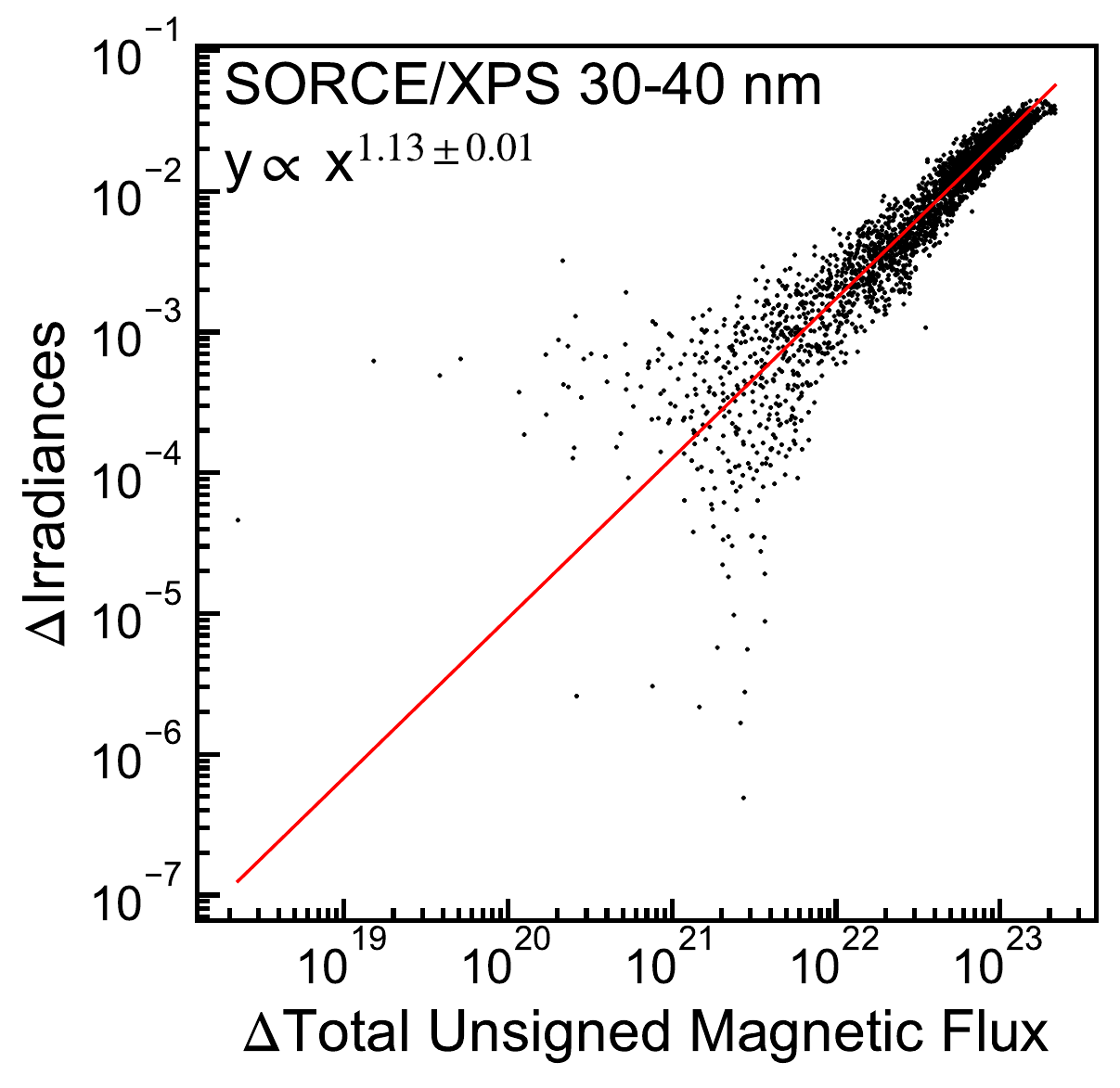}
\plotone{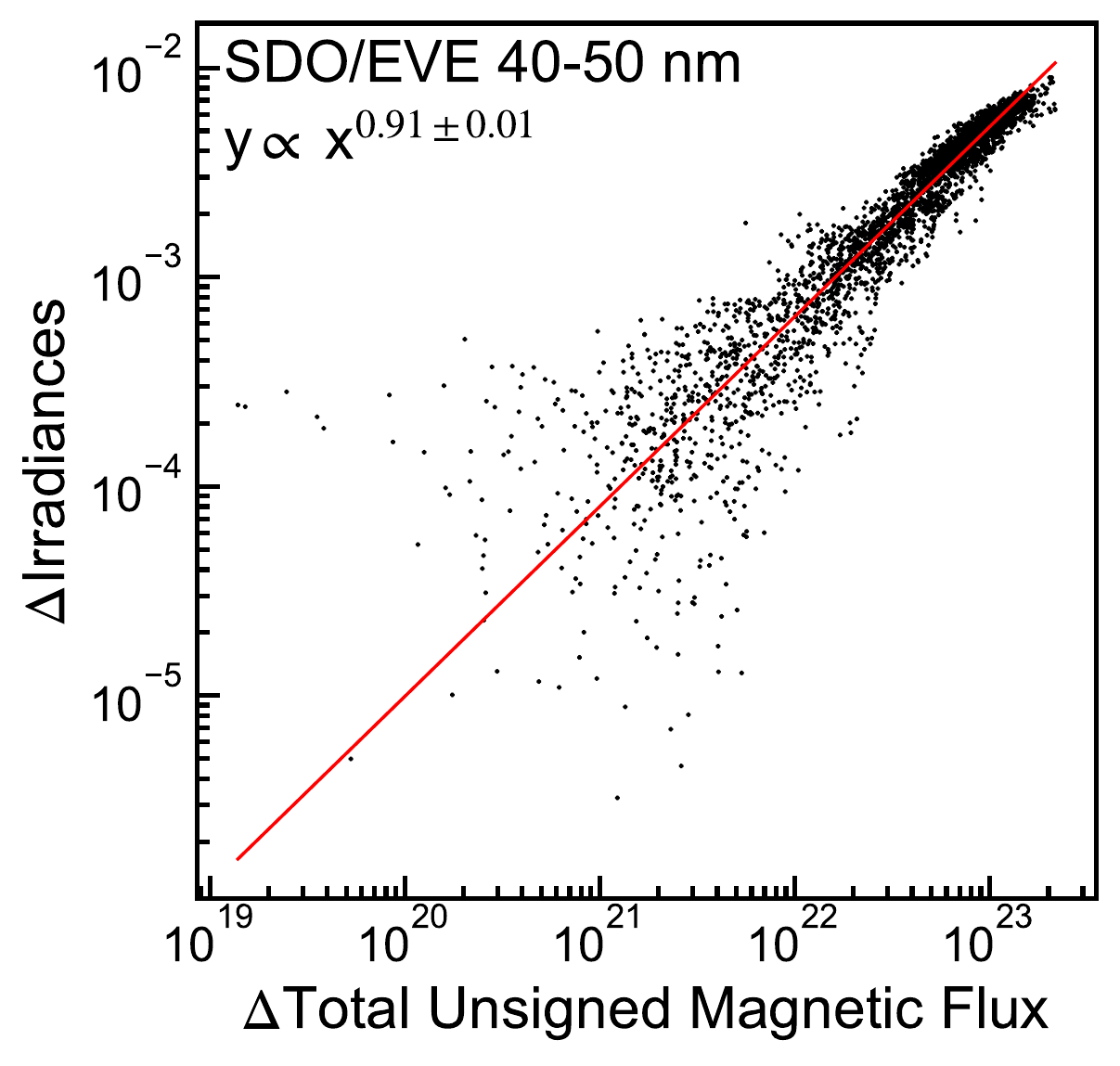}
\plotone{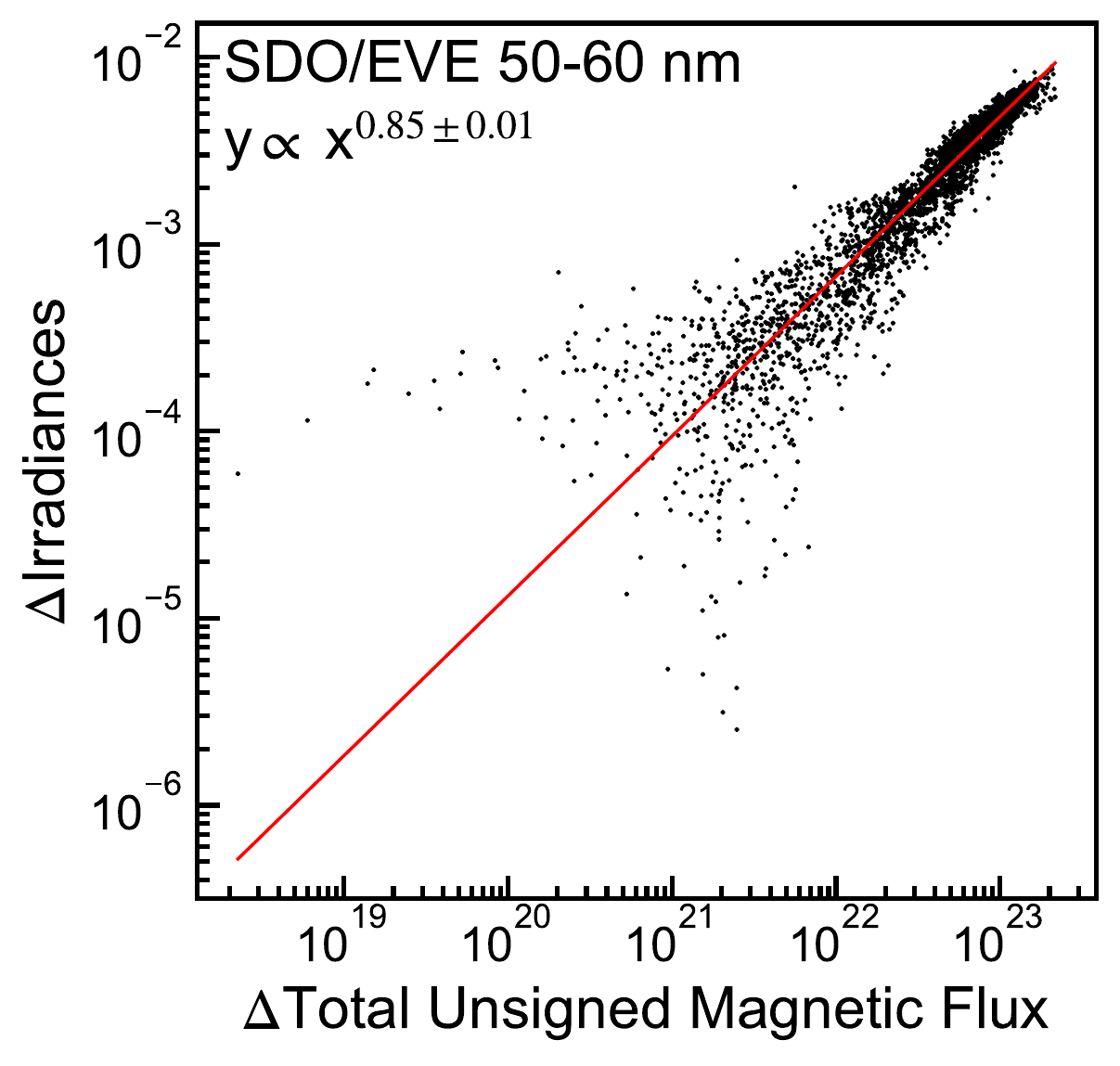}
\plotone{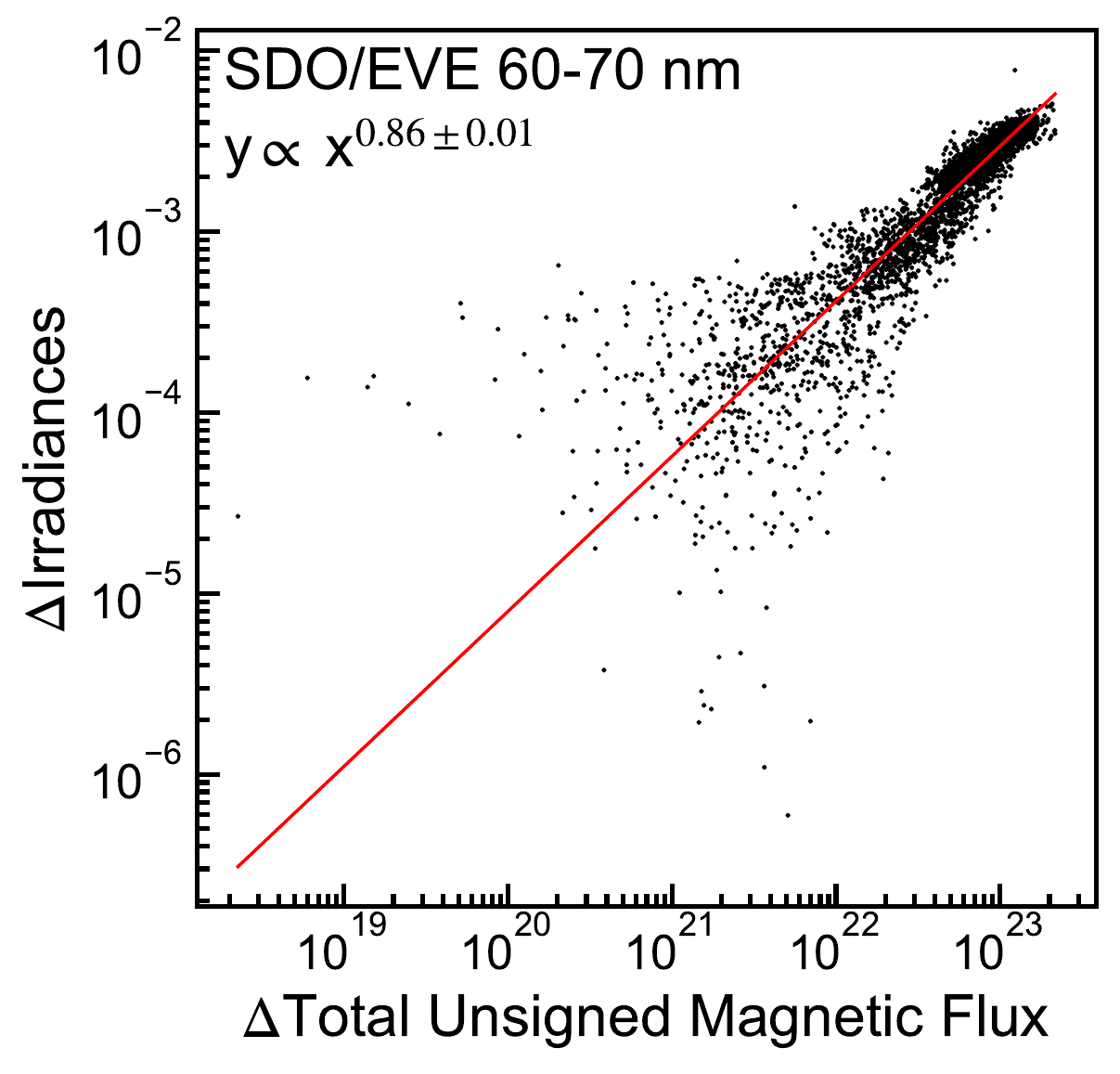}
\plotone{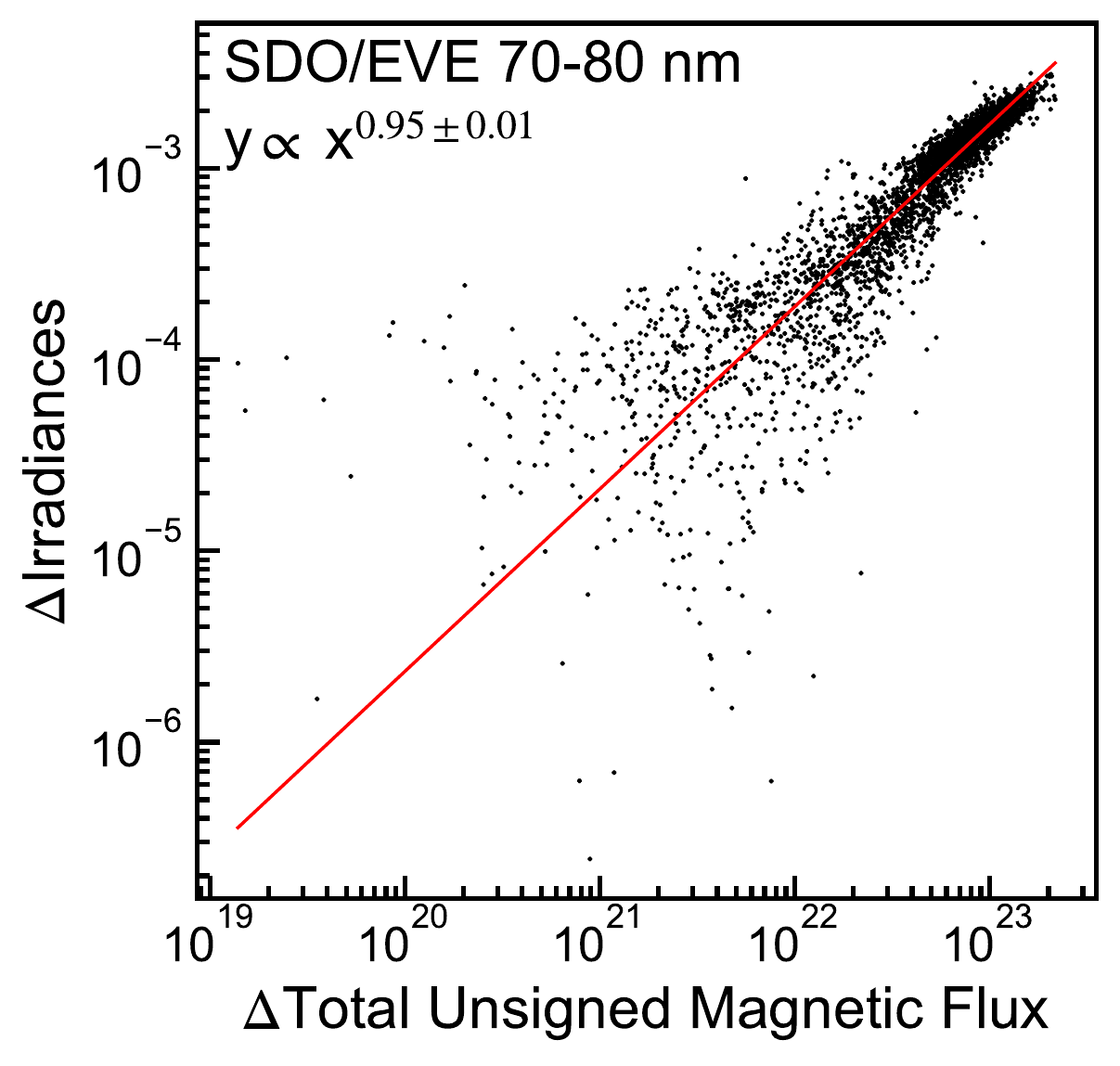}
\plotone{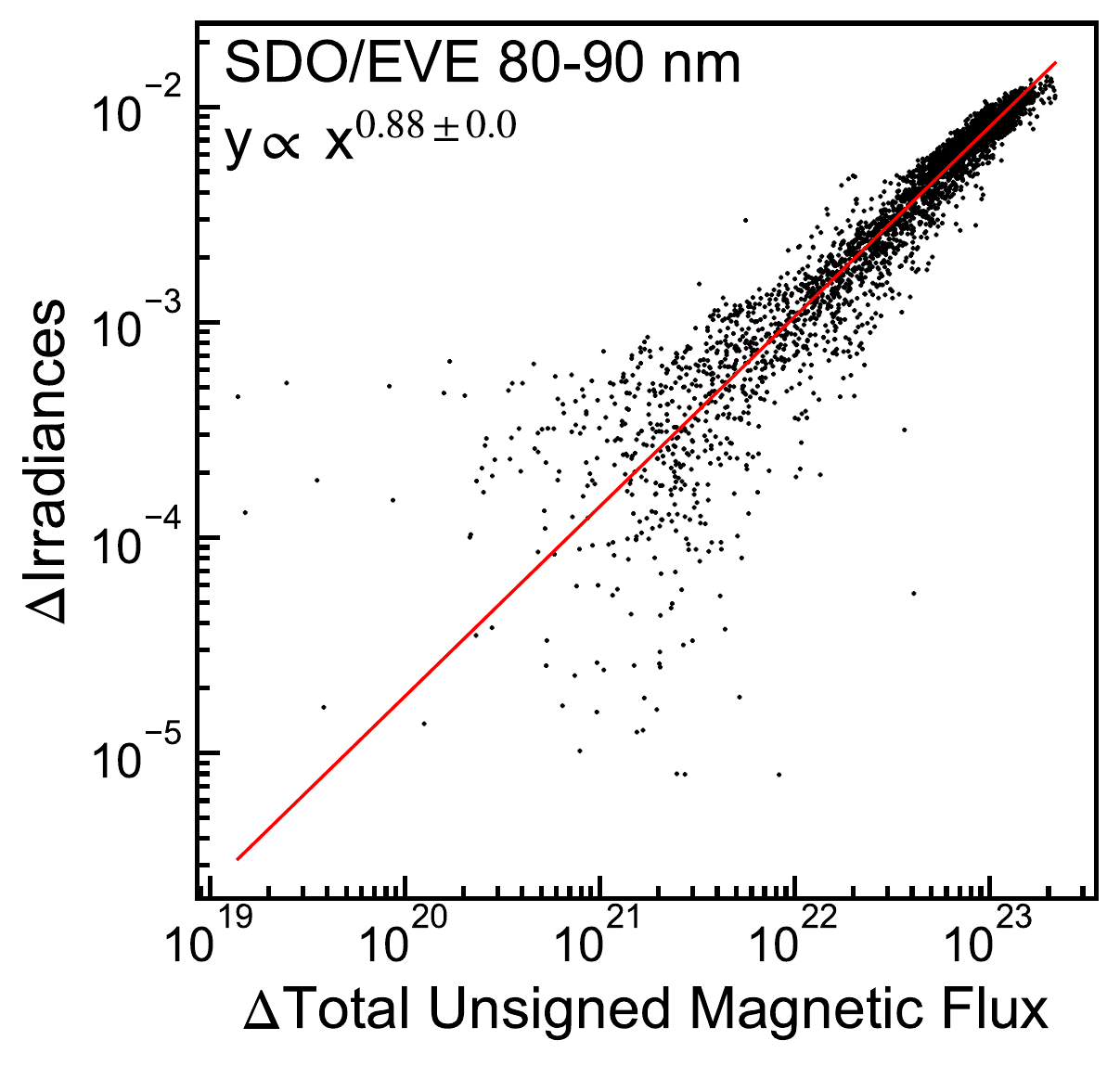}
\plotone{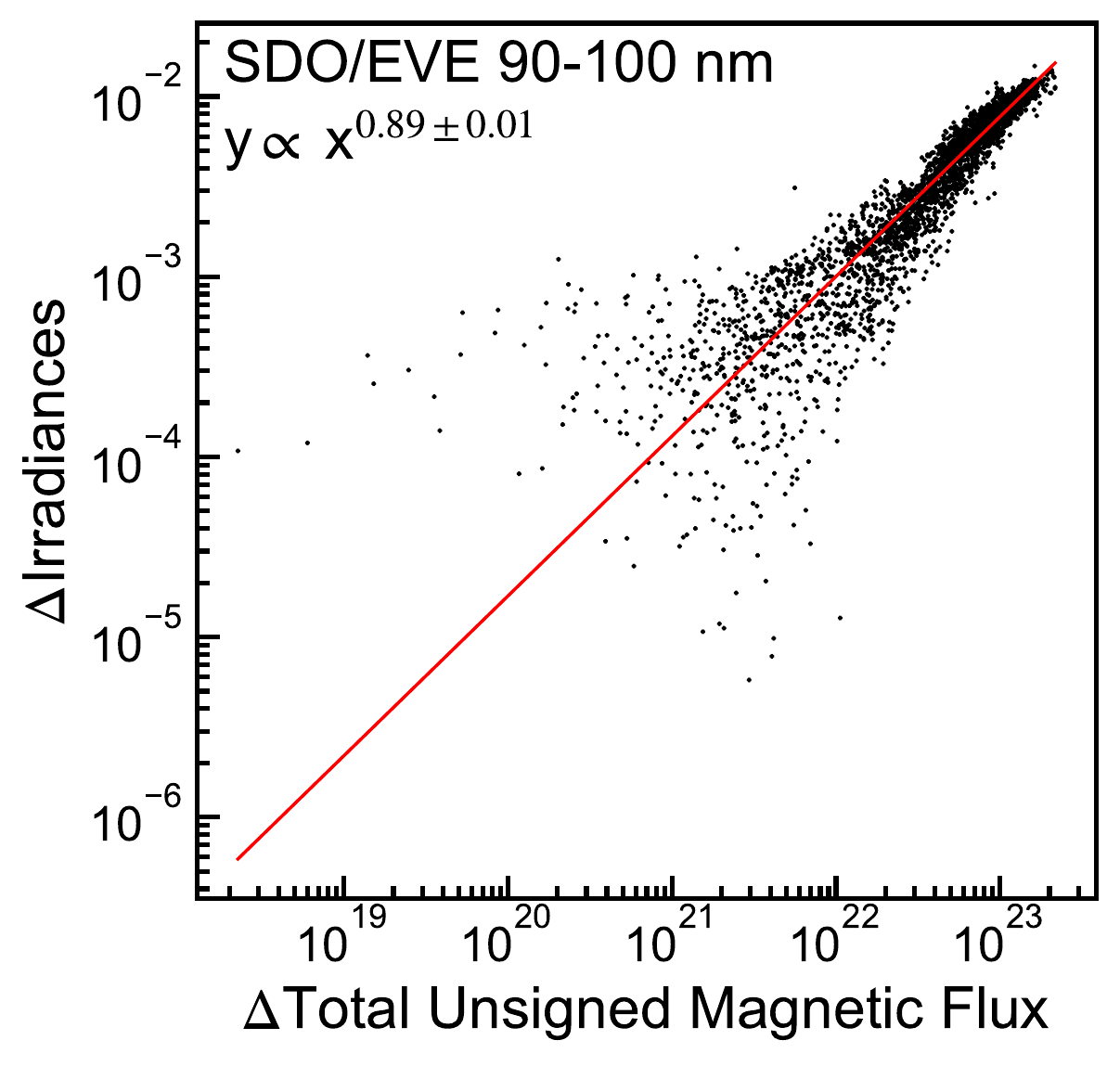}
\plotone{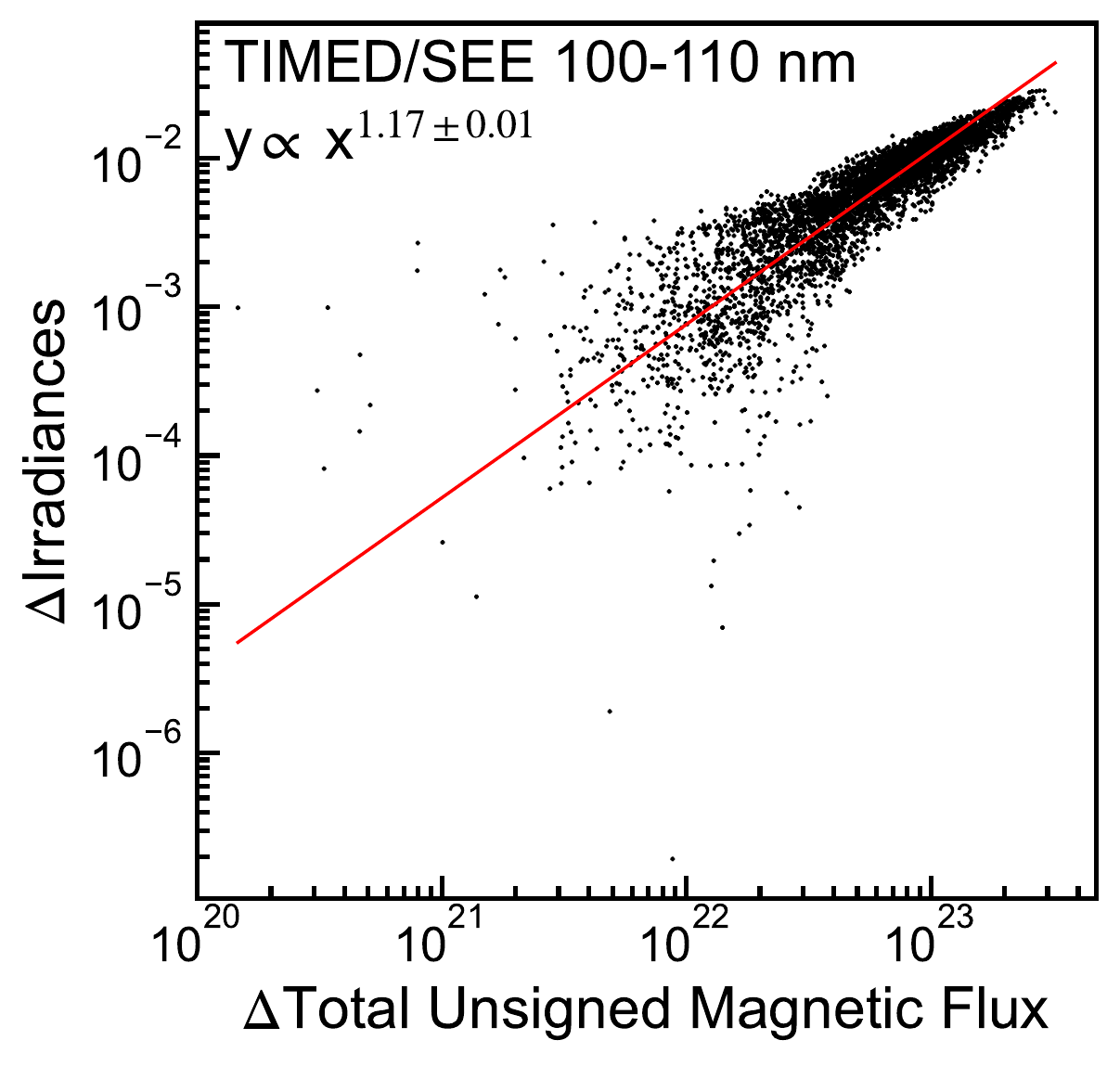}
\plotone{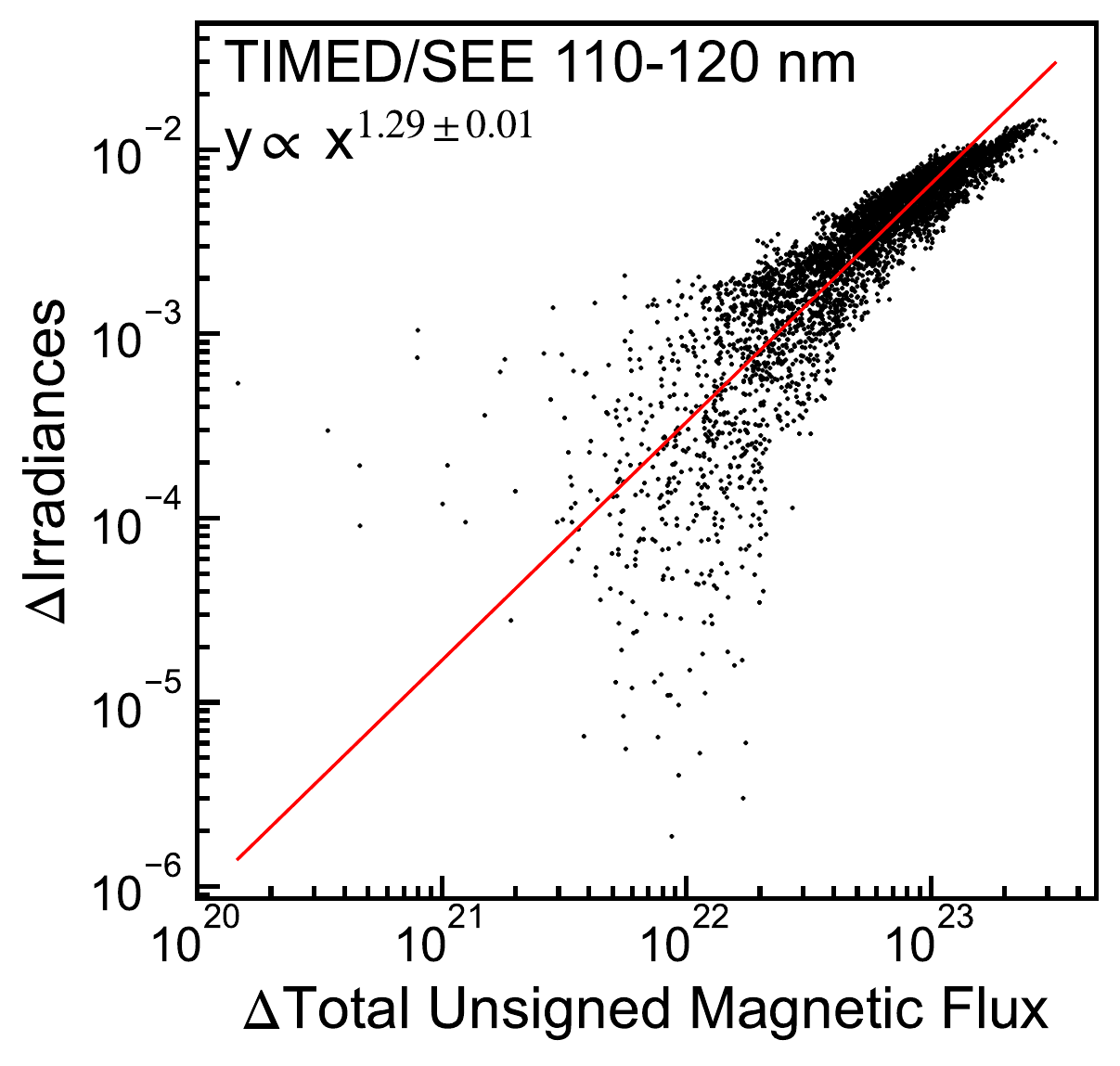}
\plotone{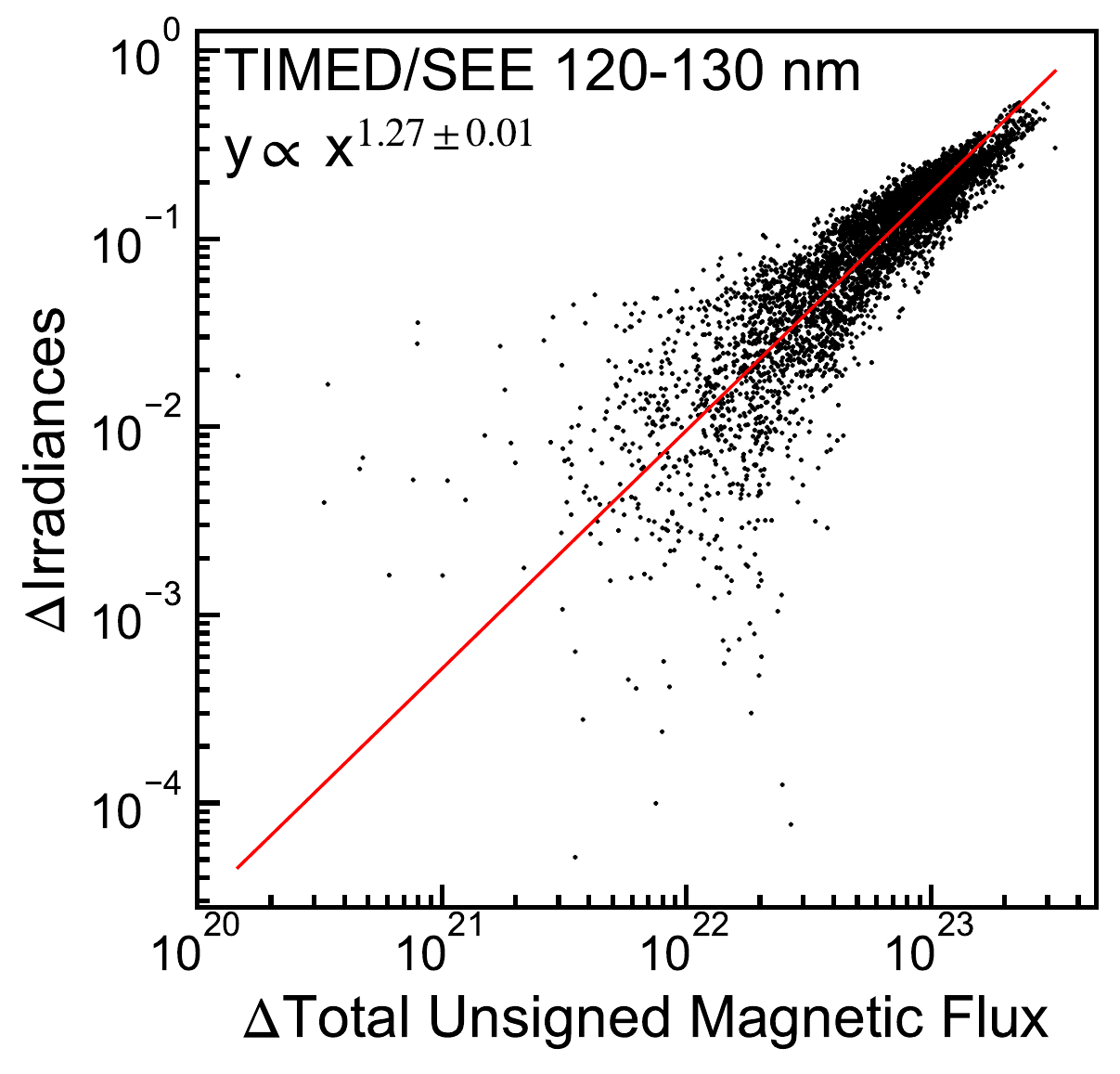}
\caption{
Relationship between total unsigned magnetic flux ($\Phi$ - $\Phi_0$) and Sun-as-a-star X-ray, EUV, and FUV fluxes averaged over each 10 nm bin ($I$ - $I_0$). Note that all values are subtracted by basal level ($\Phi_0$, $I_0$). The red line represents the fitting line, which was fitted in log-log space using \textsf{scipy.odr} installed in \textsf{python}. The fitted power-law indexes are indicated at the top left and listed in Table \ref{tab:2}.
}
\label{fig:3}
\end{figure*}

\addtocounter{figure}{-1}
\begin{figure*}
\epsscale{0.35}
\plotone{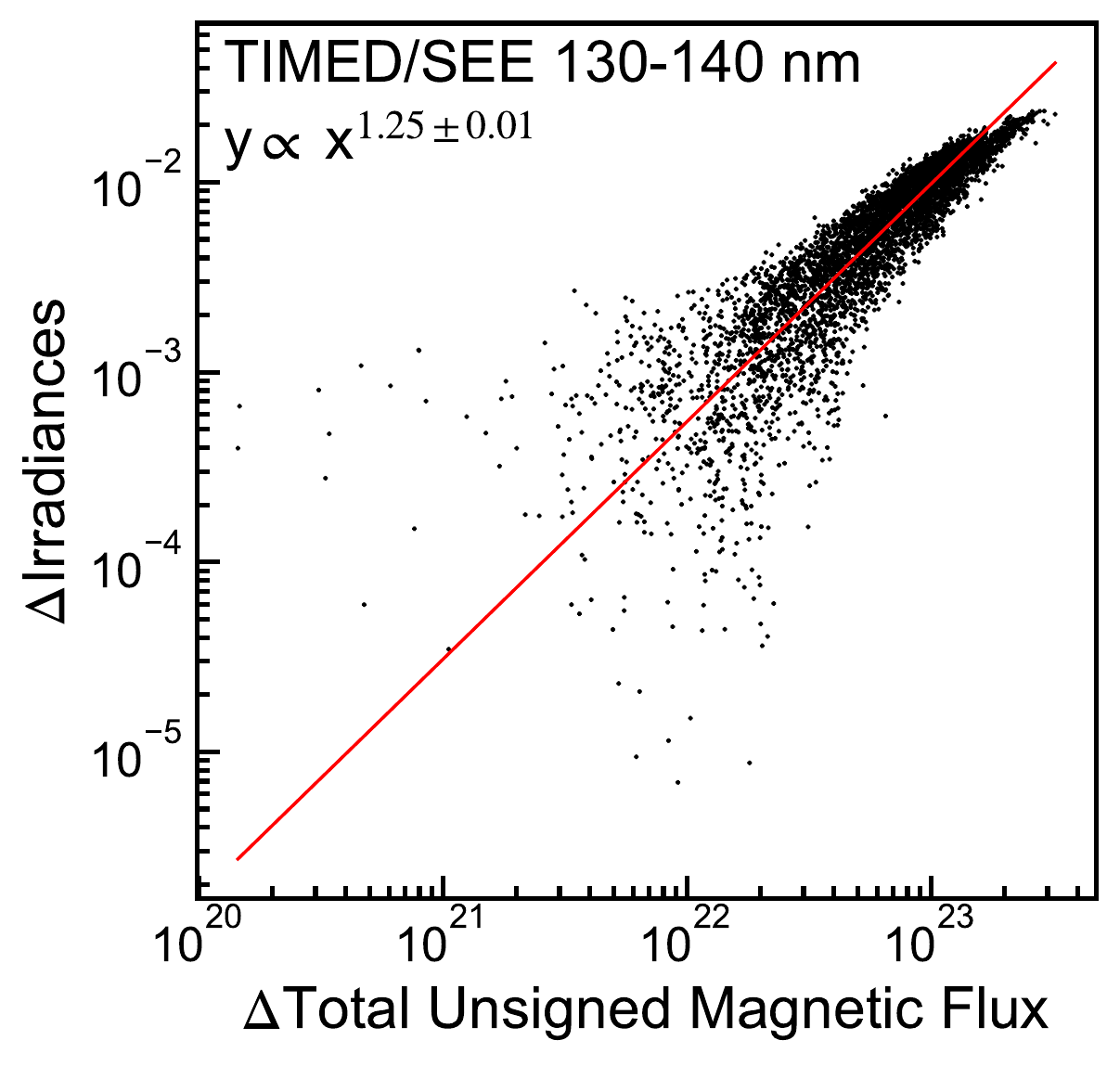}
\plotone{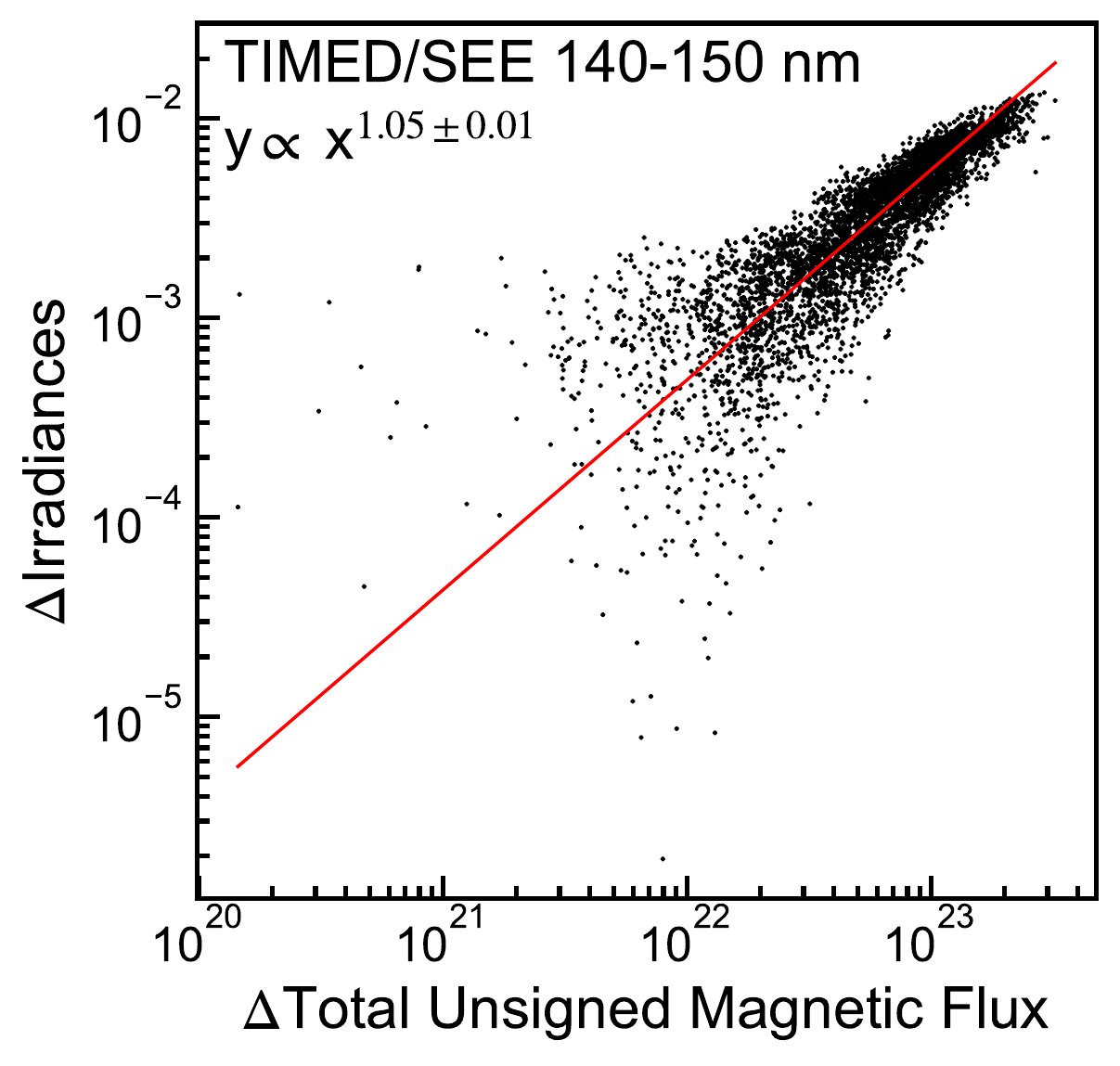}
\plotone{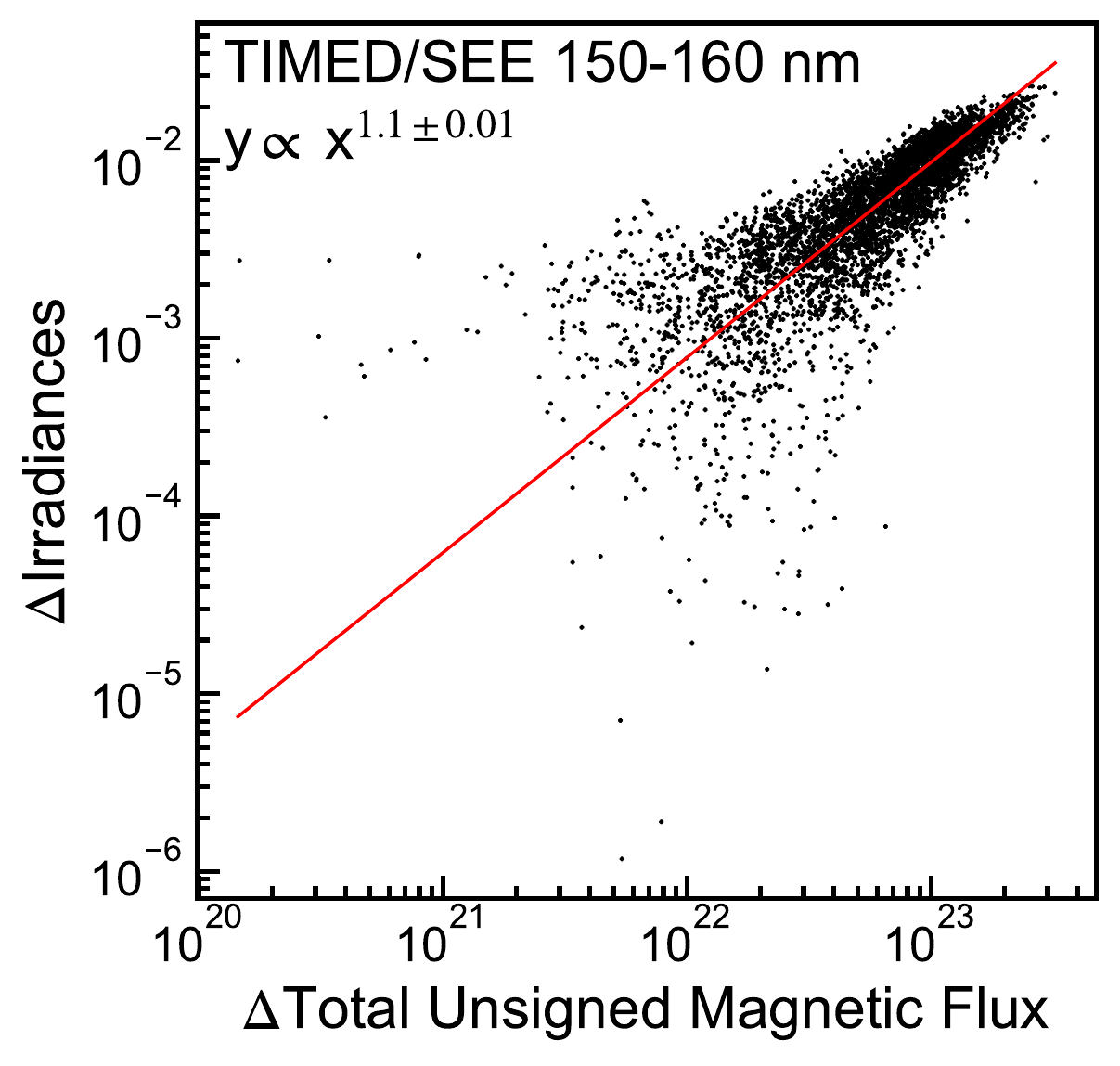}
\plotone{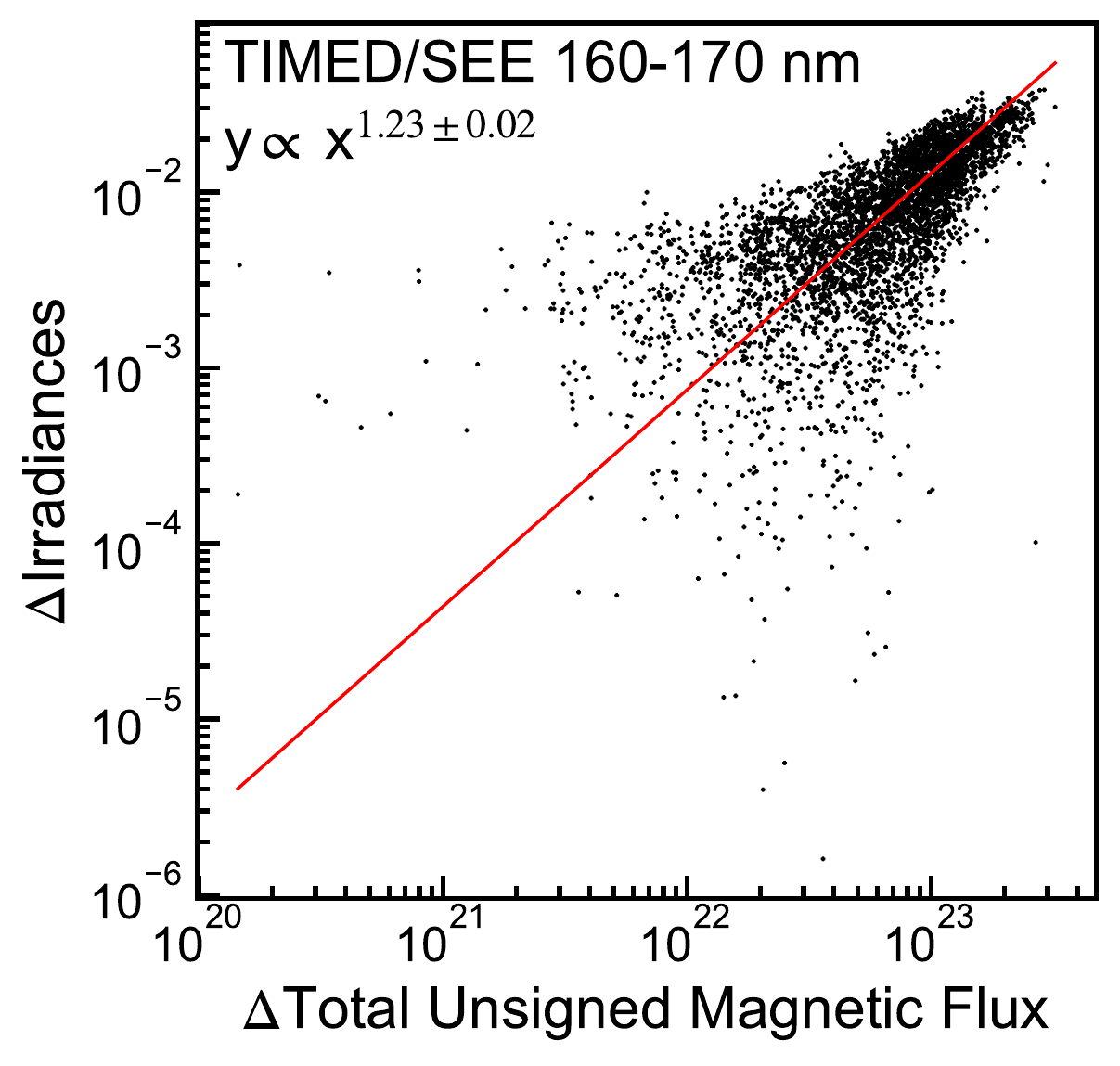}
\plotone{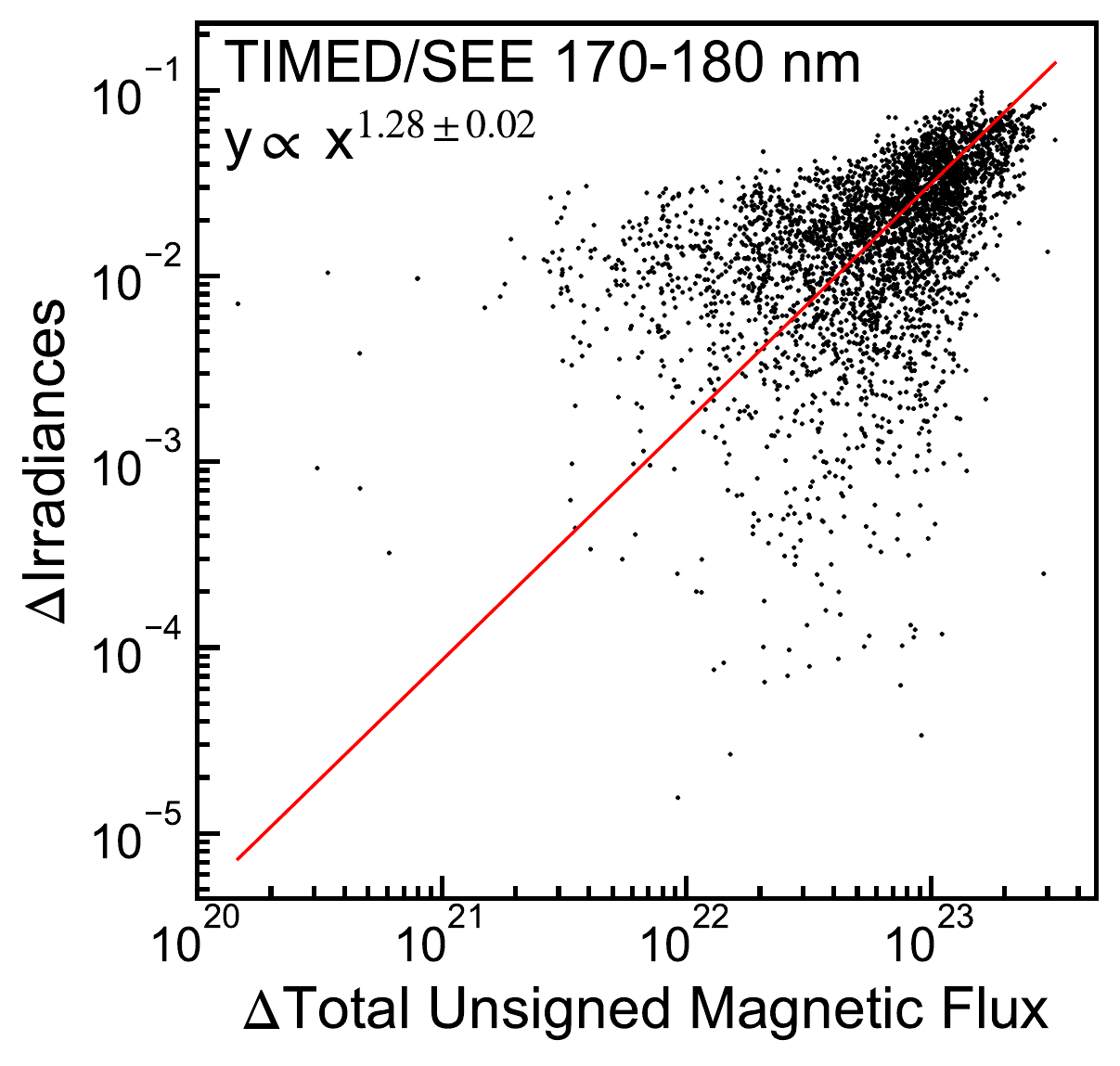}
\caption{{\it Continued.}.
}
\end{figure*}

As in Figure \ref{fig:3}, all of the Sun-as-a-star irradiance for all wavelength in X-ray, EUV, and FUV ($I$) have positive correlation with the total unsigned magnetic flux ($\Phi$).
To obtain the scaling relations between them, we follow the methods employed in \cite{2022ApJ...927..179T} and \cite{2022ApJS..262...46T}.
We first obtained the basal-level fluxes ($I_0$, $\Phi_0$) and daily variations (i.e., residuals: $\Delta I$ = $I$ - $I_0$, $\Delta \Phi$ = $\Phi$ - $\Phi_0$) and created a scatter plot of the residuals ($\Delta I$ vs. $\Delta \Phi$) in log-log space.
Figure \ref{fig:3} shows the examples of the scatter plots in 10 nm bin, but these plots can be made for each wavelength with a spectral resolution of 0.1--1 nm (and a sampling of 0.02--1 nm).
\cite{2022ApJ...927..179T} and \cite{2022ApJS..262...46T} discuss that the basal fluxes can be considered as background components that are always present. On the other hand, the residuals reflect the appearance of magnetic fields of active regions, and therefore a basal-level subtraction can produce a better response of XUV+FUV flux against the magnetic flux, regardless of the basal-level flux whose physical origin is poorly understood. 
At least, the basal-level subtraction is necessary to fit with a single power-law relation because they often do not show clear linear relations in log-log space if we use the total magnetic flux.
Another advantage of the basal-level subtraction is that the wide dynamic ranges for scatter plots in log-log space enable us to derive higher correlated power-law relations.

In this study, basal fluxes are defined as the median values around activity minimum. The definition of activity minimum depends on the instruments because each has a different available activity minimum as we mentioned above (see Table \ref{tab:1}). 
For SORCE/XPS and SDO/EVE data, we defined the activity minimum as the days with the following conditions: (1) the solar minimum from March 2019 to February 2020 (the same definition as \cite{2022ApJ...927..179T} and \cite{2022ApJS..262...46T}); (2) when the total sunspot number is 0; and (3) when the total unsigned magnetic flux $\Phi$ is less than the 5th percentile for the all data used in the analysis. 
Note that the total number of data is different for SDO/HMI\&EVE, and SORCE/XPS respectively, so each basal-level value is not completely simultaneously obtained, but this would be a slight difference and not change the result significantly.
The number of SDO/HMI\&EVE data satisfying these conditions is 94 days, while that of SORCE/XPS is 78. 
In this study, the basal value of the total unsigned magnetic flux is 1.18$\times 10^{23}$ Mx.

For TIMED/SEE data, we defined the activity minimum as the days with the following conditions: (1) the solar minimum in 2009; (2) when the total sunspot number is 0; and (3) when the total unsigned magnetic flux $\Phi$ is less than the 10th percentile for the all data of SDO/HMI and SoHO/MDI used in the analysis (from 2002 to 2021). The thresholds are adjusted to get enough number of data and the number of days satisfying these was 32.
After the above subtraction of the basal levels, we abandoned the data after 2017 January 1 due to the calibration problem.

\begin{figure*}
\epsscale{0.5}
\plotone{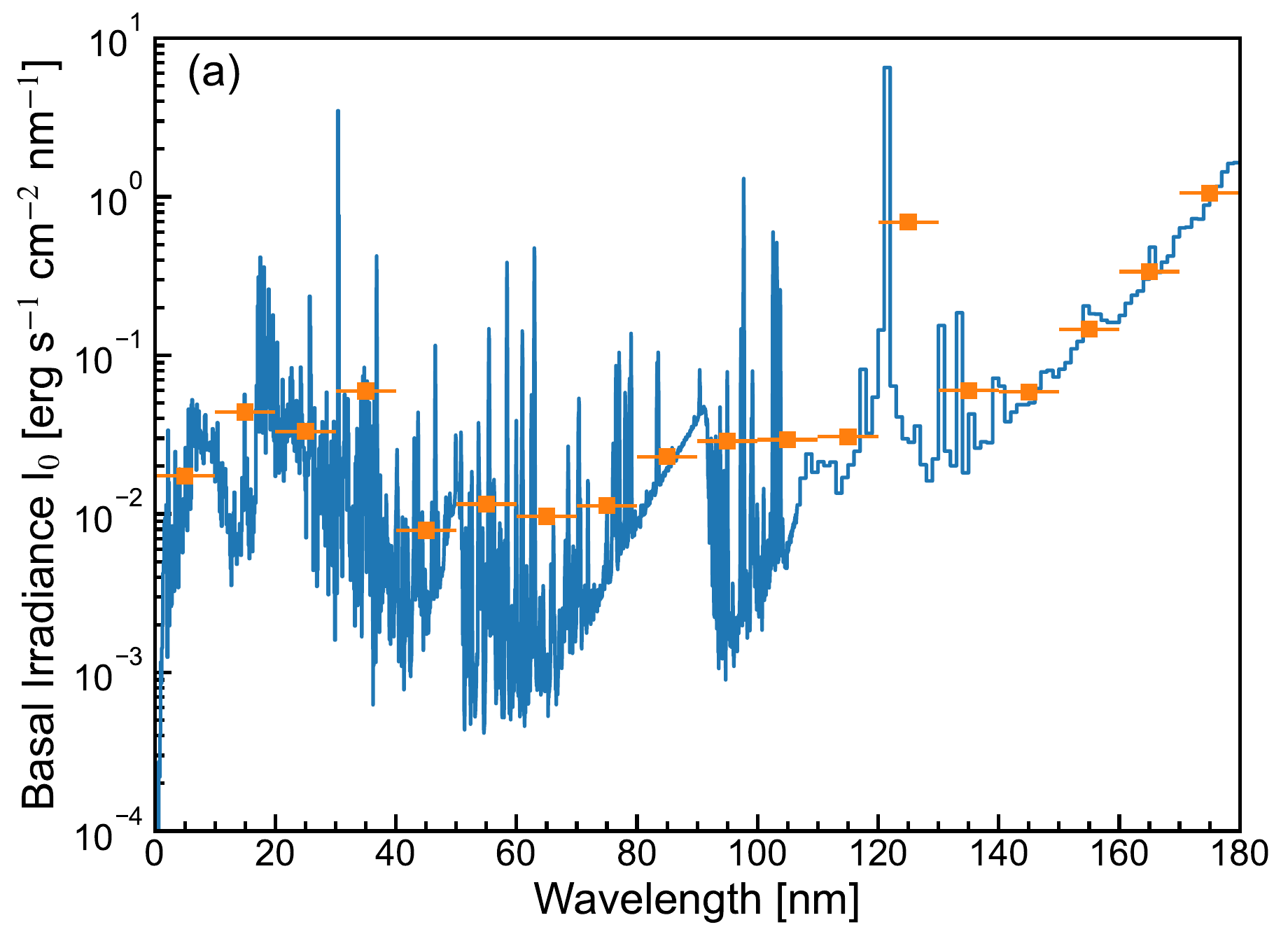}
\plotone{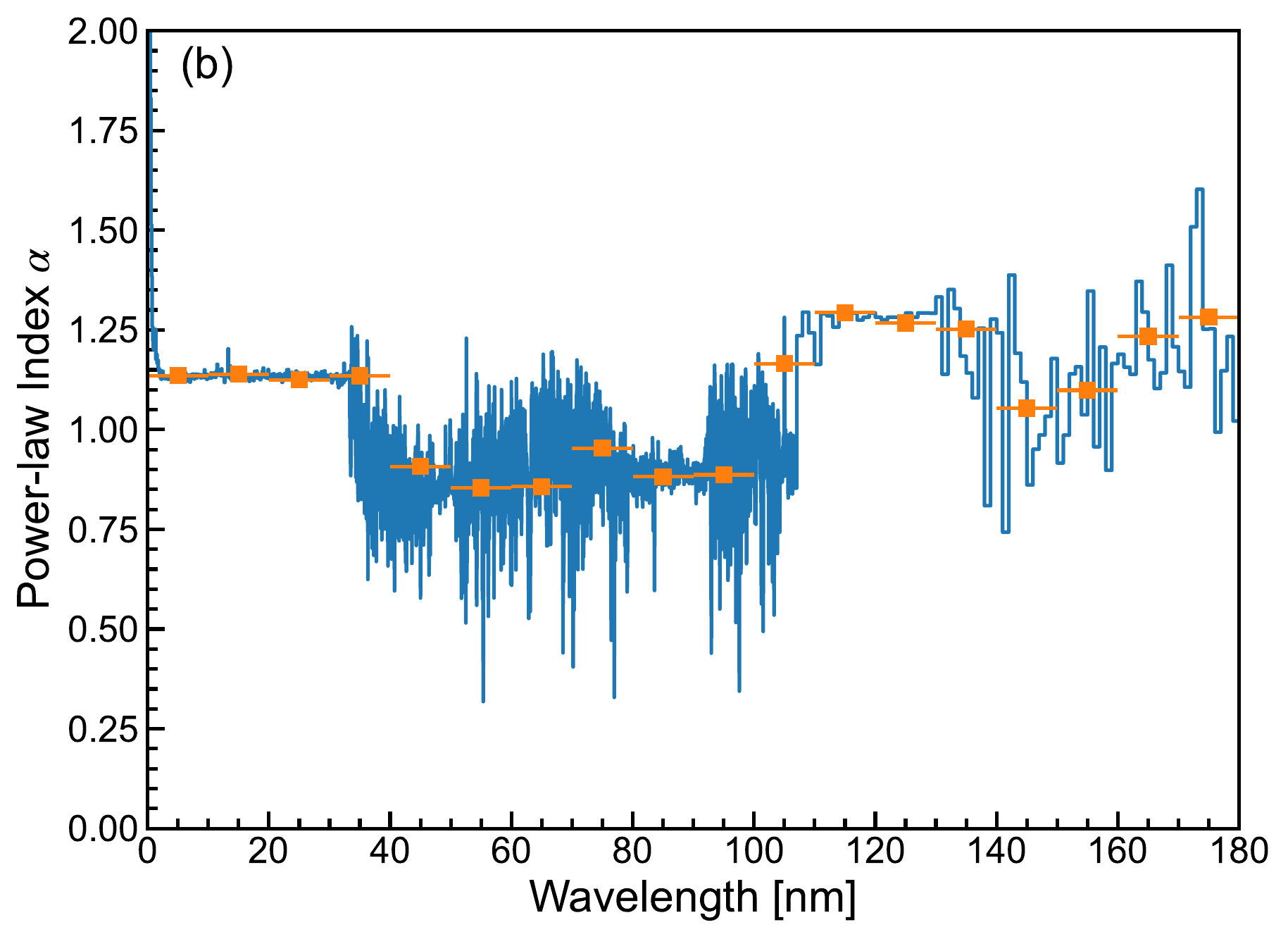}
\plotone{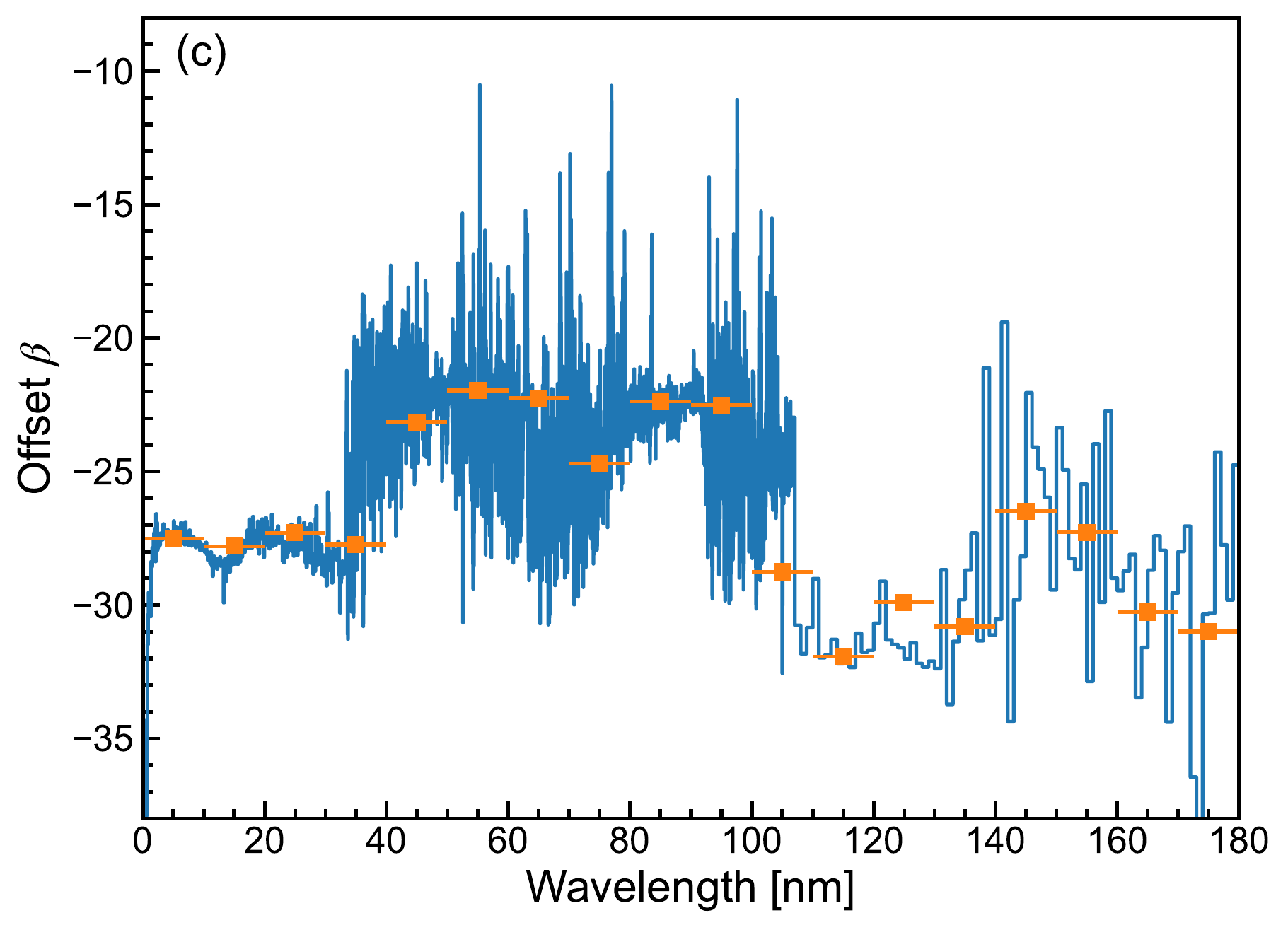}
\plotone{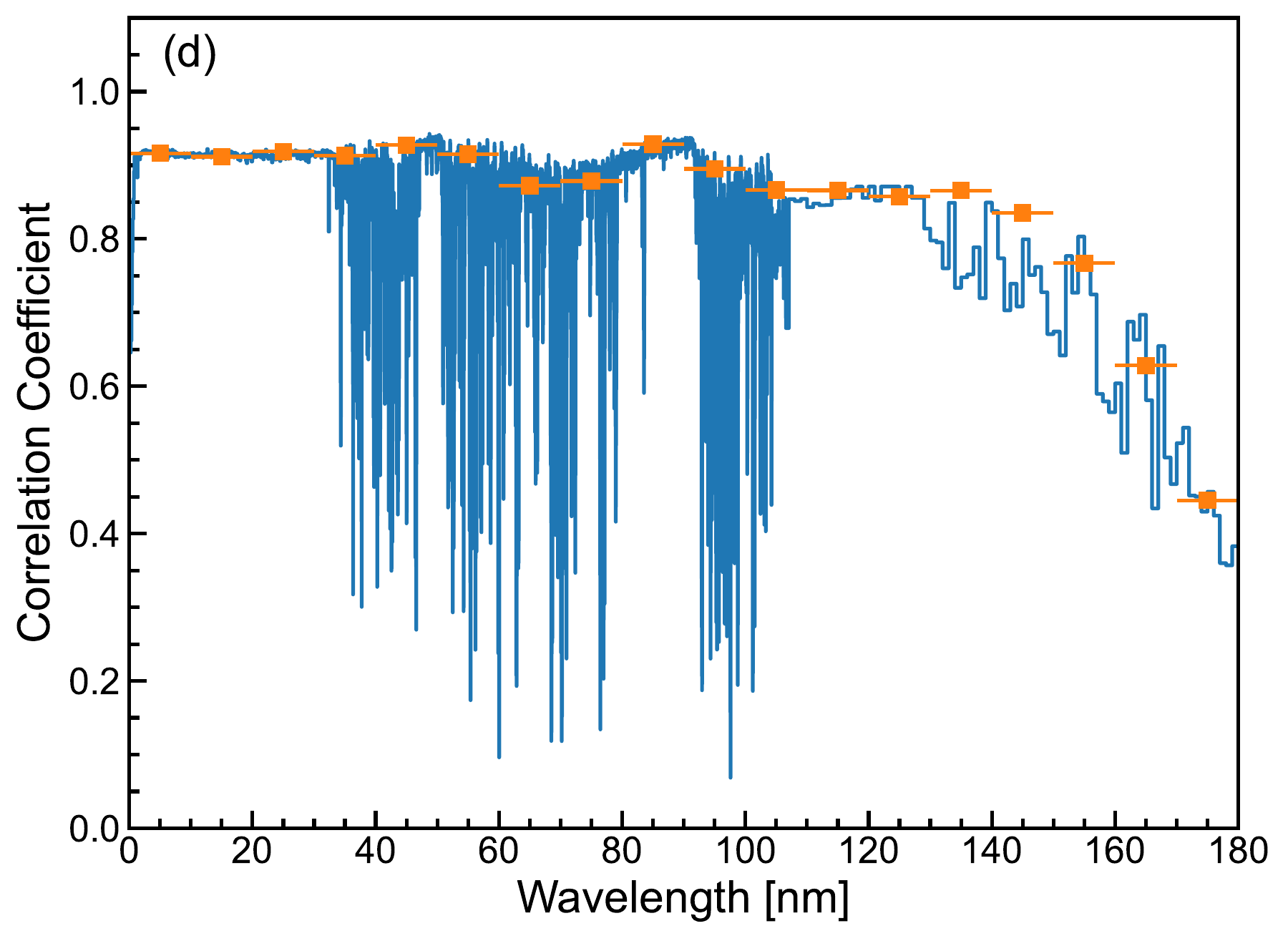}
\caption{Fitted parameters of power-law relations between the Sun-as-a-star irradiance and total unsigned magnetic flux (in the formulae of $I(\lambda) = I_{0}(\lambda) + 10^{\beta(\lambda)} (\Phi - \Phi_{0})^{\alpha(\lambda)}$) and the correlation coefficient as a function of wavelength. Each panel shows (a) the basal level irradiance I$_{0}$, (b) the power-law index $\alpha(\lambda)(\lambda)$, (c) the offset $\beta(\lambda)$, and (d) the correlation coefficient. The blue line is the fitted parameters at the best wavelength resolution and samplings (Table \ref{tab:3}), and the orange dots are the fitted parameters at the 10-nm sampling (Table \ref{tab:2}).
}
\label{fig:4}
\end{figure*}

Power-law relations between the Sun-as-a-star X-ray, EUV, and FUV irradiance ($\Delta I$($\lambda$) = $I(\lambda)$ -- $I_{0}(\lambda)$) and total unsigned magnetic flux ($\Delta \Phi$ = $\Phi$ -- $\Phi_{0}$) have been calculated for each wavelength $\lambda$ with a spectral resolution of 0.1--1 nm.
The linear fit was applied to the logarithmic data (log$\Delta I$, log$\Delta \Phi$) by using \textsf{scipy.odr} installed in \textsf{python} in the following equation
\begin{eqnarray}
{\rm log} \Delta I &=& \alpha(\lambda) {\rm log} \Delta \Phi + \beta(\lambda),
\label{eq:1}
\end{eqnarray}
where parameters $\alpha(\lambda)$ and $\beta(\lambda)$ are the fitting parameters. We applied a differential weighting method which puts more weight on larger data in log space. The weights for log$\Delta I $ and log$\Delta \Phi$ were set as log$\Delta I$ -- min(log$\Delta I$) +10$^{-6}$ and log$\Delta \Phi$ -- min(log$\Delta \Phi$) +10$^{-6}$, respectively. 
Then, Equation (\ref{eq:1}) can be expressed as
\begin{eqnarray}
I(\lambda) &=& I_{0}(\lambda) + 10^{\beta(\lambda)} (\Phi - \Phi_{0})^{\alpha(\lambda)}.
\label{eq:2}
\end{eqnarray}
This fitting was performed for all wavelengths $\lambda$ and the parameters $\alpha(\lambda)$ and $\beta(\lambda)$ for each wavelength are summarized in Table \ref{tab:3}. 
Pearson product-moment correlation coefficient (C.C.) was also calculated by using \textsf{corrcoef} installed in \textsf{numpy}.
Table \ref{tab:2} shows the fitting results for fluxes averaged over 10-nm bins and Tables \ref{tab:3} shows those for the best spectral bins of 0.1--1 nm.
Figure \ref{fig:3} show the relationship between $\Delta I$ and $\Delta \Phi$ for 10-nm bins as examples. 
The 10-nm-bin data were prepared because (1) the correlation coefficient is low at some wavelengths in the case of the best spectral resolution and (2) in some cases, when considering the effects on planetary atmospheres, such a high wavelength resolution may not be necessary, and the data in 10-nm bins can be more useful.
Figure \ref{fig:4} shows the result of the fitting in the formula of Equation (\ref{eq:2}) ($I_{0}(\lambda)$, $\beta(\lambda)$, and $\alpha(\lambda)$) and correlation coefficient for each wavelength (blue lines) and each 10-nm band (orange points).
Note that in some panels of Figure \ref{fig:3}, a trend in the largest values of flux to be below the fitted red line. In this study, we did a fit that extracts the trend that most of the data points follow. When the flux is extremely high, there are few data points, and the possibility that saturation occurs cannot be dismissed. In particular, the scattering of data in FUV is so large  that it is possible that the overall trend is not being tracked correctly.

\begin{figure*}
\epsscale{0.7}
\plotone{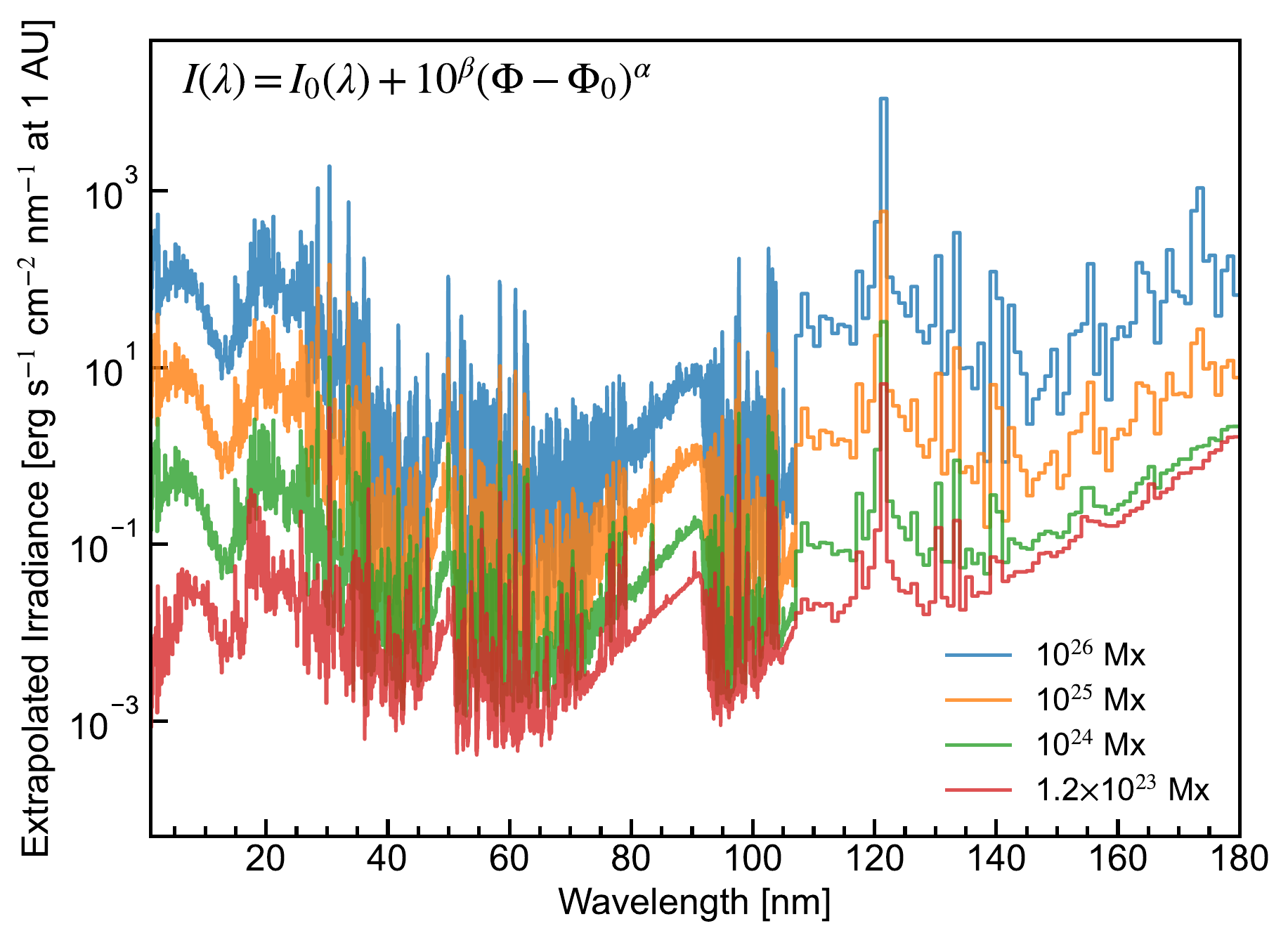}
\caption{Example of the extrapolated spectra in X-ray, EUV, and FUV estimated from the Equation (\ref{eq:2}) ($I(\lambda) = 10^{\beta(\lambda)} (\Phi - \Phi_0)^{\alpha(\lambda)}+I_0(\lambda) $) and a given total unsigned magnetic flux of 10$^{24}$, 10$^{25}$, and 10$^{26}$ Mx.  The model spectrum at solar minimum value of 1.2$\times$10$^{23}$ Mx is also plotted with red color as a reference.
}
\label{fig:5}
\end{figure*}


\subsection{Power-law index $\alpha$}\label{sec:3-3}

The most important parameter is the power-law index $\alpha(\lambda)$ of Equation (\ref{eq:2}). As in Figure \ref{fig:4}(b), the power-law indexes $\alpha(\lambda)$ appear to be generally wavelength-dependent or possibly instrument-dependent.
For X-ray and short EUV wavelength ($<$33 nm), the power-law indices are above unity and almost constant in wavelength at approximately $\sim1.14$. 
This trend was also seen in the emission lines in our previous work \citep{2022ApJS..262...46T}.
This wavelength range is based on the SORCE/XPS data, but it should be noted that SORCE/XPS is only a model spectrum based on the CHIANTI atomic database \citep{1997A&AS..125..149D}, which may lead to the constant power-law index $\alpha(\lambda)$. 
For longer EUV wavelength (33-106.6 nm), we used the SDO/EVE data. The power-law indices $\alpha(\lambda)$ drop to around or below unity in the EUV range. 
We notice that the scatter in $\alpha(\lambda)$ seen in Figure \ref{fig:4}(b) does not clearly represent the different behaviors of the emission lines or continuum.

\cite{2022ApJ...927..179T} and \cite{2022ApJS..262...46T} show that as the temperature decreases, the power-law index of the corresponding emission lines/bands also decreases; the $\alpha$ values are above unity for coronal lines ($\sim$10$^{6}$ K) while the $\alpha$ values lie below unity for chromospheric lines ($\sim$10$^{4}$ K). In this study, as for the X-ray to EUV range, we can see a similar trend that as the wavelength increases (i.e., the temperature of emission line formation decreases from coronal temperature to transition region temperature), the power-law index drops from above-unity to around-unity. 
On the other hand, as the wavelength increases to FUV range, the power-law index increases up to above-unity $\sim$1-1.3.
This is different from what we expected physically because the emission temperature of FUV range is generally lower than that of EUV range \citep[][]{2012ApJ...745...25L}.
This problem may have the following causes. First, TIMED/SEE does not have high wavelength resolution and does not resolve emission lines sufficiently. Second, when correlating TIMED/SEE flux and magnetic field, the magnetic field data is a combination of SDO/HMI and SoHO/MDI data, which is not homogeneous in time. Third, the longer the wavelength of FUV range, the lower the correlation coefficient becomes. 
For these reasons, we may need to be careful to the data of FUV range, although we did our best analysis using the currently available datasets. 
An alternative analysis of SORCE/SOLSTICE data for FUV range is also done in Appendix \ref{app:a}, but we need to keep in mind that both instruments would have their own problems in their data (see Section \ref{data:see}).



\section{Reconstruction of XUV fluxes for young Sun-like stars}\label{sec:4}

\subsection{Reconstructed vs. Observed Stellar Spectra}\label{sec:4-1}

\begin{figure*}
\epsscale{0.5}
\plotone{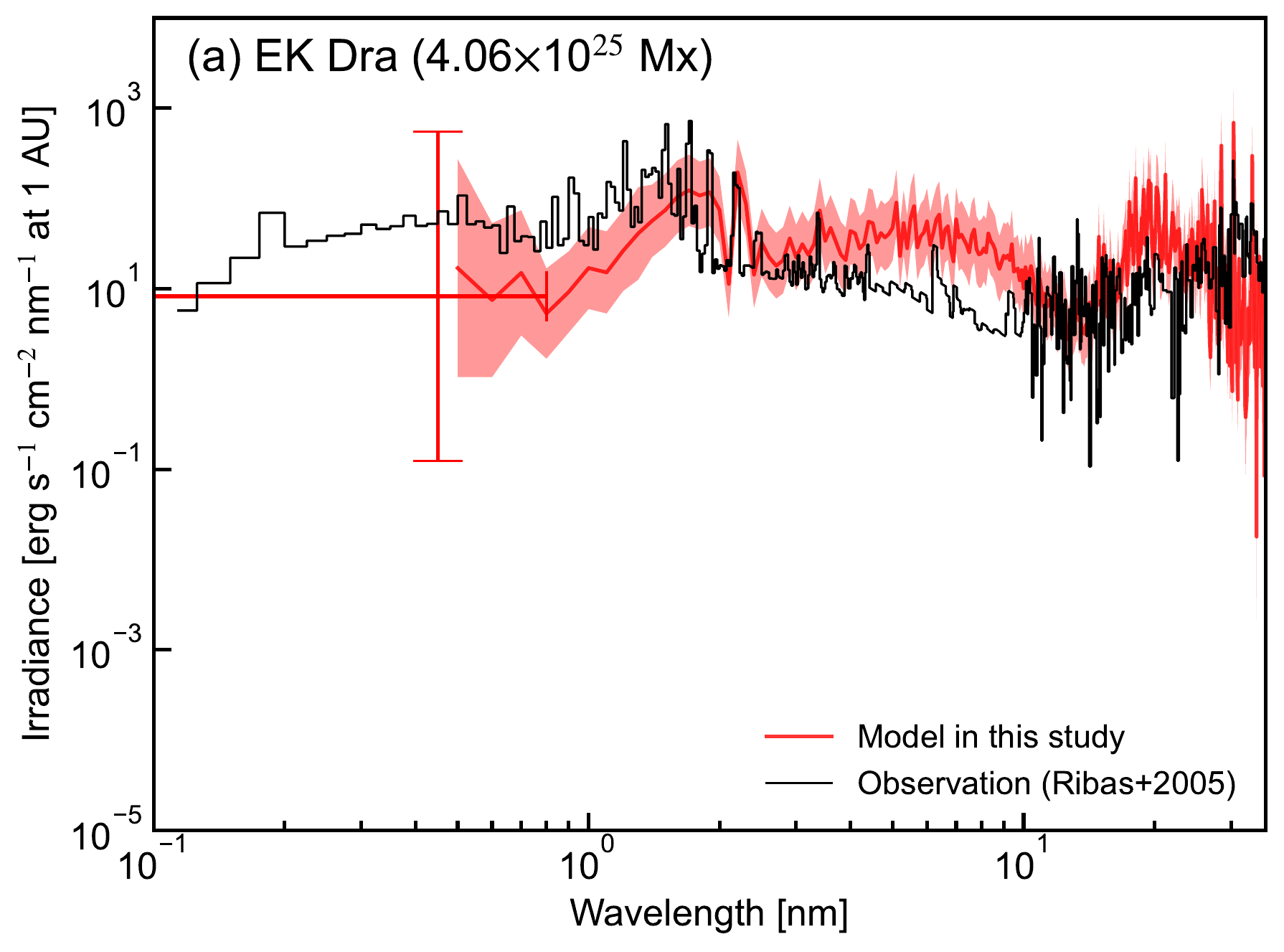}
\plotone{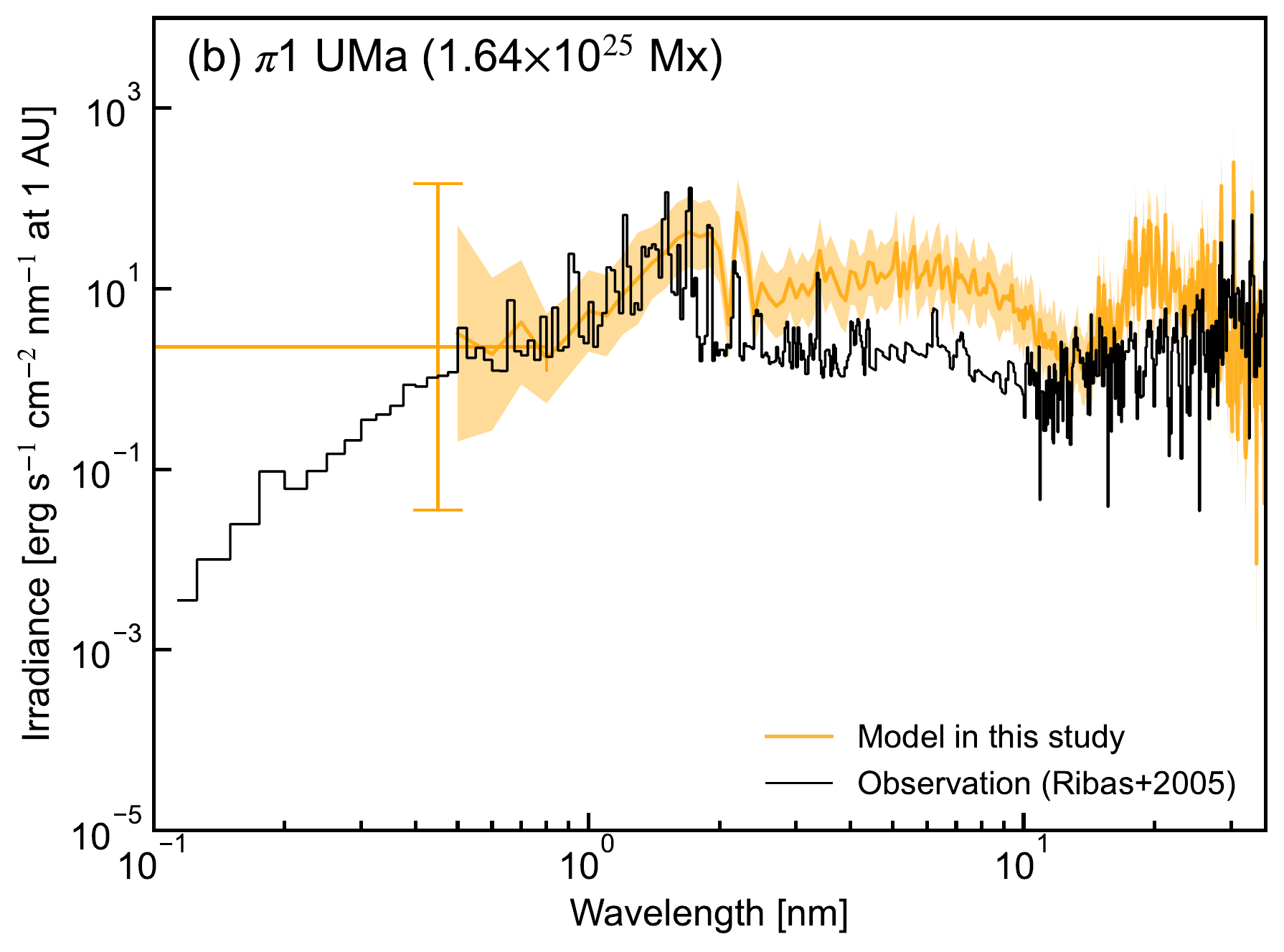}
\plotone{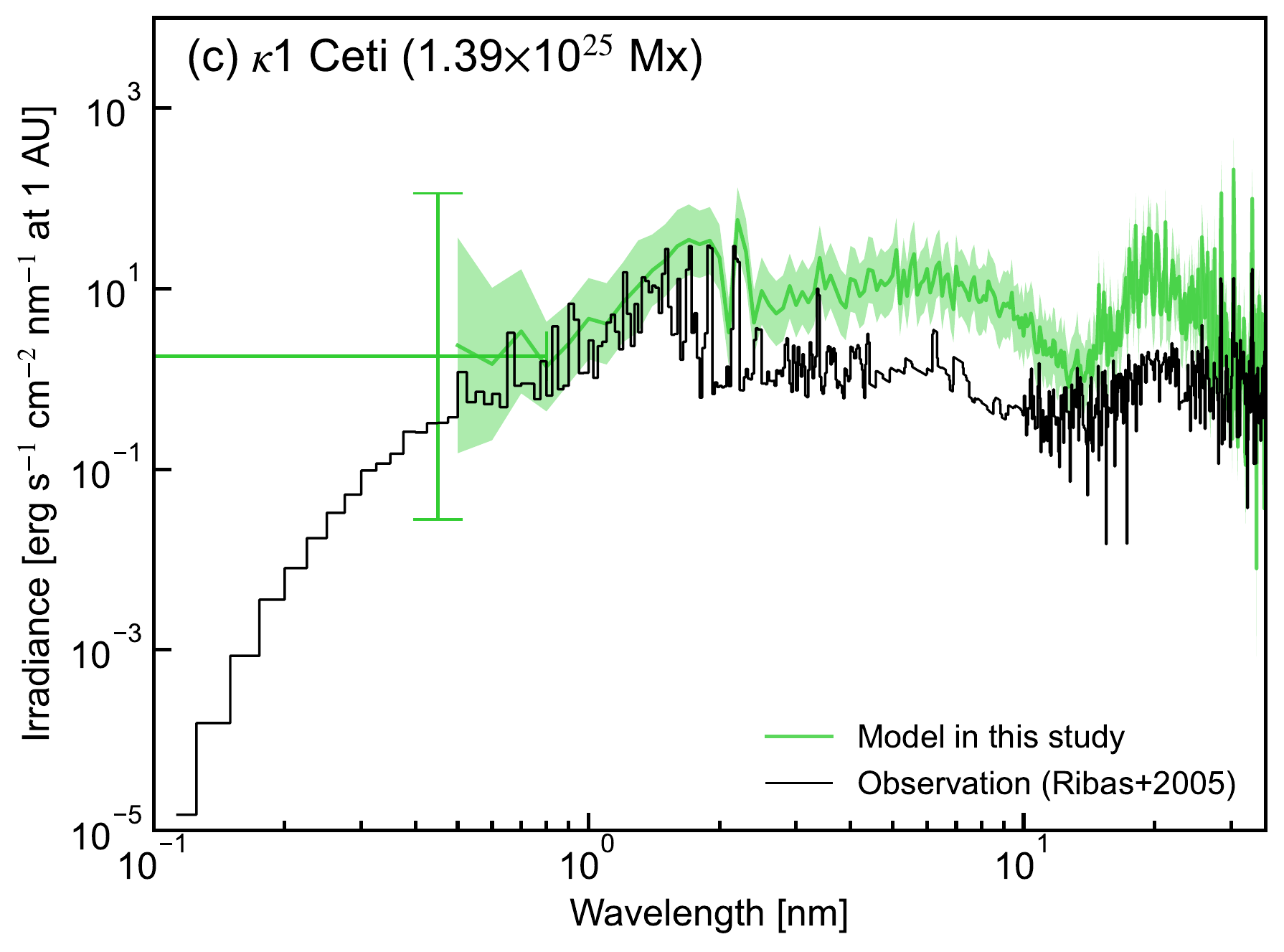}
\caption{
Comparisons between the observed XUV spectra of nearby active Sun-like stars (black lines) and the reconstructed spectra from the solar empirical laws (Equation (\ref{eq:2}); colored lines). 
The stellar name and its total unsigned magnetic fluxes is described at the top left of each panel.
The colored region covering each colored line indicates the error bars of the model which is estimated from the fitting errors of the scaling relations given in Table \ref{tab:3}.
The observed data are basically taken from \cite{2005ApJ...622..680R}.
Note that the horizontal axes are logarithmic scale.
In the panels, reconstructed data points with the scaling relation of GOES X-ray flux (1--8 {\AA}) and magnetic flux obtained by our previous study \cite{2022ApJ...927..179T} are plotted with error bars.
}
\label{fig:6}
\end{figure*}

\begin{figure*}
\epsscale{0.5}
\plotone{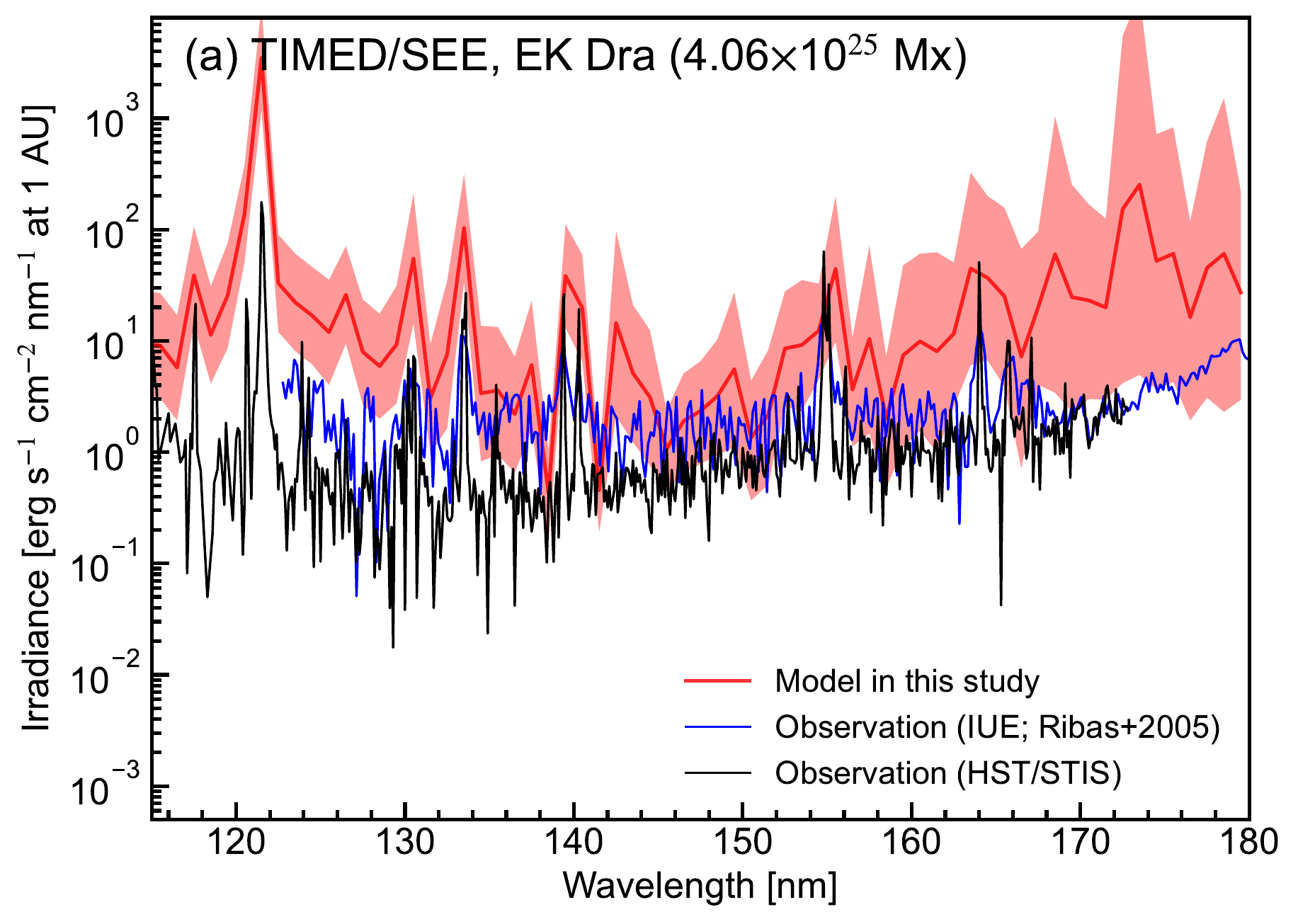}
\plotone{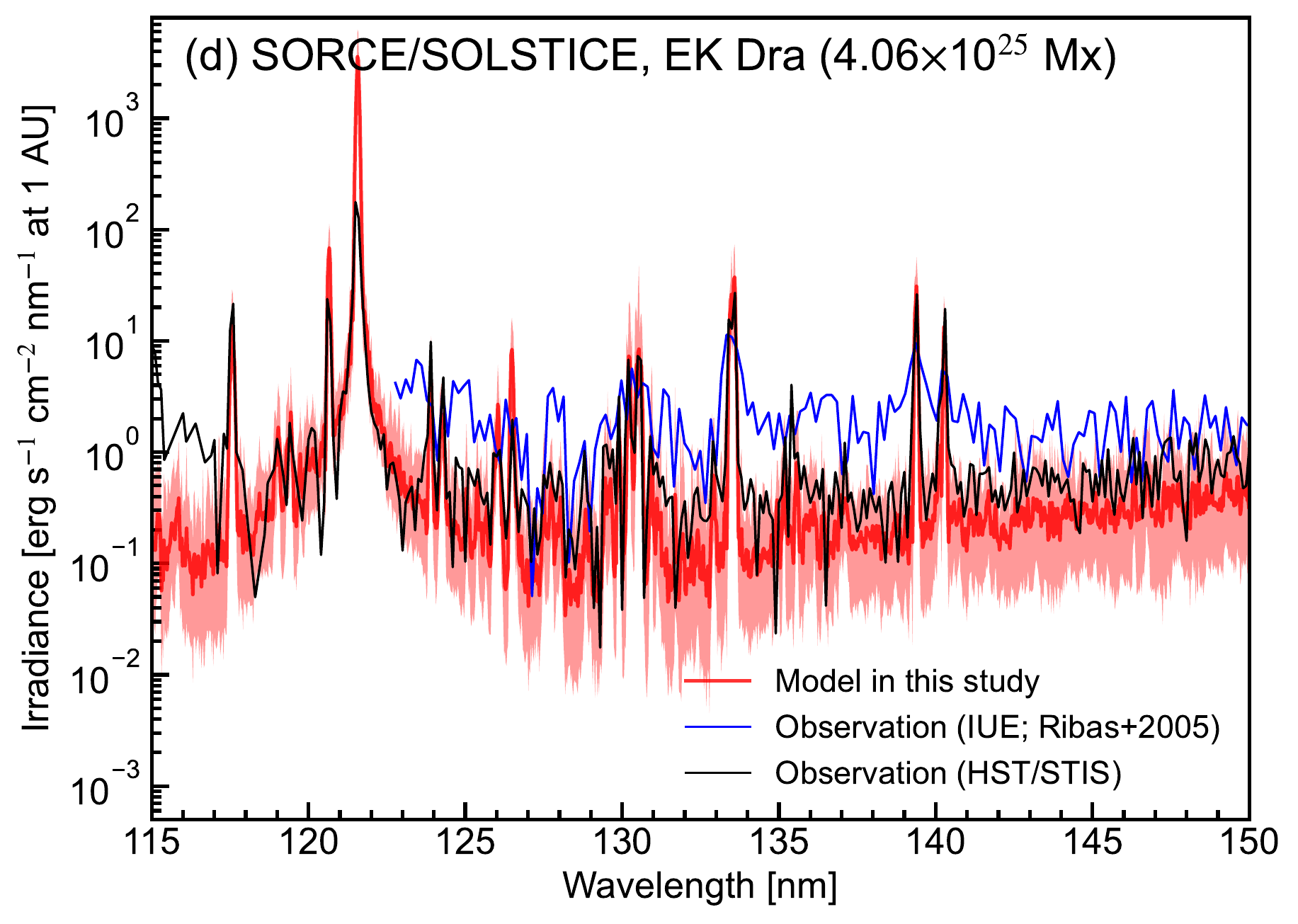}
\plotone{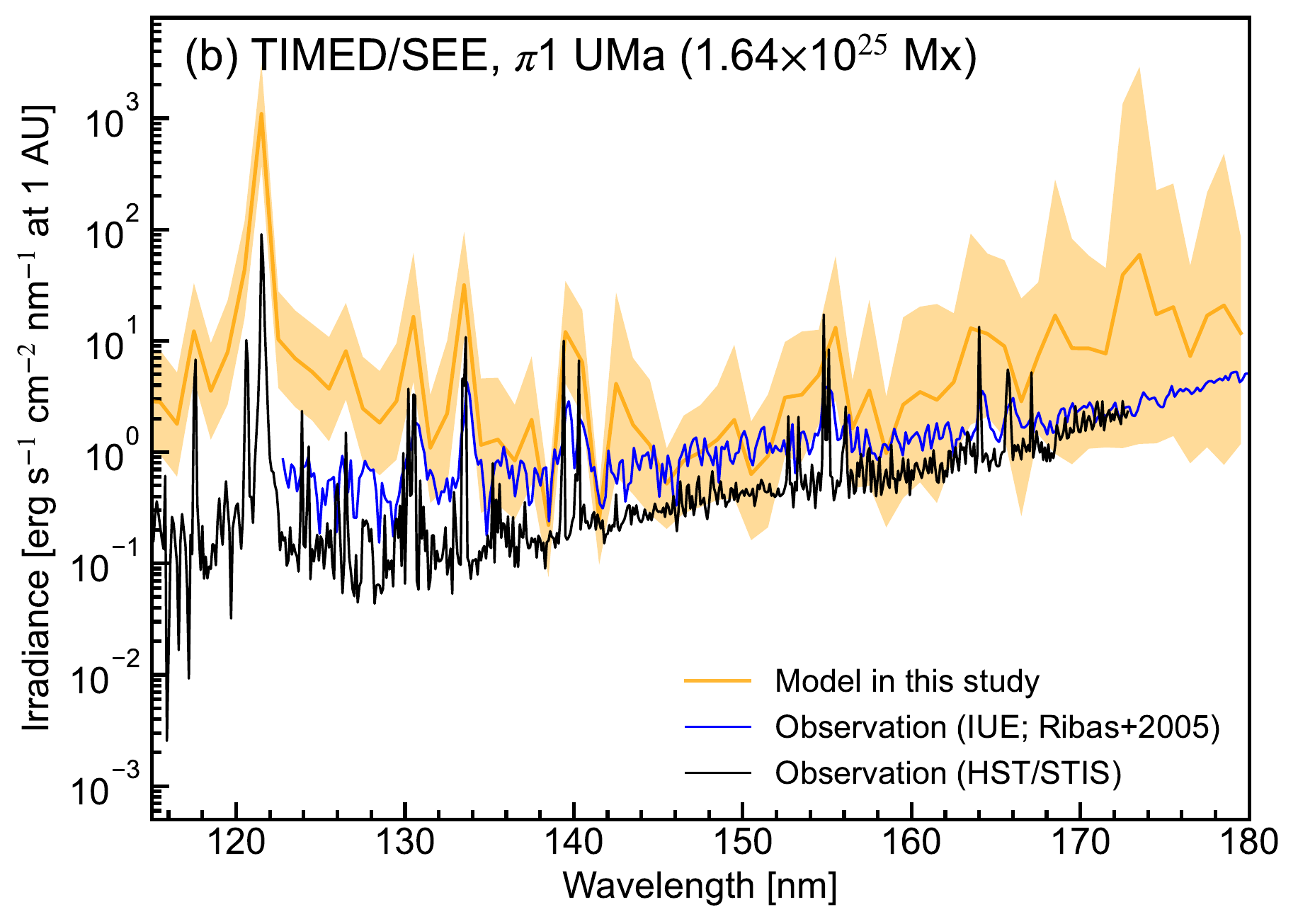}
\plotone{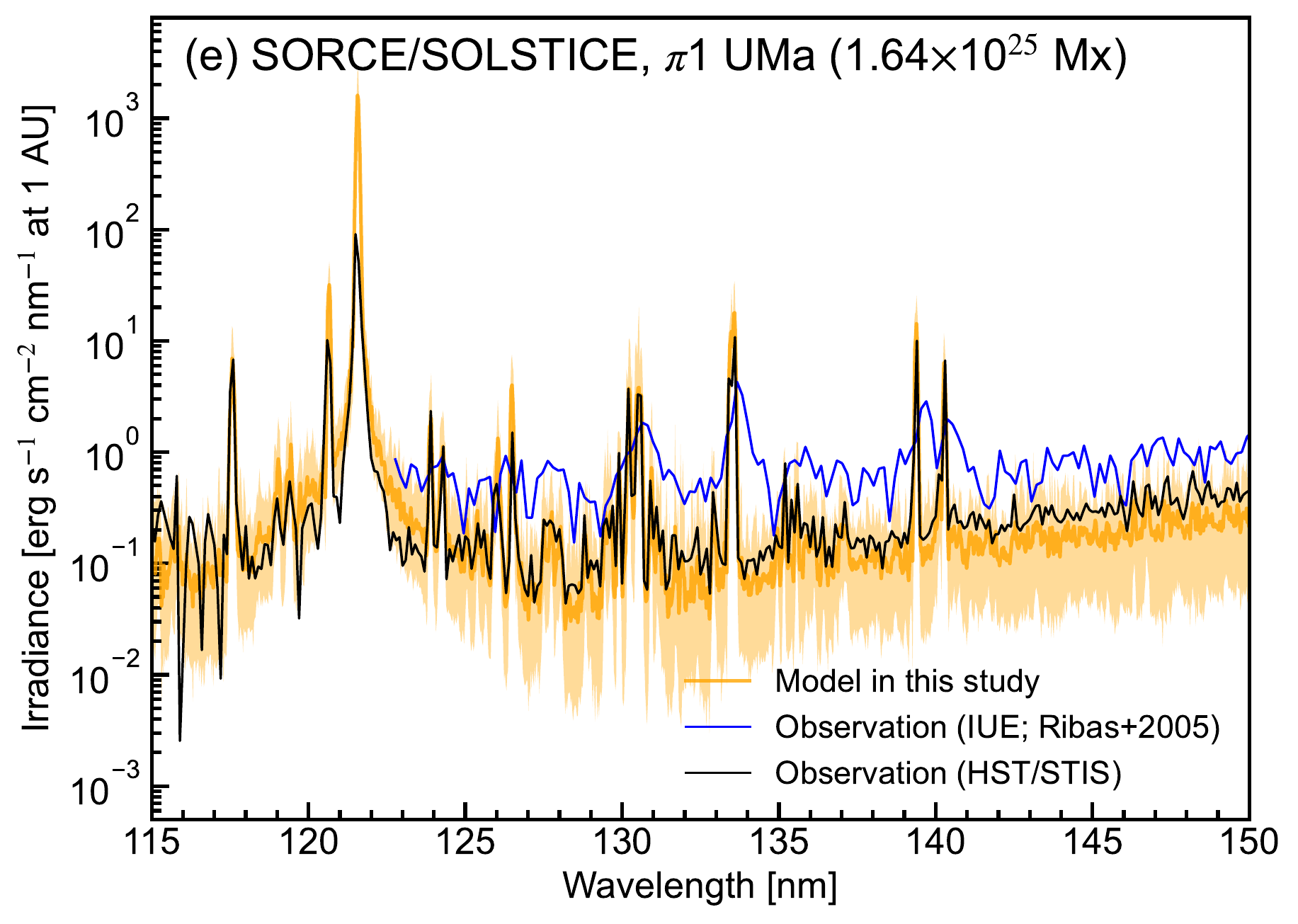}
\plotone{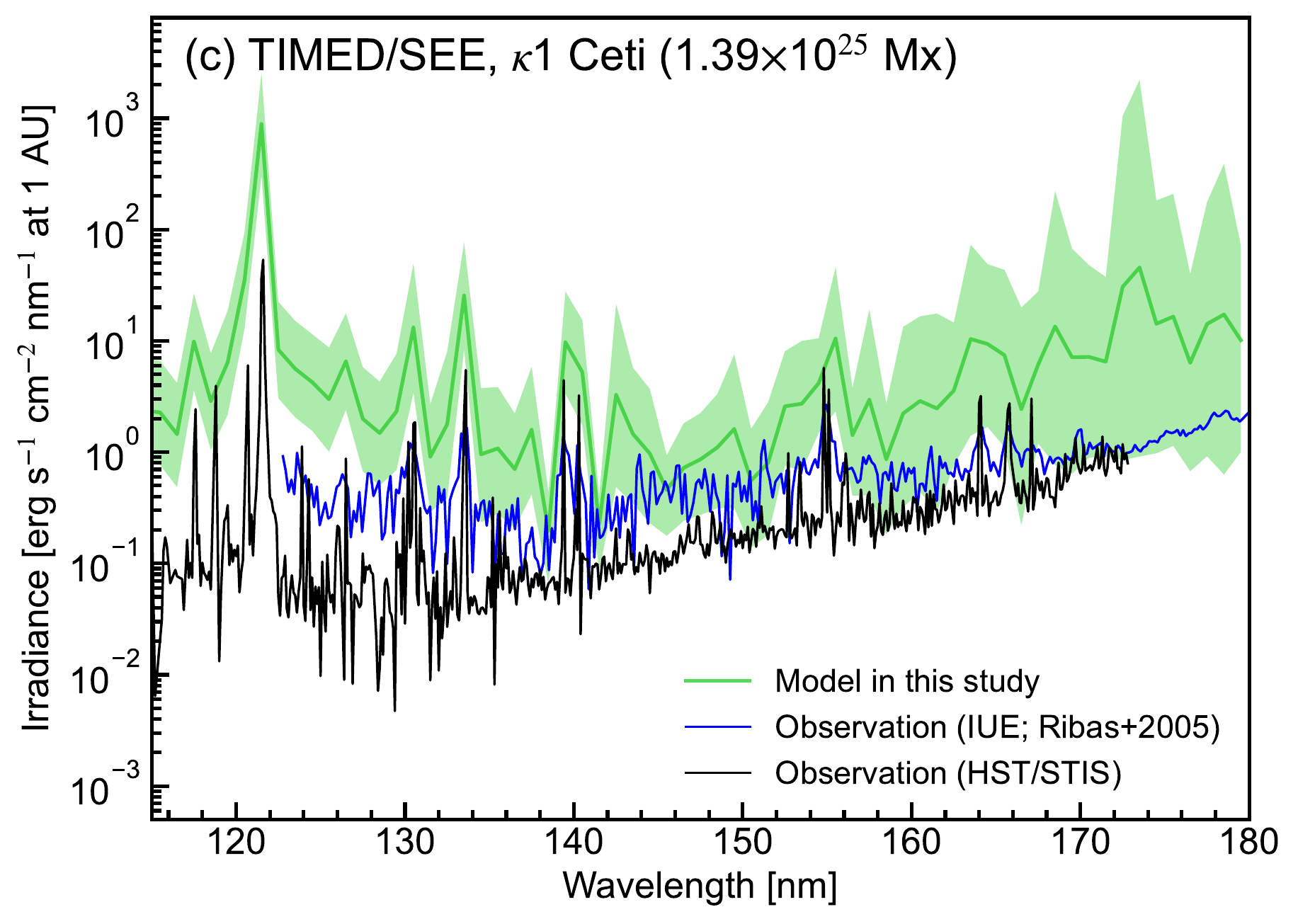}
\plotone{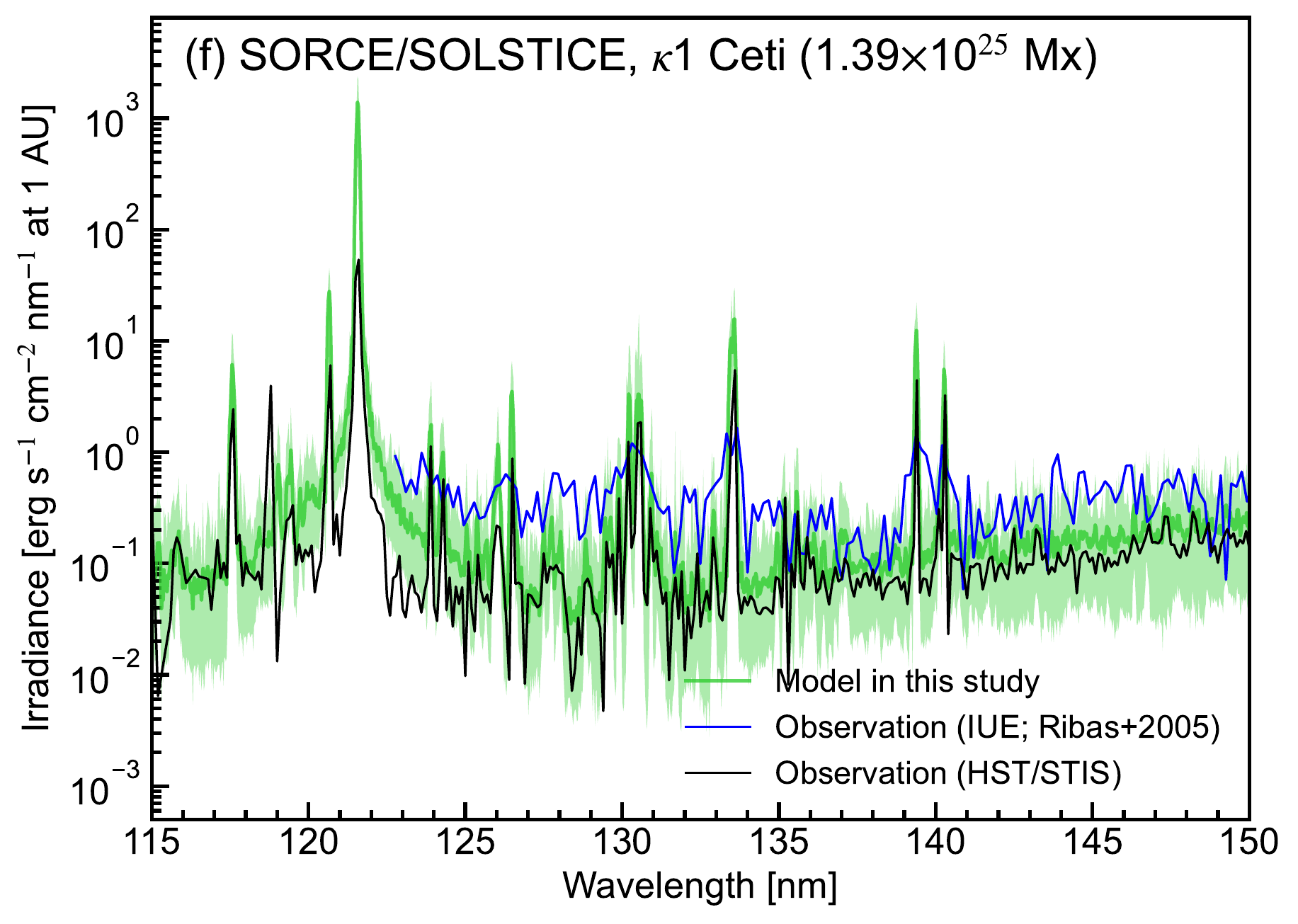}
\caption{
Comparisons between the observed FUV spectra of nearby active Sun-like stars (black and blue lines) and the reconstructed spectra from the solar empirical laws (Equation (\ref{eq:2}); colored lines). 
The stellar name and its total unsigned magnetic fluxes is described at the top left of each panel.
The colored region covering each colored line indicates the error bars of the model which is estimated from the fitting errors of the scaling relations given in Table \ref{tab:3}.
The observed data are basically taken from \cite{2005ApJ...622..680R}.
HST/STIS data were downloaded at the MAST (\url{https://archive.stsci.edu/hst/search.php}) and scaled to the values at 1 AU.
}
\label{fig:6-2}
\end{figure*}

In Section \ref{sec:power-law}, the whole XUV/FUV spectra $I(\lambda)$ are given as a function of total unsigned magnetic fluxes $\Phi$ of the Sun in the Equation (\ref{eq:2}). 
The parameters $\alpha(\lambda)$ and $\beta(\lambda)$ are given in Table \ref{tab:2} or \ref {tab:3} (all values are available online).
If we assume that this relation can be applied to stars more active than the Sun, then we can estimate the whole XUV/FUV spectra of stars using Equation (\ref{eq:2}) and Table \ref{tab:2} or \ref {tab:3}.
For example, Figure \ref{fig:5} demonstrates how the XUV spectra are extrapolated when the total unsigned magnetic fluxes of the visible hemisphere of the stars are as large as 10$^{24}$, 10$^{25}$, and 10$^{26}$ Mx (the maximum value in solar cycle 24 is approximately 3.35$\times$10$^{23}$ Mx).
This new estimation method can be meaningful because we can derive the elusive stellar EUV spectrum (longward of 36 nm) directly from total unsigned magnetic fluxes which have been already obtained for active stars from ground-based high-dispersion spectroscopic observations in optical and infrared bands \citep[e.g.,][]{2020A&A...635A.142K}.

For the purpose of the application of our scaling relations to planet-hosting stars, we examine how well our method estimates the overall XUV/FUV spectral shape, including not only the emission lines but also the continuum level.
Several nearby young Sun-like stars, such as EK Dra (G1.5V, age of 120 Myr), $\pi^1$ UMa (G1.5V, age of 500 Myr), and $\kappa^1$ Ceti (G5V, age of 600 Myr), have observations of the XUV and FUV spectra \citep[except for longward EUV ranges;][]{2005ApJ...622..680R} as well as the total unsigned magnetic flux \citep{2020A&A...635A.142K}.
These stars are much younger and more active than the Sun \citep[see Table \ref{tab:4};][]{2005ApJ...622..680R}, and thus are appropriate to test the performance of our scaling method.
Stellar parameters in Table \ref{tab:4} are taken from \cite{2020ApJ...902...36T} and stellar distances are obtained by the Gaia Early Data Release 3 \citep{2021A&A...649A...1G}.
The XUV/FUV spectra were taken from \cite{2005ApJ...622..680R}.
X-ray to EUV spectra (0.1--36 nm) were obtained by the EUVE (8--36 nm), ROSAT (0.6--12.4 nm) and ASCA  (0.1--4 nm) and are ISM-corrected assuming some reasonable values for the hydrogen column density \cite[see][for further information]{2005ApJ...622..680R}.
FUV spectra were obtained by the International Ultraviolet Explorer (IUE; 115--195 nm). 
Note that the IUE data may have some bias at the short wavelength end and the unphysical flux rise in some cases (Ribas, private communication). 
In addition to IUE data, we used the FUV data taken by the Space Telescope Imaging Spectrograph (STIS; 114--172 nm) onboard the Hubble Space Telescope (HST).
The data were downloaded at MAST\footnote{\url{https://archive.stsci.edu/hst/search.php}}.
The observation dates of HST/STIS are 2012 March 27 for EK Dra (observation ID: oboq01010), 2012 September 25 for $\pi^1$ UMa (IDs: obq202010 and obq202020), and 2000 September 19 for $\kappa^1$ Ceti (IDs: o5bn03050 and o5bn03060).
The HST spectra from multiple observation IDs were averaged over time and were smoothed over 0.1 nm to increase the signal-to-noise ratio.
No ISM correction has been applied for IUE and HST data because the correction is not significant for the overall continuum levels \citep{2015AJ....150....7A}. 
Also, the total unsigned magnetic flux of these were calculated by \cite{2020A&A...635A.142K} based on the Stokes I profiles of several photospheric lines showing the Zeeman broadening with different magnetic sensitivities.
The method can avoid the cancellation of opposite field polarities and has an advantage in estimating the total unsigned magnetic flux.

Figures \ref{fig:6} and \ref{fig:6-2} (a-c) show a comparison between the observed and reconstructed spectra for the Sun-like stars. 
The reconstructed XUV/FUV spectra $I(\lambda)$ for EK Dra, $\pi^1$ UMa, and $\kappa^1$ Ceti are estimated from Equation (\ref{eq:2}) and Table \ref{tab:3} and their observed total unsigned magnetic flux (Table \ref{tab:2}).
Note that for the FUV range (115-150 nm), we also tested the scaling laws derived by the SORCE/SOLSTICE with a careful analysis done in Appendix \ref{app:a} (see Figure \ref{fig:6-2}(d-f)).
Comparing the model spectrum (colored lines) with the observed data (black and blue lines), we see that our empirical law roughly agrees with the observations within an error of one order of magnitude, for almost all of the wavelength ranges from X-ray to FUV.
In more detail, it is about one order of magnitude accuracy at XUV (2--30 nm) and FUV range, while it reproduces very well the spectra at shortward X-ray ($<$2 nm) and longward EUV (30--36 nm).
The agreement indicates that the solar power-law scaling relations can be applicable to Sun-like stars with a variety of ages and activity levels and supports the predictive capability of our methodology, especially in X-ray ($<$2 nm) and EUV (30--36 nm) range (and possible missing EUV range of 36-92 nm).
On the other hand, it should be noted that the overestimations of flux by our method can also be seen in the other range. Further discussion for this describancy will be done in Section \ref{sec:dis}.

\subsection{Emission lines around the elusive EUV}\label{sec:4-2}

\begin{figure*}
\epsscale{0.35}
\plotone{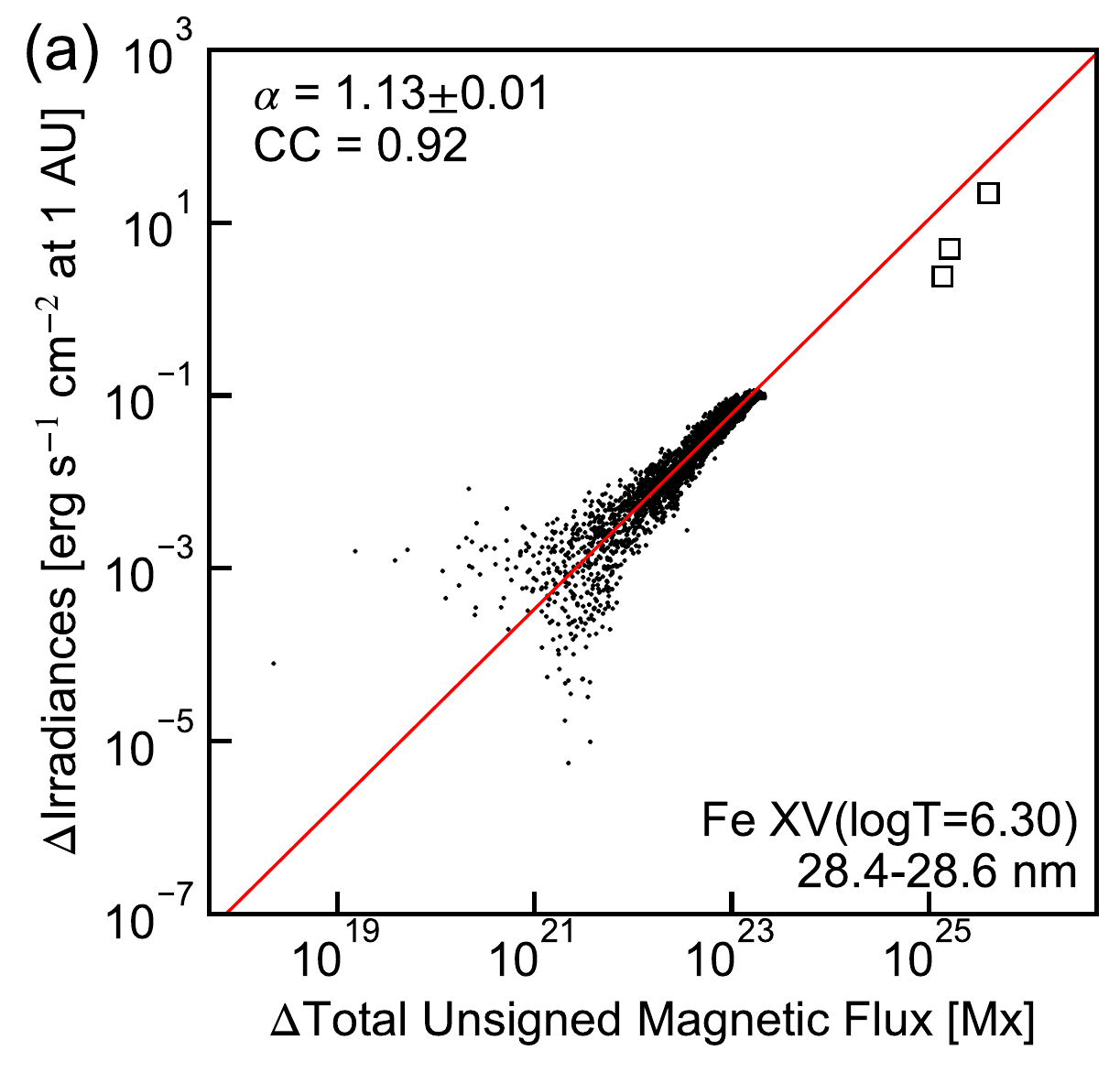}
\plotone{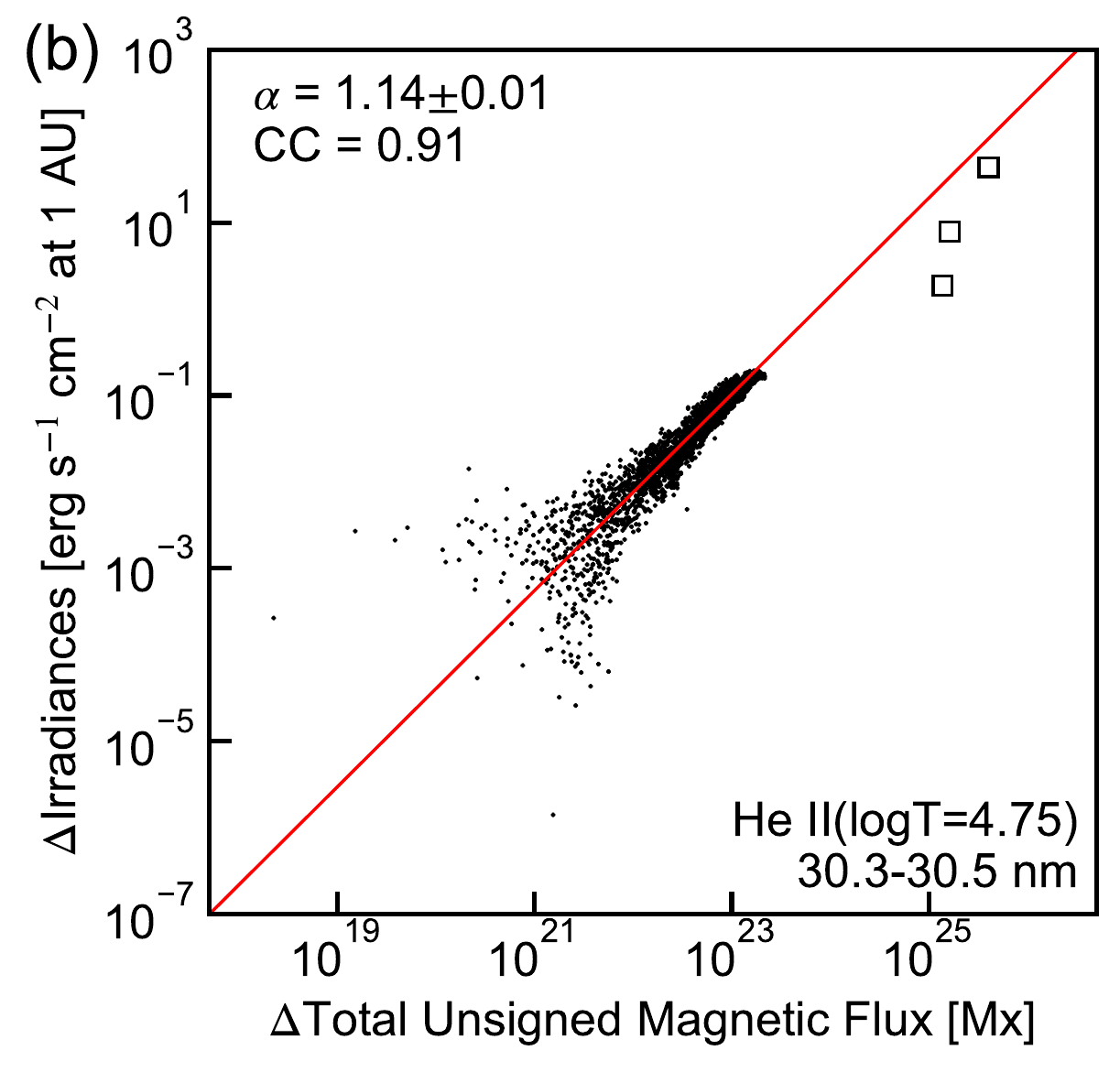}
\plotone{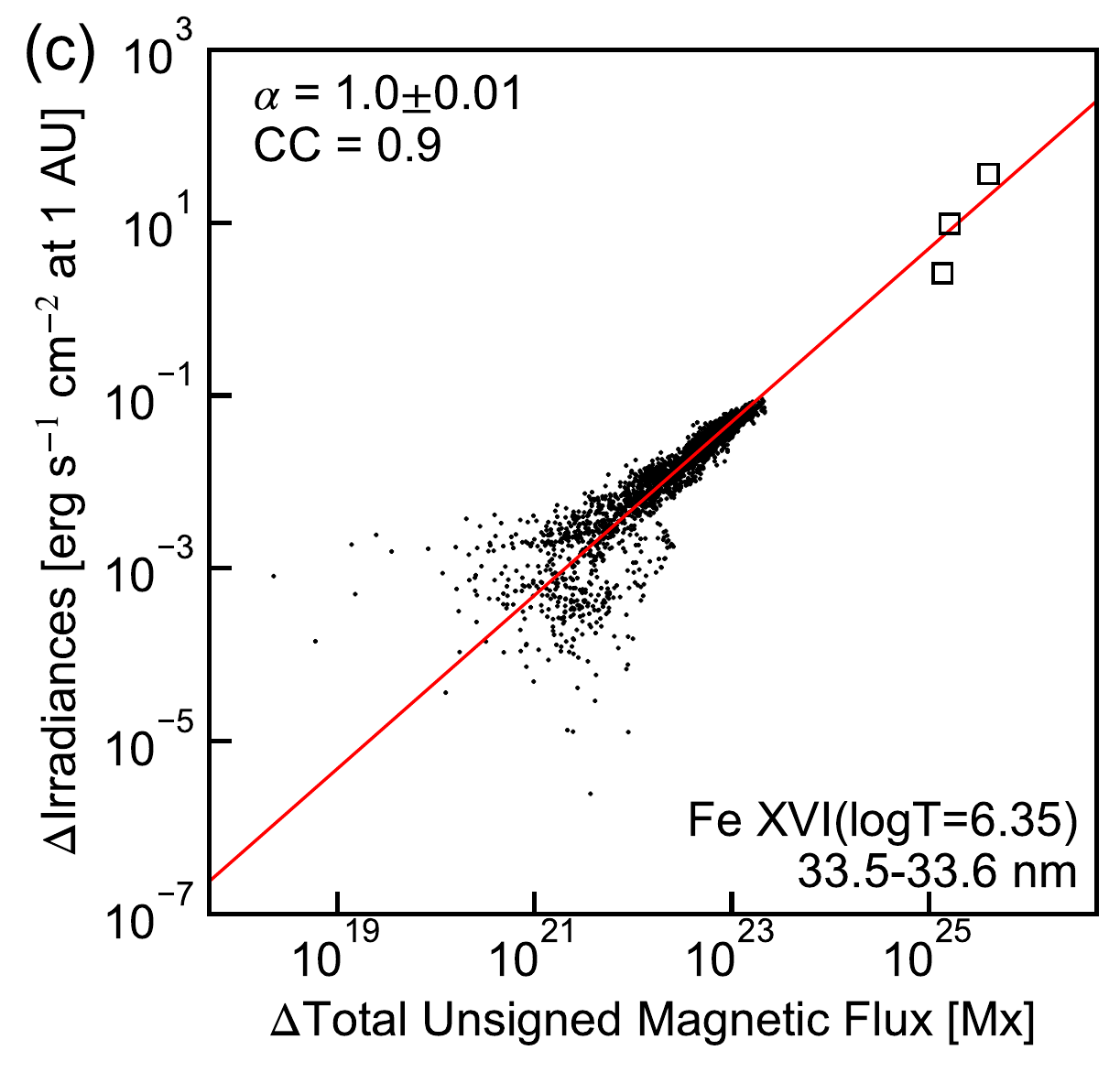}
\plotone{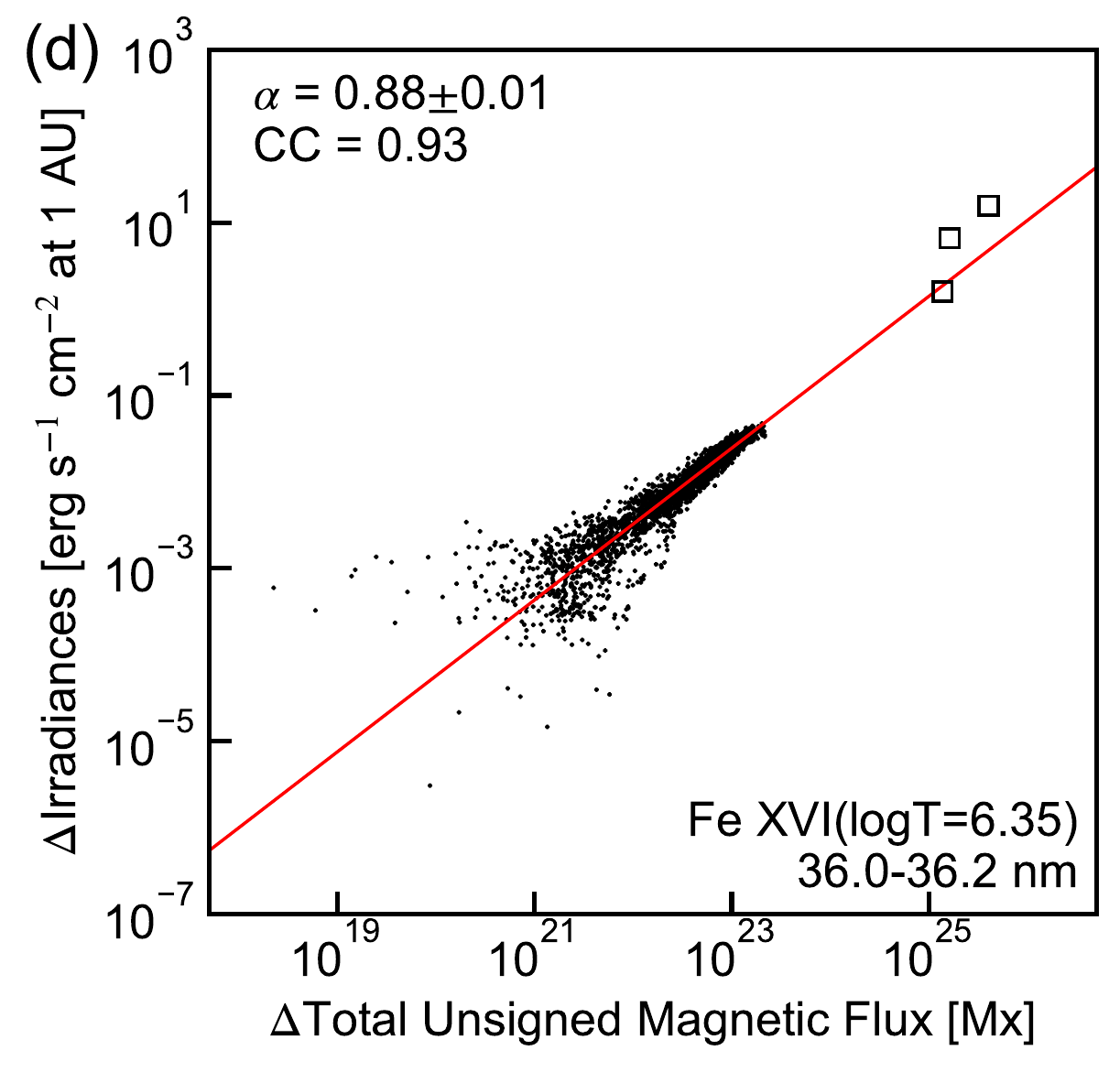}
\plotone{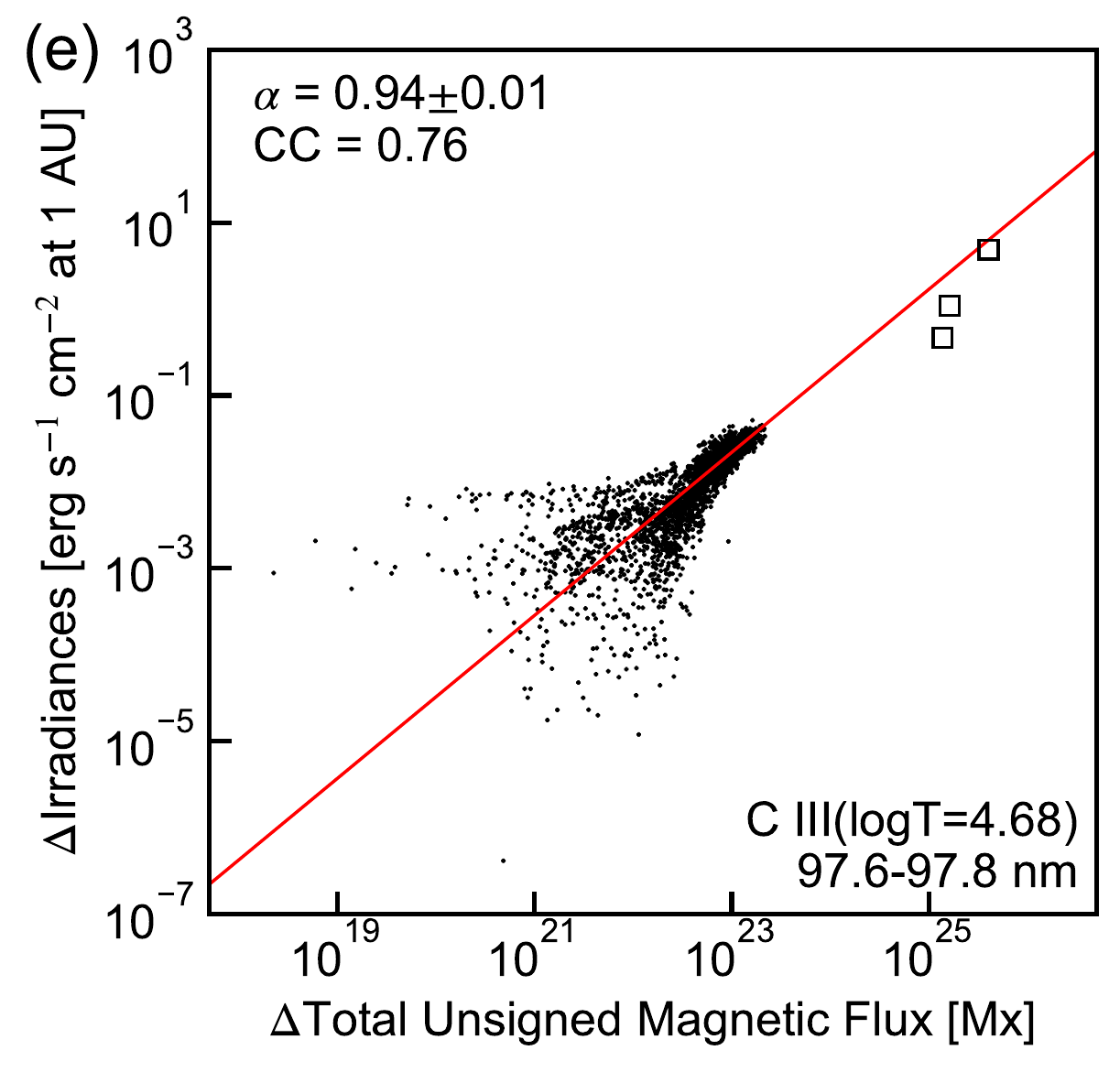}
\plotone{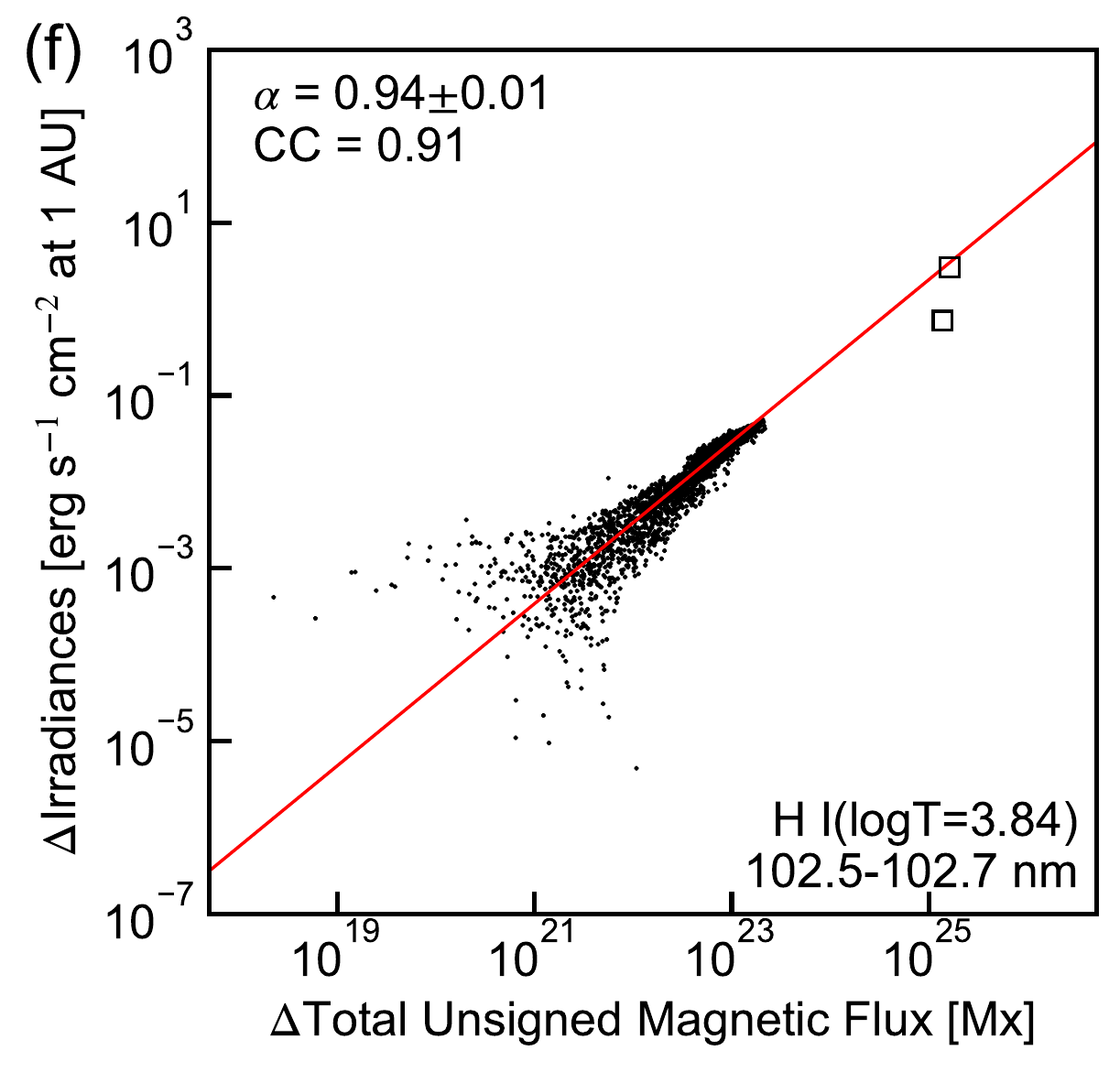}
\plotone{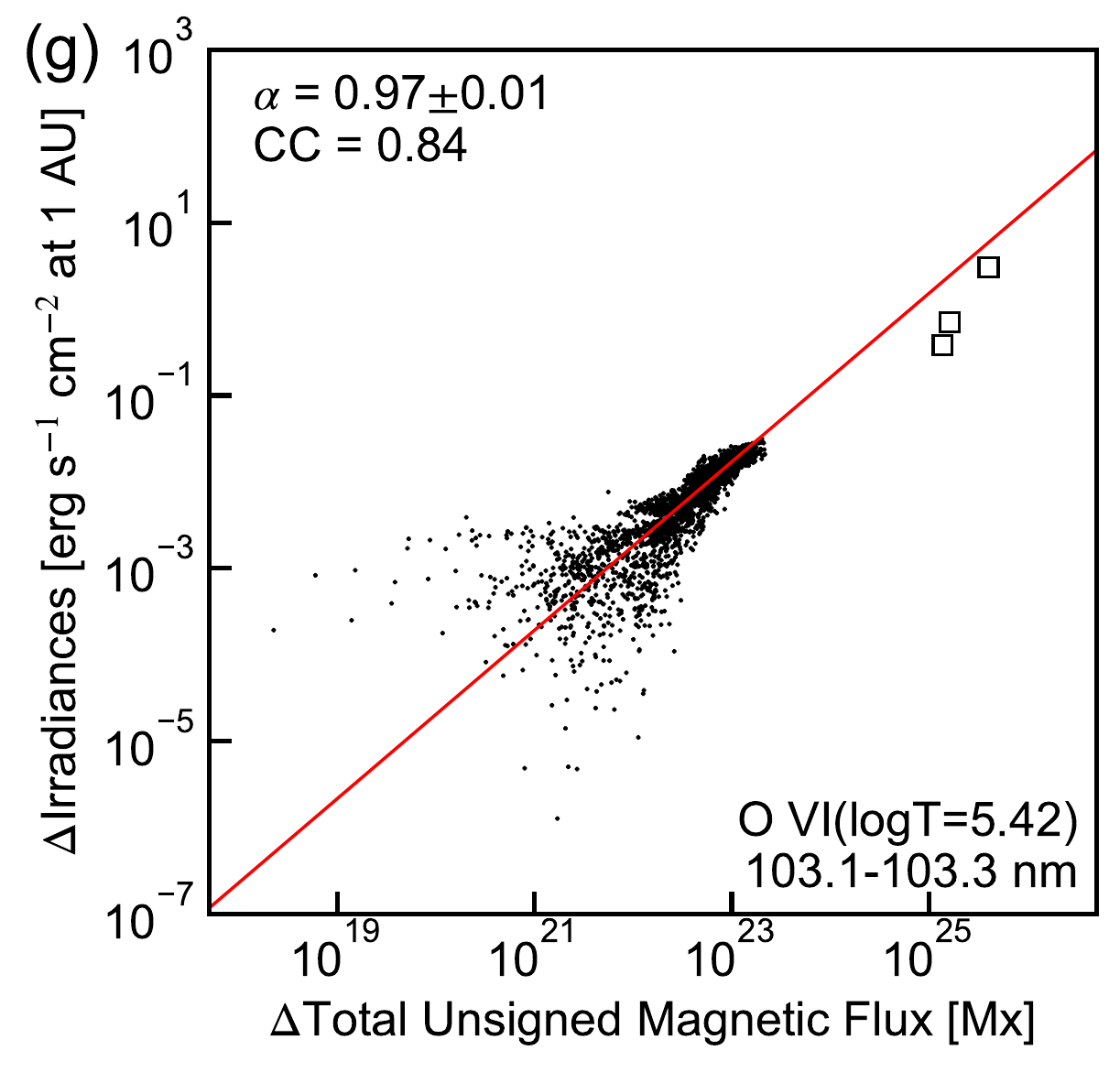}
\plotone{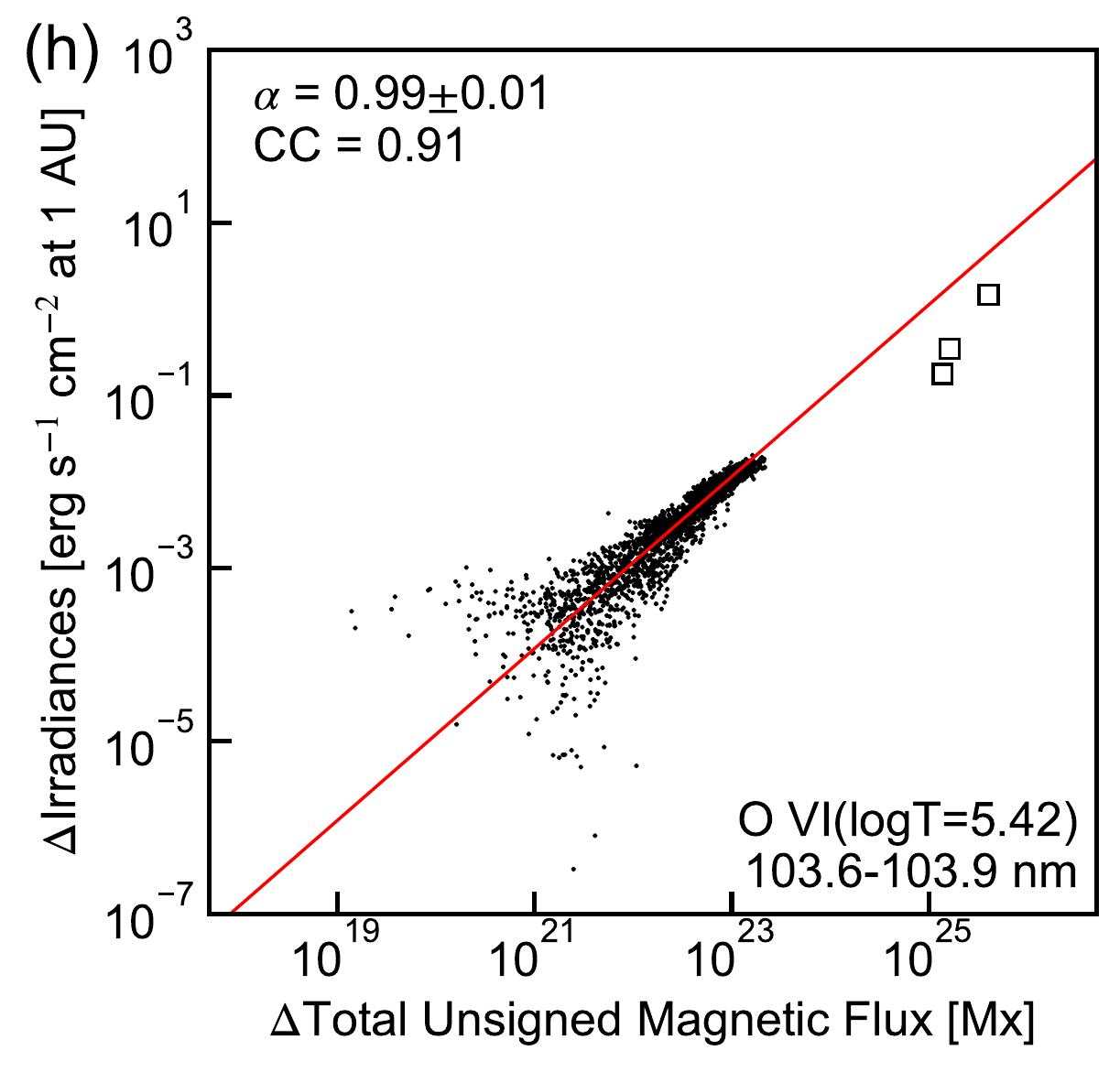}
\caption{
Comparison of solar and stellar data for the EUV emission lines. The stellar data are taken from \cite{2005ApJ...622..680R}. The red lines represent the fitting line in log-log space using \textsf{scipy.odr} installed in \textsf{python}. The fitting function is the same as Equation (\ref{eq:1}). The power-law index and correlation coefficient (C.C.) are indicated at the top left. 
}
\label{fig:7}
\end{figure*}

In Section \ref{sec:4-2}, we mentioned only the broadband spectral shape, but what about the irradiance of spectral lines?
\cite{2022ApJ...927..179T} shows that this flux-flux methodology can predict well fluxes of emission lines/bands of active Sun-like stars (G0-K1 dwarfs), but their study was limited to the selected five emission lines and bands (i.e., X-ray band 0.52--12.4 nm, Fe XV 28.4 nm, C II 133.5 nm, Ly$\alpha$, and Mg II k+h line).
Here, we additionally analyzed the scaling relations and its applications to stars for eight emission lines in the wavelength range very close to the elusive EUV wavelength (36--91.2 nm) as listed in Table \ref{tab:6} (the Fe XV 28.4 nm line was also analyzed in \cite{2022ApJ...927..179T}).
We studied spectral lines for which stellar observational data is reported by \cite{2005ApJ...622..680R}.
We focus on these emission lines because we are particularly interested in the reconstruction of this elusive EUV band and comparison with the observed stellar EUV emission.

Figure \ref{fig:7} shows scatter plots of the residuals of the solar irradiance of the selected emission lines versus that of the total unsigned magnetic flux. 
The power-law index and correlation coefficient are provided in each panel. 
Here, we derived again the scaling laws for each emission line by integrating the intensity within the full width of 1/10 maximum. The line width was calculated by fitting the median spectrum around activity maximum with the Gaussian function. 
In Figure \ref{fig:7}, the observed values of the young Sun-like stars, EK Dra, $\pi^1$ UMa, and $\kappa^1$ Ceti, are overplotted. 
For almost all panels, the stellar data are located on the extensions of the power-law relations of the solar data within an order of magnitude.
In particular, the reconstruction of fluxes from spectral lines close to the missing EUV is excellent (e.g., Fe XVI, C III, and H I).
These indicate that our solar flux-flux scaling relation can have a good predictive capability to estimate fluxes of EUV spectral lines of stars, and possibly even those in the elusive EUV range.

\begin{figure*}
\epsscale{0.7}
\plotone{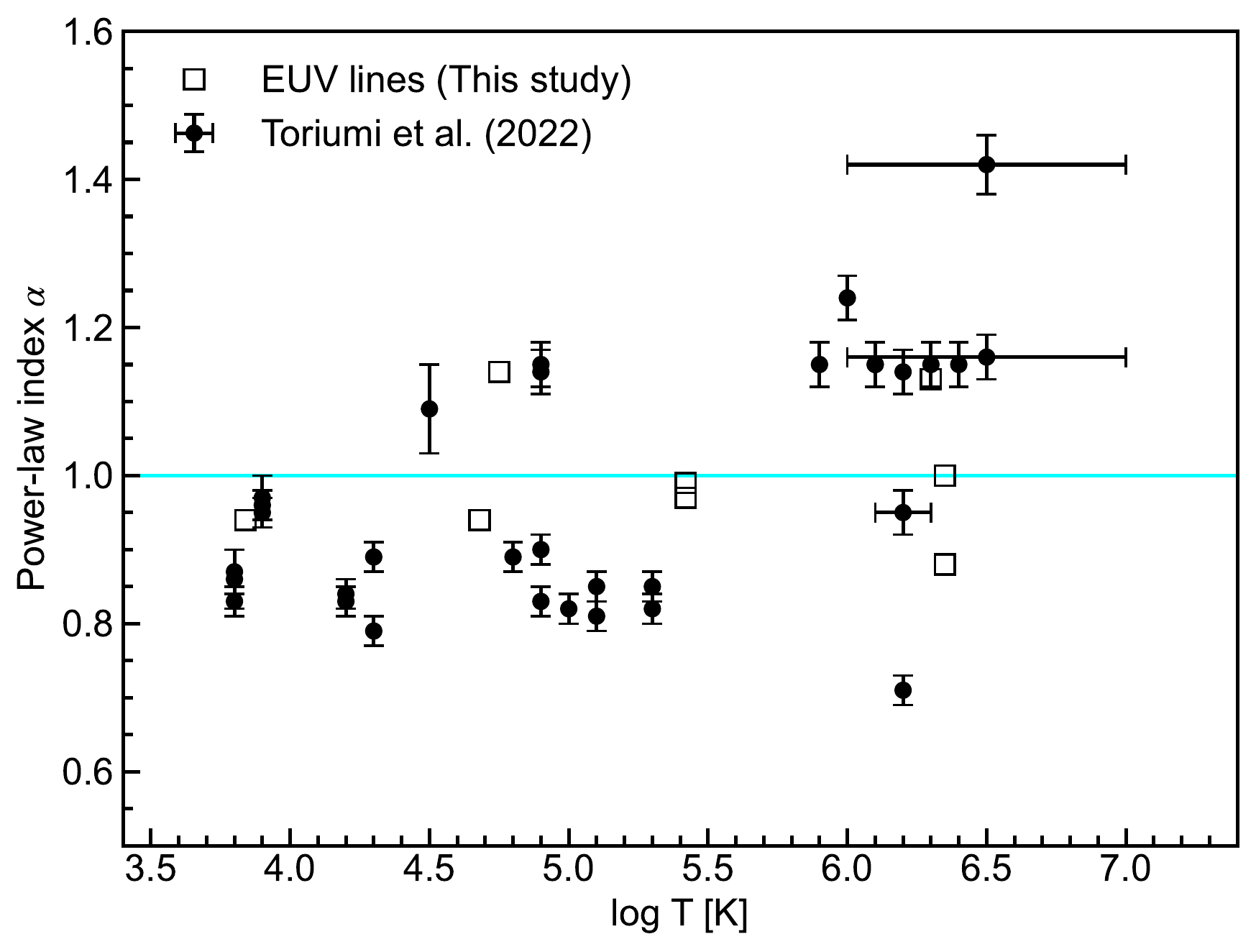}
\caption{
Relationship between the temperature of the emission lines/bands and the power-law index $\alpha$.
The black squares indicate the EUV emission near the elusive EUV wavelength (36--92 nm) analyzed in this study (see Table \ref{tab:6}) whose stellar data are also available (Figure \ref{fig:7}).
The fitting error bars are not plotted here because they are too small compared to the symbol size.
The temperatures of the emission line are taken from \cite{2005ApJ...622..680R} in this study.
The black dots are the data from X-ray to radio wavelength taken from \cite{2022ApJ...927..179T} and \cite{2022ApJS..262...46T}.
}
\label{fig:8}
\end{figure*}

Figure \ref{fig:8} shows the the power-law indices as a function of the formation temperature of emission lines. 
\cite{2022ApJ...927..179T} and \cite{2022ApJS..262...46T} show that as the temperature decreases, the power-law index also decreases.
Our power-law index data in EUV range (open squares) are almost consistent with the previous studies within the data scatter (filled circles).
Note that the He 30.4 nm line has a low formation temperature (log$T$ = 4.75$\sim$4.9) but a power-law index of above unity ($\alpha$ = 1.23$\pm$0.01). 
This is consistent with the value reported in \cite{2022ApJS..262...46T}, $\alpha$ = 1.15$\pm$0.03, and similar properties can be seen in other helium lines, such as He II 25.6 nm and He I 1083.0 nm.
The power-law indexes of these lines are irregular from the general description above. 
\cite{2022ApJS..262...46T} argued that the inconsistency of the He I 1083.0 nm  emission line flux may be due to the photoionization effect of coronal emission, and \cite{2017A&A...597A.102G} suggests that a similar effect occurs for He II 30.4 nm line.
While the cause of this above-unity power-law index of He lines is not clear, it is worth mentioning that the stellar fluxes of He 30.4 nm lines are consistent with the extension of the scaling law of above-unity power-law index even as shown in Figure \ref{fig:7}(b). 
This may indicate that the non-LTE driven formation mechanism of these He lines can be universal for the active stars \citep{2016A&A...594A.104L,2017A&A...597A.102G}.

\subsection{X-ray vs. EUV luminosity relation}

\begin{figure*}
\epsscale{0.5}
\plotone{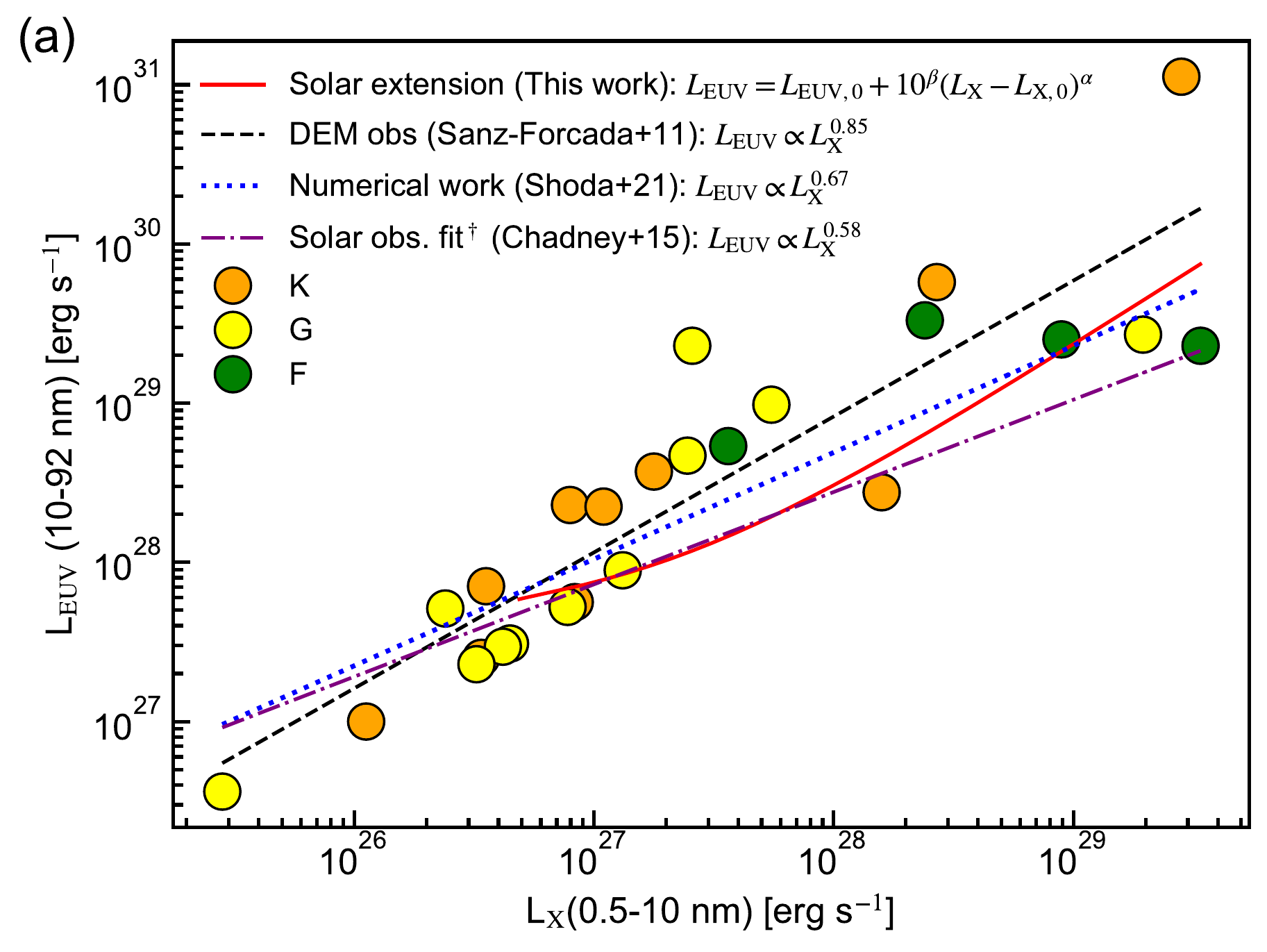}
\plotone{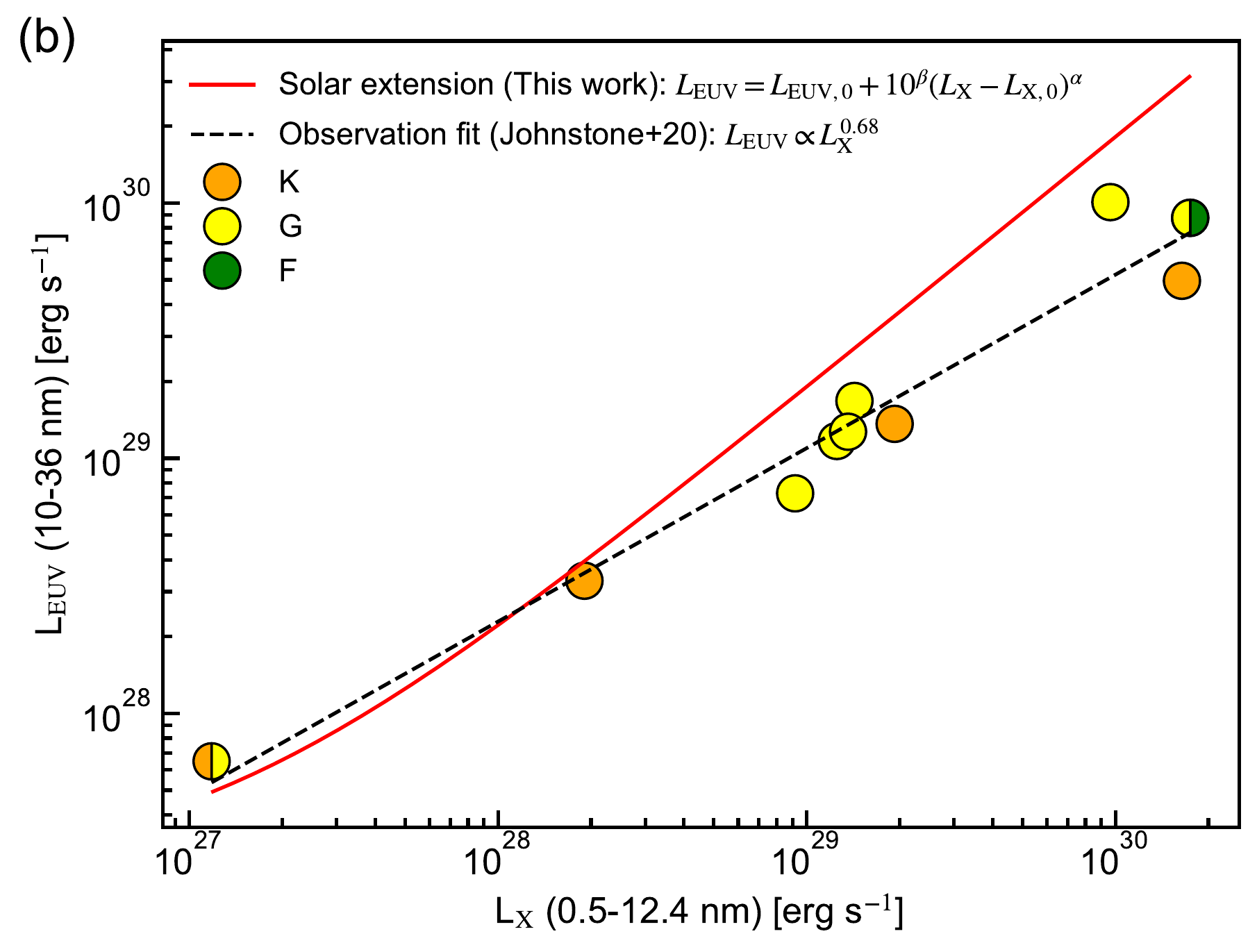}
\caption{X-ray vs. EUV luminosity for FGK dwarfs. (a) Symbols shows the observed X-ray luminosity (0.5-10 nm) and EUV (10-92 nm) luminosity estimated by differential emission measure (DEM) analysis \citep{2011A&A...532A...6S}. The black dashed line shows the results of a linear fitting to the double logarithmic plots of the DEM analysis data. The red solid line shows the extension of the solar scaling relation $L_{\rm EUV} = L_{\rm EUV,0}+10^{\beta}(L_{\rm X}-L_{\rm X,0})^{\alpha}$ obtained in this study ($\alpha = 0.95$, $\beta = 1.9$, $L_{\rm X,0}$ =   4.9$\times10^{26}$ erg s$^{-1}$, and $L_{\rm EUV,0}$ =  5.9$\times 10^{27}$ erg s$^{-1}$). The blue line is a theoretical scaling relation obtained by \cite{2021A&A...656A.111S}. Solar empirical scaling relation derived in \cite{2015Icar..250..357C} is also plotted with a purple dash-dotted line as a reference, but note that their definition of boundary wavelength between EUV and X-ray  is a bit different (12.4 nm) from that of other scaling laws in this panel. (b) Symbols indicate the relation between the observed X-ray luminosity (0.5-12.4 nm)  and the observed short EUV (10-36 nm) luminosity \citep{2021A&A...649A..96J}. The red solid line shows the extension of the solar scaling relation $L_{\rm EUV} = L_{\rm EUV,0}+10^{\beta}(L_{\rm X}-L_{\rm X,0})^{\alpha}$ obtained in this study ($\alpha = 0.98$, $\beta = 0.85$, $L_{\rm X,0}$ =   5.9$\times10^{26}$ erg s$^{-1}$, and $L_{\rm EUV,0}$ =  3.7$\times 10^{27}$ erg s$^{-1}$)
}
\label{fig:9}
\end{figure*}

We also investigated the relation of X-ray ($L_{\rm X}$) and EUV luminosities ($L_{\rm EUV}$) based on our scaling relation. 
Several previous studies reported this scaling relation observationally and theoretically.
\cite{2011A&A...532A...6S} reported the stellar EUV luminosity (10-92 nm) estimated from differential emission measure (DEM) analysis of stars and its relationship with X-ray luminosity (0.5-10 nm).
Figure \ref{fig:9}(a) shows the plot of X-ray and EUV relations for GKF dwarfs reported in \cite{2011A&A...532A...6S} (see Figure 2 in their paper). 
\cite{2021A&A...656A.111S} numerically investigated the theoretical relation between $L_{\rm X}$ and $L_{\rm EUV}$ fluxes for the same wavelength bands, and their scaling relation ($L_{\rm EUV}$ $\propto$ $L_{\rm X}^{0.67}$) is also overplotted with a blue dotted line.
\cite{2015Icar..250..357C} derived a solar empirical scaling relation between EUV band (12.4--91.2 nm) and X-ray band (0.5--12.4 nm) as $L_{\rm EUV}$ $\propto$ $L_{\rm X}^{0.58}$ by using TIMED/SEE without subtracting the basal level.

In this study, we derived an empirical scaling relation in the same energy band as stellar observations (i.e., EUV is 10--92 nm and X-ray is 0.5-10 nm) as
\begin{equation}\label{eq:3}
L_{\rm EUV} = L_{\rm EUV,0}+10^{\beta}(L_{\rm X}-L_{\rm X,0})^{\alpha}, 
\end{equation}
where $\alpha = 0.95$, $\beta = 1.9$, $L_{\rm X,0}$ =   4.9$\times10^{26}$ erg s$^{-1}$, and $L_{\rm EUV,0}$ =  5.9$\times 10^{27}$ erg s$^{-1}$.
We found that our scaling relation indicated with red line in Figure \ref{fig:9}(a) is almost consistent with the data and theoretical model. 
However, the power-law relation is a bit steeper than the fit of the observations and theoretical model (a blue line) and as a result, our scaling laws is underestimating the EUV luminosity for the inactive stars ($L_{\rm X}<10^{28}$ erg s$^{-1}$).
It should be noted here that the DEM data is just observation-based model spectra and may have some uncertainties.

\cite{2021A&A...649A..96J} reported the short-band (10--36 nm) EUV luminosity observed by EUVE, as a function of X-ray luminosity (0.5--12.4 nm).
Figure \ref{fig:9}(b) shows the observational data and the fitting results.
Here, we also derived an empirical scaling relation for the same energy band (i.e., EUV is 10--36 nm and X-ray is 0.5-12.4 nm) in the formulae of Equation (\ref{eq:3}), and the parameters are obtained as follows: $\alpha = 0.98$, $\beta = 0.85$, $L_{\rm X,0}$ =   5.9$\times10^{26}$ erg s$^{-1}$, and $L_{\rm EUV,0}$ =  3.7$\times 10^{27}$ erg s$^{-1}$.
We found that our scaling relation is almost consistent with the observational data, but the power-law relation appears to be steeper.
This trend is actually consistent with the results in Section \ref{sec:4-1} where the X-ray luminosity is underestimated for a very active star EK Dra, so this slightly large power-law index could be due to the underestimations of the X-ray luminosity for very active stars (not the overestimation of EUV luminosity).

\section{Discussion and Conclusion}\label{sec:dis}

Here we have proposed a new reconstruction method of stellar XUV/FUV spectra by using empirical scaling flux-flux relations (see, the description of Equation (\ref{eq:2}) and parameters in Tables \ref{tab:2} and \ref{tab:3}) and measurements of magnetic fluxes of stars.
This method can be more accurate than other estimation methods of stellar XUV/FUV spectra (especially, the missing longward range of EUV fluxes) because total unsigned magnetic flux of stars can be derived directly from ground-based optical and infrared observations.
One of the advantages of our estimation method of XUV/FUV spectrum is that the power-law index in the relation with magnetic flux can be physically understood as described in Section \ref{sec:3-3} (see \cite{2022ApJS..262...46T} for the detail physics explanations).
The wavelength dependence of power-law index $\alpha(\lambda)$ could also be a strong constraint to the models of the coronal heating \citep[see, e.g.,][]{2021A&A...656A.111S}.
Another advantage of our method is in relating to the physical parameters that can be directly coupled with various numerical magnetohydrodynamic models of solar/stellar dynamo and sunspot/starspots.
If the evolution of magnetic fluxes on stars are derived from numerical modeling \citep[e.g.,][]{2017LRSP...14....4B,2019ApJ...886L..21T,2021NatAs...5.1100H}, one can evaluate the XUV/FUV fluxes and its impact on the astrospheric space weather.
Furthermore, \cite{2022ApJS..262...46T} derived the power-law relations between fluxes of many emission lines/bands from X-ray to radio (F) in the equation of $F = 10^{\beta_{\rm T}} (\Phi-\Phi_0)^{\alpha_{\rm T}} + F_{0}$. 
Combining the power-law relations in \cite{2022ApJS..262...46T} and this study, estimations of stellar XUV/FUV spectrum $I$($\lambda$) can be derived from the fluxes of emission lines $F$ as 
\begin{equation}
I(\lambda) = 10^{\beta(\lambda) - \beta_{\rm T}\alpha(\lambda)/\alpha_{\rm T}} (F - F_0)^{\alpha(\lambda) - \alpha_{\rm T}} + I_{0}(\lambda),
\end{equation}
where $\alpha(\lambda)$ and $\beta(\lambda)$ are parameters given in this study (Tables \ref{tab:2} and \ref{tab:3}) and $\alpha_{\rm T}$ and $\beta_{\rm T}$ are those given in \cite{2022ApJS..262...46T} (Tables 1, 2, and 3 therein). 

We tested the efficacy of our estimation method for three nearby young Sun-like stars with known total unsigned magnetic fluxes. The estimated spectra for these stars are available at \url{https://github.com/KosukeNamekata/StellarXUV.git}. We found that the observations and empirical extensions agree over a wide range of wavelengths within an-order-of-magnitude error bars.
In particular, the X-ray bands ($<$2 nm) and EUV bands (30--36 nm) have good agreement between the extrapolations and observations not only for continuum but also for emission lines. 
The consistency around the missing EUV range (36-92 nm) supports the prediction capability of our method in this EUV wavelength. 
Also, we noticed that the extrapolated FUV spectra with SORCE/SOLSTICE data shows better consistency with stellar FUV data (Figure \ref{fig:6-2}(d-f)) rather than those with TIMED/SEE which is used as a primary focus in our study (see Section \ref{sec:2}), although the valid wavelength coverage of SORCE/SOLSTICE is limited (115-150 nm) and some cautions are needed with the usage of SORCE data (see Appendix \ref{app:a}).


We should note that in this verification with three young Sun-like stars, the scaling laws tend to overestimate stellar fluxes by a factor of 5 or up to one order of magnitude at some wavelengths (especially for the 2-30 nm XUV range).
This does not necessarily suggest the deficiency in the scaling law, since the problem can be divided into three possible causes: (1) uncertainties in stellar observations, (2) uncertainties in solar observations, and (3) the limits of solar extension.
First, the stellar fluxes in the X-ray to EUV bands are provided by three different satellites (ASCA, ROSAT, and EUVE), and they were observed at the different epochs (e.g., 1994 May 24 for ASCA, 1993 October 19 for ROSAT, and 1995 December 6 for EUVE for EK Dra). 
Therefore, the broad spectra may have varied due to the various phases of stellar activity, and thus cannot be consistently compared. 
Note also that the measurements of the magnetic flux were taken 10 to 20 years later as compared to the XUV observations of the star used in our study \citep[from 2006 to 2016 for EK Dra;][]{2020A&A...635A.142K}.
Since the magnetic flux often varies by a factor of 2--3 depending on the activity phase as in the case of the Sun (see Figure \ref{fig:2}), the magnetic flux when the XUV spectra were observed may be smaller, and the difference between observation and model may be narrower.
In addition, the magnetic fluxes derived from spectral lines in the optical band of active stars may suffer from the rotational broadening of stars \citep{2020A&A...635A.142K}. Thus, it is possible that the magnetic flux could be a bit overestimated for the rapidly rotating stars in our target, so this may contribute to the excess of the reconstructed flux in the band of 2--30 nm.
Observations of the diagnostic lines in the near infrared band would be required for more reliable reconstruction of the surface magnetic flux because the Zeeman broadening is proportional to wavelengths \citep[e.g.,][]{2022A&A...662A..41R}.

Second, as it follows from Figure \ref{fig:2}, emission fluxes from 2--30 nm and the FUV bands derived from TIMED and SORCE data do not match to each other. 
Note that the level-4 XPS spectra used in our analysis are  calculated with the combination of the CHIANTI model spectra, and there would be some calibration issues even in solar spectra (see Section \ref{sec:2}), and more well-defined long-term spectra may be required for further accurate extrapolations to active stars.

Third, there are two factors of uncertainties, stellar coronal abundance and contribution of flare emission to various spectral bands, as the line fluxes are proportional to the abundance of with low first ionization potential \citep{2015RSPTA.37340259T}.
The differences in stellar coronal abundance between the Sun and other stars can affect the XUV fluxes to some extent. 
The reported coronal abundance (e.g., C, N, O, Ne, Mg, Si, Fe) of EK Dra is, however, smaller than the solar one, according to XMM-Newton observations \citep[see Table 3 in ][]{2005A&A...432..671S}, so this factor may contribute to the discrepancy to some extent because small coronal Fe abundance result in weak XUV emission in general \citep{1993ApJS...88..253S}.
In addition, the contribution of frequent giant flares to the quiescent XUV component, especially for the highly magnetically active star EK Dra and X-ray bright stars, can possibly account for the change in SED in the X-ray band.
\cite{2000ApJ...541..396A} concluded that the occurrence frequency of X-ray flares ($>10^{32}$ erg) is proportional to the X-ray luminosity of the star.
Also, observations of solar X-class flares (energy $\sim$10$^{32}$ ergs) suggest that the typical enhancement in X-ray flux at the flare peak by two orders of magnitude corresponds to the increase in the EUV flux by a factor of 2. 
Thus, large flares have greater contribution in the increase of the X-ray flux than to the EUV flux \citep[][]{2010JGRA..11512330M,2018SpWea..16.1470C}. 


Based on the discussion presented above, we conclude that our method based on the magnetic flux presented here can be useful to evaluate the XUV/FUV with an one-order-of-magnitude errors, but further validations are necessary by using nearly simultaneous observations of stars.
In particular, our scaling relations consistently predict the flux around missing EUV range, so they can work well to estimate the missing EUV spectrum that is especially important for exoplanet studies.
Note that our method is tested for Sun-like stars (G-dwarfs), and more careful validation is necessary for the application to cooler K/M-dwarfs because of the photospheric contribution to the emission in the FUV.
Finally, we would like to summarize the studies that need to be performed in the near future. 
First, XUV/FUV spectra used in this study are derived from the data taken by different instruments at different epochs \citep{2005ApJ...622..680R}, and thus, nearly simultaneous observations of XUV/FUV spectra and measurements of magnetic fluxes are needed for further validation of our method.
Next, in order to investigate the contribution of flares to the XUV spectrum, how the XUV spectrum of flares changes with respect to the energy scale should be tested on the Sun and applied to stars \citep[cf.,][]{2014ApJ...793...70M,2018SpWea..16.1470C,2021NatAs...5..697V}. 
Also, in the present study, our method is validated only for Sun-like (G-type) stars. 
It is important to verify the applicability to K and M-dwarfs as well for the evaluation of the impacts on the planets around cool dwarfs, which has been the focus of recent studies \citep[e.g.,][]{2016ApJ...820...89F,2019SPIE11118E..08F,2016ApJ...824..101Y,2017ApJ...843...31Y,2016ApJ...824..102L,2018ApJ...867...71L,2019ApJ...871L..26F,2021ApJ...911...18W}.
Early study by \cite{2003ApJ...598.1387P} have shown the universal correlation between X-ray and magnetic fluxes that seems to be common for Sun-like stars and M-dwarfs, indicating the applicability of the solar scaling laws to M-dwarfs. However, \cite{2022A&A...662A..41R} derived much steeper log($L_{\rm X}$) vs log($\Phi$) index , $\alpha$ = 1.58 for 292 M-dwarfs than that of the Sun, $\alpha$ = 1.16, as derived by \cite{2022ApJS..262...46T}. In addition, \cite{2014ApJ...780...61L,2020ApJ...902....3L} showed that the relations between X-ray and Ly$\alpha$ emission are spectral class dependent, and thus some physics-based adjustments may be required to apply our method to M-dwarfs (e.g., basal-level fluxes). 
Specifically, recent study by \cite{2021ApJ...919...29S} suggests that the chromospheres, transition regions and coronae of M dwarfs are strongly magnetized and stratified environments as compared to earlier type (G and K) stars, which can possibly reflect various contributions of the wave heating with respect to magnetic reconnection (or nanoflare) driven atmospheric heating.
Finally, a more detailed comparison between our empirical method and MHD numerical simulations \citep[e.g.,][]{2021A&A...656A.111S} should be performed to refine the physical interpretation of the empirical methods and stellar atmospheric heating mechanisms in the near future.

\begin{acknowledgments}
The authors thank Dr. Thomas Woods for assistance with the solar XUV and FUV datasets.
The authors thank Dr. Ignasi Ribas for providing us the observed XUV and FUV spectra of Sun-like stars.
The authors thank Dr. Thomas R. Ayres for assistance with the usage of Hubble data.
The authors thank Dr. Yang Liu for the analysis of the MDI data.
The authors thank Dr. Shinsuke Takasao for fruitful comments.
Data are courtesy of the science teams of SDO, SoHO, SORCE, TIMED, and Hubble Space Telescope.
HMI and EVE are instruments on board SDO, a mission for NASA's Living With a Star program.
SOHO is a project of international cooperation between ESA and NASA.
The results presented in this document rely on data measured from SORCE and TIMED which were accessed via the LASP Interactive Solar Irradiance Datacenter (LISIRD) (\url{https://lasp.colorado.edu/lisird/}).
This research is based on observations made with the NASA/ESA Hubble Space Telescope obtained from the Space Telescope Science Institute, which is operated by the Association of Universities for Research in Astronomy, Inc., under NASA contract NAS 5-26555.
Some of the data presented in this paper were obtained from the Mikulski Archive for Space Telescopes (MAST) at the Space Telescope Science Institute. The specific observations analyzed can be accessed via \dataset[10.17909/33p8-6v73]{https://doi.org/10.17909/33p8-6v73}.
This work was supported by JSPS KAKENHI Grant Nos. JP21J00316 (PI: K. Namekata), JP20KK0072 (PI: S. Toriumi), JP21H01124 (PI: T. Yokoyama), JP21H04492 (PI: K. Kusano), JP21J00106 (PI: Y. Notsu), and JP22K14077 (PI: M. Shoda). V.S.A. was supported by the GSFC Sellers Exoplanet Environments Collaboration (SEEC), which is funded by the NASA Planetary Science Division's Internal Scientist Funding Model (ISFM), NASA's TESS Cycle 1, HST Cycle 27 and NICER Cycle 2 project funds.
Y.N. was also supported by NASA ADAP award program Number 80NSSC21K0632 (PI: Adam Kowalski).
\end{acknowledgments}

\begin{deluxetable*}{lcccc}
\tabletypesize{\footnotesize}
\tablecaption{Formula for estimating stellar XUV fluxes in 10 nm bin as a function of total unsigned flux $\Phi$ of stars (in the unit of erg s$^{-1}$ cm$^{-2}$ nm$^{-1}$ at 1 AU). The data are plotted in Figure \ref{fig:4} with orange dots.}
\tablewidth{0pt}
\tablehead{
\colhead{Wavelength} & \colhead{C.C.} & \multicolumn{3}{c}{$I(\lambda) = I_{0}(\lambda) + 10^{\beta(\lambda)}  (\Phi - \Phi_{0})^{\alpha(\lambda)} $ } \\
\cline{3-5} \\
\colhead{Band [nm]} &  & $I_{0}(\lambda)$ & $\alpha(\lambda)$ & $\beta(\lambda)$ 
}
\startdata
0 -- 10  nm &  0.92 & 1.74$\times$10$^{-2}$ & 1.14 $_{\pm 0.01 }$ & -27.5 $_{\pm 0.01 }$\\
10 -- 20  nm &  0.91 & 4.39$\times$10$^{-2}$ & 1.14 $_{\pm 0.01 }$ & -27.8 $_{\pm 0.01 }$\\
20 -- 30  nm &  0.92 & 3.31$\times$10$^{-2}$ & 1.12 $_{\pm 0.01 }$ & -27.3 $_{\pm 0.01 }$\\
30 -- 40  nm &  0.91 & 5.96$\times$10$^{-2}$ & 1.13 $_{\pm 0.01 }$ & -27.7 $_{\pm 0.01 }$\\
40 -- 50  nm &  0.93 & 7.87$\times$10$^{-3}$ & 0.91 $_{\pm 0.01 }$ & -23.2 $_{\pm 0.01 }$\\
50 -- 60  nm &  0.92 & 1.15$\times$10$^{-2}$ & 0.85 $_{\pm 0.01 }$ & -22.0 $_{\pm 0.01 }$\\
60 -- 70  nm &  0.87 & 9.64$\times$10$^{-3}$ & 0.86 $_{\pm 0.01 }$ & -22.2 $_{\pm 0.01 }$\\
70 -- 80  nm &  0.88 & 1.13$\times$10$^{-2}$ & 0.95 $_{\pm 0.01 }$ & -24.7 $_{\pm 0.01 }$\\
80 -- 90  nm &  0.93 & 2.30$\times$10$^{-2}$ & 0.88 $_{\pm 0.00 }$ & -22.4 $_{\pm 0.00 }$\\
90 -- 100  nm &  0.90 & 2.88$\times$10$^{-2}$ & 0.89 $_{\pm 0.01 }$ & -22.5 $_{\pm 0.01 }$\\
100 -- 110  nm &  0.87 & 2.94$\times$10$^{-2}$ & 1.17 $_{\pm 0.01 }$ & -28.8 $_{\pm 0.01 }$\\
110 -- 120  nm &  0.87 & 3.07$\times$10$^{-2}$ & 1.29 $_{\pm 0.01 }$ & -31.9 $_{\pm 0.01 }$\\
120 -- 130  nm &  0.86 & 6.93$\times$10$^{-1}$ & 1.27 $_{\pm 0.01 }$ & -29.9 $_{\pm 0.01 }$\\
130 -- 140  nm &  0.87 & 6.00$\times$10$^{-2}$ & 1.25 $_{\pm 0.01 }$ & -30.8 $_{\pm 0.01 }$\\
140 -- 150  nm &  0.84 & 5.88$\times$10$^{-2}$ & 1.05 $_{\pm 0.01 }$ & -26.5 $_{\pm 0.01 }$\\
150 -- 160  nm &  0.77 & 1.46$\times$10$^{-1}$ & 1.10 $_{\pm 0.01 }$ & -27.3 $_{\pm 0.01 }$\\
160 -- 170  nm &  0.63 & 3.37$\times$10$^{-1}$ & 1.23 $_{\pm 0.02 }$ & -30.3 $_{\pm 0.02 }$\\
170 -- 180  nm &  0.45 & 1.06$\times$10$^{0}$ & 1.28 $_{\pm 0.02 }$ & -31.0 $_{\pm 0.02 }$\\
\enddata
\tablecomments{$\Phi$ is given in the unit of Mx. $\Phi_0$ is the basal level of the magnetic flux which is given as 1.18$\times 10^{23}$ Mx. The scaling relations of the wavelength bands above 100 nm are derived only from TIMED/SEE data, and those of the wavelength bands below 40 nm are derived only from SORCE/XPS data.
}
\label{tab:2}
\end{deluxetable*}

\begin{deluxetable*}{ccccc}
\tabletypesize{\footnotesize}
\tablecaption{Formula for estimating stellar XUV fluxes in best wavelength resolution as a function of total unsigned flux $\Phi$ of stars (in the unit of erg s$^{-1}$ cm$^{-2}$ nm$^{-1}$ at 1 AU). The data are plotted in Figure \ref{fig:4} with blue lines.}
\tablewidth{0pt}
\tablehead{
\colhead{Wavelength} & \colhead{C.C.} & \multicolumn{3}{c}{$I(\lambda) = I_{0}(\lambda) + 10^{\beta(\lambda)} (\Phi - \Phi_{0})^{\alpha(\lambda)} $ } \\
\cline{3-5} \\
\colhead{[nm]} &  & $I_{0}(\lambda)$ & $\alpha(\lambda)$ & $\beta(\lambda)$ 
}
\startdata
0.10  &  0.65 & 1.76$\times$10$^{-15}$ & 11.30 $_{\pm 0.27 }$ & -266.9 $_{\pm 6.20 }$\\
0.20  &  0.65 & 2.58$\times$10$^{-10}$ & 7.20 $_{\pm 0.16 }$ & -170.2 $_{\pm 3.67 }$\\
0.30  &  0.66 & 8.67$\times$10$^{-8}$ & 4.11 $_{\pm 0.08 }$ & -98.9 $_{\pm 1.92 }$\\
0.40  &  0.71 & 1.84$\times$10$^{-6}$ & 2.72 $_{\pm 0.05 }$ & -66.3 $_{\pm 1.08 }$\\
0.50  &  0.78 & 1.19$\times$10$^{-5}$ & 1.83 $_{\pm 0.03 }$ & -45.6 $_{\pm 0.57 }$\\
0.60  &  0.83 & 5.45$\times$10$^{-5}$ & 1.51 $_{\pm 0.02 }$ & -37.9 $_{\pm 0.40 }$\\
0.70  &  0.85 & 2.74$\times$10$^{-4}$ & 1.38 $_{\pm 0.01 }$ & -34.3 $_{\pm 0.33 }$\\
0.80  &  0.89 & 2.20$\times$10$^{-4}$ & 1.26 $_{\pm 0.01 }$ & -31.5 $_{\pm 0.23 }$\\
0.90  &  0.91 & 4.06$\times$10$^{-4}$ & 1.22 $_{\pm 0.01 }$ & -30.4 $_{\pm 0.21 }$\\
1.00  &  0.90 & 1.16$\times$10$^{-3}$ & 1.20 $_{\pm 0.01 }$ & -29.5 $_{\pm 0.21 }$\\
$\cdots$ & $\cdots$ & $\cdots$ & $\cdots$ & $\cdots$ \\
170.50  &  0.52 & 6.37$\times$10$^{-1}$ & 1.15 $_{\pm 0.02 }$ & -28.0 $_{\pm 0.41 }$\\
171.50  &  0.54 & 6.43$\times$10$^{-1}$ & 1.11 $_{\pm 0.02 }$ & -27.0 $_{\pm 0.38 }$\\
172.50  &  0.45 & 7.28$\times$10$^{-1}$ & 1.51 $_{\pm 0.03 }$ & -36.4 $_{\pm 0.74 }$\\
173.50  &  0.45 & 7.23$\times$10$^{-1}$ & 1.60 $_{\pm 0.04 }$ & -38.6 $_{\pm 0.81 }$\\
174.50  &  0.43 & 8.85$\times$10$^{-1}$ & 1.25 $_{\pm 0.02 }$ & -30.3 $_{\pm 0.54 }$\\
175.50  &  0.46 & 1.08$\times$10$^{0}$ & 1.25 $_{\pm 0.02 }$ & -30.3 $_{\pm 0.54 }$\\
176.50  &  0.42 & 1.16$\times$10$^{0}$ & 0.99 $_{\pm 0.02 }$ & -24.3 $_{\pm 0.42 }$\\
177.50  &  0.36 & 1.43$\times$10$^{0}$ & 1.15 $_{\pm 0.02 }$ & -27.7 $_{\pm 0.54 }$\\
178.50  &  0.36 & 1.62$\times$10$^{0}$ & 1.23 $_{\pm 0.03 }$ & -29.8 $_{\pm 0.66 }$\\
179.50  &  0.38 & 1.64$\times$10$^{0}$ & 1.02 $_{\pm 0.02 }$ & -24.8 $_{\pm 0.44 }$\\
\enddata
\tablecomments{$\Phi$ is given in the unit of Mx. $\Phi_0$ is the basal level of the magnetic flux which is given as 1.18$\times 10^{23}$ Mx. The full table is available online.
}
\label{tab:3}
\end{deluxetable*}

\begin{deluxetable*}{ccccccccccc}
\tabletypesize{\footnotesize}
\tablecaption{Characteristics of the Sun-like stars whose total unsigned magnetic fluxes and XUV/FUV spectra are available.}
\tablewidth{0pt}
\tablehead{
\colhead{Name} & \colhead{Sp. type$^\S$} & \colhead{$T_{\rm eff}$$^\S$} & \colhead{log $g$$^\S$} & \colhead{Age$^\S$}  & \colhead{$P_{\rm rot}$$^\S$} & \colhead{$R$$^\S$} & \colhead{distance$^{\ast}$} & \colhead{$\Phi$$^{\S,\dagger}$} &  \colhead{ref. of XUV data}  \\
& & (K) & & (Myr) & (d) & ($R_{\odot}$) & (pc) & (Mx) &
}
\startdata
EK Dra & G1.5V & 5845 & 4.47 & 120 & 2.6 & 0.97$^{+0.02}_{-0.03}$ & 34.4 & 4.06$^{+0.44}_{-0.44}\times10^{25}$ & \cite{2005ApJ...622..680R} \\
$\pi^1$ UMa & G1.5V & 5873 & 4.44 & 500 & 4.9 & 0.95$^{+0.01}_{-0.01}$ & 14.4 & 1.64$^{+0.20}_{-0.22}\times10^{25}$ & \cite{2005ApJ...622..680R} \\
$\kappa^1$ Ceti &  G5V &  5742 &  4.49 &  600 &  9.3 &  0.95$^{+0.03}_{-0.02}$ & 9.13 & 1.39$^{+0.24}_{-0.22}\times10^{25}$ & \cite{2005ApJ...622..680R} \\
\hline
Sun (min) & G2V & 5777 & 4.44 & 4600 & 25.4 & 1.00 & (1 AU) & 1.18$\times10^{23}$$^\ddag$ & This study  \\
Sun (max) & -- & -- & -- &-- & -- & -- & -- &  3.35$\times$10$^{23}$ & -- 
\enddata
\tablecomments{
$^\S$These values are taken from \cite{2022ApJ...927..179T}. The total unsigned magnetic fluxes $\Phi$ were re-calculated from surface-averaged magnetic flied strength and stellar radii, and its errors were calculated from the errors of magnetic field strength. The radius error bars are taken from \cite{2002NewA....7..211G} and \cite{2007ApJS..168..297T}.
$^{\ast}$The distances to stars are calculated by the parallaxes provided by the Gaia Early Data Release 3.
$^\dagger$The stellar magnetic flux is calculated as a hemispheric value.
$^\ddag$Only the basal level of total unsigned magnetic flux of the Sun was re-calculated in this paper.
}
\label{tab:4}
\end{deluxetable*}

\begin{deluxetable*}{cccccccccc}
\tabletypesize{\footnotesize}
\tablecaption{Power-law indexes and correlation coefficients (C.C.) for the relation between irradiances of EUV emission lines and total unsigned magnetic flux.}
\tablewidth{0pt}
\tablehead{
\colhead{Wavelength [nm]} & \colhead{Atoms} & \colhead{log($T$/K])$^\S$} & \colhead{C.C.}& \colhead{$\alpha$}& \colhead{data}\\
}
\startdata
28.4 & Fe XV & 6.30 & 0.92 & 1.13 $_{\pm 0.01 }$ & (1)\\
30.4 & He II & 4.75 & 0.91 & 1.14 $_{\pm 0.01 }$ & (1)\\
33.5 & Fe XVI & 6.35 & 0.90 & 1.00 $_{\pm 0.01 }$ & (2)\\
36.1 & Fe XVI & 6.35 & 0.93 & 0.88 $_{\pm 0.01 }$ & (2)\\
97.7 & C III & 4.68 & 0.76 & 0.94 $_{\pm 0.01 }$ & (2)\\
102.6 & H I & 3.84 & 0.91 & 0.94 $_{\pm 0.01 }$ & (2)\\
103.2 & O VI & 5.42 & 0.84 & 0.97 $_{\pm 0.01 }$ & (2)\\
103.8 & O VI & 5.42 & 0.91 & 0.99 $_{\pm 0.01 }$ & (2)\\
\enddata
\tablecomments{(1) SORCE/XPS data. (2) SDO/EVE data. $^\S$The temperature values are taken from \cite{2005ApJ...622..680R}.
}
\label{tab:6}
\end{deluxetable*}

\clearpage

\appendix

\section{Alternative empirical relations for high-resolution FUV Spectrum derived from the data of SORCE/SOLSTICE}\label{app:a}

This section describes the results of the analysis using SORCE/SOLSTICE data that covers 115--310 nm.
Only data below 150 nm were analyzed due to long-term calibration issues (see, Section \ref{data:see}).
This analysis with the SORCE/SOLSTICE data does not provide whole FUV spectrum (92--180 nm) but enough cover some strong emission lines in FUV range, such as Ly$\alpha$, with higher spectral resolution (0.1 nm) than that of TIMED/SEE (1 nm).
Therefore, the result obtained with the SORCE/SOLSTICE data will enable us to investigate solar and stellar spectral lines in detail.

Using the method described in Section \ref{sec:power-law}, power-law relations between FUV irradiance (150--150 nm) and total unsigned magnetic fluxes are derived for each wavelength $\lambda$ in the formula of Equation (\ref{eq:1}) or (\ref{eq:2}).
The periods of data used for our analysis are summarized in Table \ref{tab:1}.
Table \ref{tab:app1} summarizes the fitted parameter $\alpha(\lambda)$ and $\beta(\lambda)$ for each wavelength.
We found that the power-law indexes $\alpha(\lambda)$ obtained by using SORCE/SOLSTICE are less than unity for almost all wavelength at 150--150 nm, while those by TIMED/SEE are mostly unity or above-unity (see, Figure \ref{fig:4} and Tables \ref{tab:2} and \ref{tab:3}).


\section{Alternative empirical relations for low-resolution XUV spectrum derived from the data of TIMED/SEE}\label{app:b}

In this section, we describe results of a supplementary analysis of the empirical scaling relations of XUV spectrum at 0.5--106.5 nm obtained by TIMED/SEE.
The data is not used in the main text because of the reason described in Section \ref{data:xps} and \ref{data:eve}.
The fitted parameters in the power-law relations (Equation (\ref{eq:1}) or (\ref{eq:2})) for each 10 nm bin and each 1 nm wavelength are summarized in Tables \ref{tab:app3} and \ref{tab:app3}, respectively.
By combining the data in Tables \ref{tab:2} and \ref{tab:3}, the scaling relations for the wavelength between 0.5 and 180 nm can be obtained only from the TIMED/SEE data.
However, the power-law indexes in XUV range is relatively large $\sim$1.5, and we found that the scaling laws largely overestimate the stellar XUV irradiance by an order of magnitude, especially in 0--30 nm.
For this reason, we used the more recent and calibrated XUV spectra obtained by SORCE/XPS and SDO/EVE for our primary focus.

\clearpage

\begin{deluxetable}{ccccc}
\tabletypesize{\footnotesize}
\tablecaption{Formula for estimating stellar FUV fluxes in 115--150 nm range as a function of total unsigned flux $\Phi$ of stars (in the unit of erg s$^{-1}$ cm$^{-2}$ nm$^{-1}$ at 1 AU)  derived only from SORCE/SOLSTICE data.}
\tablewidth{0pt}
\tablehead{
\colhead{Wavelength} & \colhead{C.C.} & \multicolumn{3}{c}{$I(\lambda) = I_{0}(\lambda) + 10^{\beta(\lambda)} (\Phi - \Phi_{0})^{\alpha(\lambda)} $ } \\
\cline{3-5} \\
\colhead{[nm]} &  & $I_{0}(\lambda)$ & $\alpha(\lambda)$ & $\beta(\lambda)$ 
}
\startdata
115.013  &  0.42 & 1.19$\times$10$^{-2}$ & 0.60 $_{\pm 0.01 }$ & -16.3 $_{\pm 0.32 }$\\
115.037  &  0.47 & 1.28$\times$10$^{-2}$ & 0.59 $_{\pm 0.01 }$ & -16.0 $_{\pm 0.30 }$\\
115.062  &  0.46 & 1.31$\times$10$^{-2}$ & 0.61 $_{\pm 0.01 }$ & -16.5 $_{\pm 0.31 }$\\
115.088  &  0.45 & 1.22$\times$10$^{-2}$ & 0.62 $_{\pm 0.01 }$ & -16.8 $_{\pm 0.33 }$\\
115.112  &  0.42 & 1.18$\times$10$^{-2}$ & 0.60 $_{\pm 0.01 }$ & -16.4 $_{\pm 0.34 }$\\
115.138  &  0.53 & 1.40$\times$10$^{-2}$ & 0.64 $_{\pm 0.01 }$ & -17.1 $_{\pm 0.29 }$\\
115.162  &  0.57 & 1.83$\times$10$^{-2}$ & 0.59 $_{\pm 0.01 }$ & -16.0 $_{\pm 0.26 }$\\
115.188  &  0.59 & 2.23$\times$10$^{-2}$ & 0.63 $_{\pm 0.01 }$ & -16.7 $_{\pm 0.25 }$\\
115.213  &  0.65 & 2.38$\times$10$^{-2}$ & 0.59 $_{\pm 0.01 }$ & -15.7 $_{\pm 0.22 }$\\
115.237  &  0.59 & 2.29$\times$10$^{-2}$ & 0.62 $_{\pm 0.01 }$ & -16.4 $_{\pm 0.25 }$\\
$\cdots$ & $\cdots$ & $\cdots$ & $\cdots$ & $\cdots$ \\
149.738  &  0.61 & 6.82$\times$10$^{-2}$ & 0.63 $_{\pm 0.01 }$ & -16.6 $_{\pm 0.25 }$\\
149.762  &  0.65 & 6.91$\times$10$^{-2}$ & 0.71 $_{\pm 0.01 }$ & -18.7 $_{\pm 0.27 }$\\
149.787  &  0.63 & 6.98$\times$10$^{-2}$ & 0.70 $_{\pm 0.01 }$ & -18.3 $_{\pm 0.28 }$\\
149.812  &  0.63 & 7.03$\times$10$^{-2}$ & 0.71 $_{\pm 0.01 }$ & -18.7 $_{\pm 0.27 }$\\
149.838  &  0.64 & 7.06$\times$10$^{-2}$ & 0.69 $_{\pm 0.01 }$ & -18.1 $_{\pm 0.27 }$\\
149.863  &  0.64 & 7.09$\times$10$^{-2}$ & 0.69 $_{\pm 0.01 }$ & -18.1 $_{\pm 0.26 }$\\
149.887  &  0.61 & 7.09$\times$10$^{-2}$ & 0.68 $_{\pm 0.01 }$ & -17.9 $_{\pm 0.26 }$\\
149.912  &  0.64 & 7.10$\times$10$^{-2}$ & 0.74 $_{\pm 0.01 }$ & -19.3 $_{\pm 0.27 }$\\
149.938  &  0.61 & 7.14$\times$10$^{-2}$ & 0.67 $_{\pm 0.01 }$ & -17.6 $_{\pm 0.27 }$\\
149.963  &  0.59 & 7.34$\times$10$^{-2}$ & 0.65 $_{\pm 0.01 }$ & -17.1 $_{\pm 0.25 }$\\
\enddata
\tablecomments{$\Phi$ is given in the unit of Mx. $\Phi_0$ is the basal level of the magnetic flux which is given as 1.18$\times 10^{23}$ Mx. The full table is available online.
}
\label{tab:app1}
\end{deluxetable}


\begin{deluxetable}{ccccc}
\tabletypesize{\footnotesize}
\tablecaption{Formula for estimating stellar XUV fluxes in 10--100 nm range as a function of total unsigned flux $\Phi$ of stars (in the unit of erg s$^{-1}$ cm$^{-2}$ nm$^{-1}$ at 1 AU) derived only from TIMED/SEE data.}
\tablewidth{0pt}
\tablehead{
\colhead{Wavelength} & \colhead{C.C.} & \multicolumn{3}{c}{$I(\lambda) = I_{0}(\lambda) + 10^{\beta(\lambda)} (\Phi - \Phi_{0})^{\alpha(\lambda)} $ } \\
\cline{3-5} \\
\colhead{[nm]} &  & $I_{0}(\lambda)$ & $\alpha(\lambda)(\lambda)$ & $\beta(\lambda)$ 
}
\startdata
0 -- 10  nm &  0.85 & 1.67$\times$10$^{-2}$ & 1.53 $_{\pm 0.01 }$ & -36.6 $_{\pm 0.01}$\\
10 -- 20  nm &  0.86 & 4.54$\times$10$^{-2}$ & 1.52 $_{\pm 0.01 }$ & -36.1 $_{\pm 0.01}$\\
20 -- 30  nm &  0.85 & 3.20$\times$10$^{-2}$ & 1.51 $_{\pm 0.01 }$ & -35.9 $_{\pm 0.01}$\\
30 -- 40  nm &  0.88 & 5.75$\times$10$^{-2}$ & 1.24 $_{\pm 0.01 }$ & -29.9 $_{\pm 0.01}$\\
40 -- 50  nm &  0.87 & 8.61$\times$10$^{-3}$ & 1.30 $_{\pm 0.01 }$ & -32.3 $_{\pm 0.01}$\\
50 -- 60  nm &  0.88 & 1.21$\times$10$^{-2}$ & 0.95 $_{\pm 0.01 }$ & -24.0 $_{\pm 0.01}$\\
60 -- 70  nm &  0.85 & 1.13$\times$10$^{-2}$ & 0.91 $_{\pm 0.01 }$ & -23.4 $_{\pm 0.01}$\\
70 -- 80  nm &  0.81 & 9.44$\times$10$^{-3}$ & 1.10 $_{\pm 0.01 }$ & -28.0 $_{\pm 0.01}$\\
80 -- 90  nm &  0.87 & 2.25$\times$10$^{-2}$ & 1.06 $_{\pm 0.01 }$ & -26.5 $_{\pm 0.01}$\\
90 -- 100  nm &  0.87 & 2.78$\times$10$^{-2}$ & 0.98 $_{\pm 0.01 }$ & -24.4 $_{\pm 0.01}$\\
\enddata
\tablecomments{$\Phi$ is given in the unit of Mx. $\Phi_0$ is the basal level of the magnetic flux which is given as 1.18$\times 10^{23}$ Mx.
}
\label{tab:app2}
\end{deluxetable}

\begin{deluxetable}{ccccc}
\tabletypesize{\footnotesize}
\tablecaption{Formula for estimating stellar XUV fluxes in 0.5--106.5 nm range as a function of total unsigned flux $\Phi$ of stars (in the unit of erg s$^{-1}$ cm$^{-2}$ nm$^{-1}$ at 1 AU)  derived only from TIMED/SEE data.}
\tablewidth{0pt}
\tablehead{
\colhead{Wavelength} & \colhead{C.C.} & \multicolumn{3}{c}{$I(\lambda) = I_{0}(\lambda) + 10^{\beta(\lambda)} (\Phi - \Phi_{0})^{\alpha(\lambda)} $ } \\
\cline{3-5} \\
\colhead{[nm]} &  & $I_{0}(\lambda)$ & $\alpha(\lambda)$ & $\beta(\lambda)$ 
}
\startdata
0.50  &  0.85 & 5.20$\times$10$^{-4}$ & 1.55 $_{\pm 0.01 }$ & -37.7 $_{\pm 0.27 }$\\
1.50  &  0.85 & 1.38$\times$10$^{-2}$ & 1.56 $_{\pm 0.01 }$ & -36.8 $_{\pm 0.27 }$\\
2.50  &  0.86 & 1.61$\times$10$^{-2}$ & 1.52 $_{\pm 0.01 }$ & -36.5 $_{\pm 0.26 }$\\
3.50  &  0.86 & 1.09$\times$10$^{-2}$ & 1.52 $_{\pm 0.01 }$ & -36.7 $_{\pm 0.26 }$\\
4.50  &  0.86 & 1.98$\times$10$^{-2}$ & 1.52 $_{\pm 0.01 }$ & -36.4 $_{\pm 0.26 }$\\
5.50  &  0.86 & 1.85$\times$10$^{-2}$ & 1.52 $_{\pm 0.01 }$ & -36.5 $_{\pm 0.26 }$\\
6.50  &  0.85 & 1.99$\times$10$^{-2}$ & 1.53 $_{\pm 0.01 }$ & -36.6 $_{\pm 0.27 }$\\
7.50  &  0.86 & 2.32$\times$10$^{-2}$ & 1.51 $_{\pm 0.01 }$ & -36.0 $_{\pm 0.26 }$\\
8.50  &  0.86 & 2.25$\times$10$^{-2}$ & 1.51 $_{\pm 0.01 }$ & -36.1 $_{\pm 0.26 }$\\
9.50  &  0.86 & 2.15$\times$10$^{-2}$ & 1.51 $_{\pm 0.01 }$ & -36.2 $_{\pm 0.26 }$\\
$\cdots$ & $\cdots$ & $\cdots$ & $\cdots$ & $\cdots$ \\
97.50  &  0.81 & 1.04$\times$10$^{-1}$ & 0.81 $_{\pm 0.01 }$ & -20.1 $_{\pm 0.16 }$\\
98.50  &  0.83 & 1.36$\times$10$^{-2}$ & 1.09 $_{\pm 0.01 }$ & -27.4 $_{\pm 0.20 }$\\
99.50  &  0.84 & 1.96$\times$10$^{-2}$ & 1.24 $_{\pm 0.01 }$ & -30.7 $_{\pm 0.22 }$\\
100.50  &  0.86 & 8.62$\times$10$^{-3}$ & 1.29 $_{\pm 0.01 }$ & -32.1 $_{\pm 0.22 }$\\
101.50  &  0.85 & 1.17$\times$10$^{-2}$ & 1.19 $_{\pm 0.01 }$ & -29.7 $_{\pm 0.20 }$\\
102.50  &  0.83 & 8.92$\times$10$^{-2}$ & 1.15 $_{\pm 0.01 }$ & -27.8 $_{\pm 0.21 }$\\
103.50  &  0.83 & 8.40$\times$10$^{-2}$ & 1.12 $_{\pm 0.01 }$ & -27.2 $_{\pm 0.20 }$\\
104.50  &  0.85 & 1.38$\times$10$^{-2}$ & 1.25 $_{\pm 0.01 }$ & -30.9 $_{\pm 0.22 }$\\
105.50  &  0.87 & 1.32$\times$10$^{-2}$ & 1.13 $_{\pm 0.01 }$ & -28.3 $_{\pm 0.18 }$\\
106.50  &  0.86 & 1.44$\times$10$^{-2}$ & 1.16 $_{\pm 0.01 }$ & -29.1 $_{\pm 0.19 }$\\
\enddata
\tablecomments{$\Phi$ is given in the unit of Mx. $\Phi_0$ is the basal level of the magnetic flux which is given as 1.18$\times 10^{23}$ Mx. The full table is available online.
}
\label{tab:app3}
\end{deluxetable}

\clearpage
\bibliography{stellarXUV}{}

\newcommand{\noop}[1]{}
\begin{thebibliography}{}
\expandafter\ifx\csname natexlab\endcsname\relax\def\natexlab#1{#1}\fi
\providecommand{\url}[1]{\href{#1}{#1}}
\providecommand{\dodoi}[1]{doi:~\href{http://doi.org/#1}{\nolinkurl{#1}}}
\providecommand{\doeprint}[1]{\href{http://ascl.net/#1}{\nolinkurl{http://ascl.net/#1}}}
\providecommand{\doarXiv}[1]{\href{https://arxiv.org/abs/#1}{\nolinkurl{https://arxiv.org/abs/#1}}}

\bibitem[{{Airapetian} {et~al.}(2016){Airapetian}, {Glocer}, {Gronoff},
  {H{\'e}brard}, \& {Danchi}}]{2016NatGe...9..452A}
{Airapetian}, V.~S., {Glocer}, A., {Gronoff}, G., {H{\'e}brard}, E., \&
  {Danchi}, W. 2016, Nature Geoscience, 9, 452, \dodoi{10.1038/ngeo2719}

\bibitem[{{Airapetian} {et~al.}(2017){Airapetian}, {Glocer}, {Khazanov},
  {Loyd}, {France}, {Sojka}, {Danchi}, \& {Liemohn}}]{2017ApJ...836L...3A}
{Airapetian}, V.~S., {Glocer}, A., {Khazanov}, G.~V., {et~al.} 2017, \apjl,
  836, L3, \dodoi{10.3847/2041-8213/836/1/L3}

\bibitem[{{Airapetian} {et~al.}(2020){Airapetian}, {Barnes}, {Cohen},
  {Collinson}, {Danchi}, {Dong}, {Del Genio}, {France}, {Garcia-Sage},
  {Glocer}, {Gopalswamy}, {Grenfell}, {Gronoff}, {G{\"u}del}, {Herbst},
  {Henning}, {Jackman}, {Jin}, {Johnstone}, {Kaltenegger}, {Kay}, {Kobayashi},
  {Kuang}, {Li}, {Lynch}, {L{\"u}ftinger}, {Luhmann}, {Maehara}, {Mlynczak},
  {Notsu}, {Osten}, {Ramirez}, {Rugheimer}, {Scheucher}, {Schlieder},
  {Shibata}, {Sousa-Silva}, {Stamenkovi{\'c}}, {Strangeway}, {Usmanov},
  {Vergados}, {Verkhoglyadova}, {Vidotto}, {Voytek}, {Way}, {Zank}, \&
  {Yamashiki}}]{2020IJAsB..19..136A}
{Airapetian}, V.~S., {Barnes}, R., {Cohen}, O., {et~al.} 2020, International
  Journal of Astrobiology, 19, 136, \dodoi{10.1017/S1473550419000132}

\bibitem[{{Audard} {et~al.}(2000){Audard}, {G{\"u}del}, {Drake}, \&
  {Kashyap}}]{2000ApJ...541..396A}
{Audard}, M., {G{\"u}del}, M., {Drake}, J.~J., \& {Kashyap}, V.~L. 2000, \apj,
  541, 396, \dodoi{10.1086/309426}

\bibitem[{{Ayres}(1997)}]{1997JGR...102.1641A}
{Ayres}, T.~R. 1997, \jgr, 102, 1641, \dodoi{10.1029/96JE03306}

\bibitem[{{Ayres}(2015)}]{2015AJ....150....7A}
---. 2015, \aj, 150, 7, \dodoi{10.1088/0004-6256/150/1/7}

\bibitem[{{Brun} \& {Browning}(2017)}]{2017LRSP...14....4B}
{Brun}, A.~S., \& {Browning}, M.~K. 2017, Living Reviews in Solar Physics, 14,
  4, \dodoi{10.1007/s41116-017-0007-8}

\bibitem[{{Chadney} {et~al.}(2015){Chadney}, {Galand}, {Unruh}, {Koskinen}, \&
  {Sanz-Forcada}}]{2015Icar..250..357C}
{Chadney}, J.~M., {Galand}, M., {Unruh}, Y.~C., {Koskinen}, T.~T., \&
  {Sanz-Forcada}, J. 2015, \icarus, 250, 357,
  \dodoi{10.1016/j.icarus.2014.12.012}

\bibitem[{{Chamberlin} {et~al.}(2018){Chamberlin}, {Woods}, {Didkovsky},
  {Eparvier}, {Jones}, {Machol}, {Mason}, {Snow}, {Thiemann}, {Viereck}, \&
  {Woodraska}}]{2018SpWea..16.1470C}
{Chamberlin}, P.~C., {Woods}, T.~N., {Didkovsky}, L., {et~al.} 2018, Space
  Weather, 16, 1470, \dodoi{10.1029/2018SW001866}

\bibitem[{{Claire} {et~al.}(2012){Claire}, {Sheets}, {Cohen}, {Ribas},
  {Meadows}, \& {Catling}}]{2012ApJ...757...95C}
{Claire}, M.~W., {Sheets}, J., {Cohen}, M., {et~al.} 2012, \apj, 757, 95,
  \dodoi{10.1088/0004-637X/757/1/95}

\bibitem[{{Clette}(2021)}]{2021JSWSC..11....2C}
{Clette}, F. 2021, Journal of Space Weather and Space Climate, 11, 2,
  \dodoi{10.1051/swsc/2020071}

\bibitem[{{Cruddace} {et~al.}(1974){Cruddace}, {Paresce}, {Bowyer}, \&
  {Lampton}}]{1974ApJ...187..497C}
{Cruddace}, R., {Paresce}, F., {Bowyer}, S., \& {Lampton}, M. 1974, \apj, 187,
  497, \dodoi{10.1086/152659}

\bibitem[{{Dere} {et~al.}(1997){Dere}, {Landi}, {Mason}, {Monsignori Fossi}, \&
  {Young}}]{1997A&AS..125..149D}
{Dere}, K.~P., {Landi}, E., {Mason}, H.~E., {Monsignori Fossi}, B.~C., \&
  {Young}, P.~R. 1997, \aaps, 125, 149, \dodoi{10.1051/aas:1997368}

\bibitem[{{Duvvuri} {et~al.}(2021){Duvvuri}, {Sebastian Pineda},
  {Berta-Thompson}, {Brown}, {France}, {Kowalski}, {Redfield}, {Tilipman},
  {Vieytes}, {Wilson}, {Youngblood}, {Froning}, {Linsky}, {Parke Loyd},
  {Mauas}, {Miguel}, {Newton}, {Rugheimer}, \& {Christian
  Schneider}}]{2021ApJ...913...40D}
{Duvvuri}, G.~M., {Sebastian Pineda}, J., {Berta-Thompson}, Z.~K., {et~al.}
  2021, \apj, 913, 40, \dodoi{10.3847/1538-4357/abeaaf}

\bibitem[{{Edl{\'e}n}(1943)}]{1943ZA.....22...30E}
{Edl{\'e}n}, B. 1943, \zap, 22, 30

\bibitem[{{France} {et~al.}(2018){France}, {Arulanantham}, {Fossati}, {Lanza},
  {Loyd}, {Redfield}, \& {Schneider}}]{2018ApJS..239...16F}
{France}, K., {Arulanantham}, N., {Fossati}, L., {et~al.} 2018, \apjs, 239, 16,
  \dodoi{10.3847/1538-4365/aae1a3}

\bibitem[{{France} {et~al.}(2016){France}, {Loyd}, {Youngblood}, {Brown},
  {Schneider}, {Hawley}, {Froning}, {Linsky}, {Roberge}, {Buccino},
  {Davenport}, {Fontenla}, {Kaltenegger}, {Kowalski}, {Mauas}, {Miguel},
  {Redfield}, {Rugheimer}, {Tian}, {Vieytes}, {Walkowicz}, \&
  {Weisenburger}}]{2016ApJ...820...89F}
{France}, K., {Loyd}, R.~O.~P., {Youngblood}, A., {et~al.} 2016, \apj, 820, 89,
  \dodoi{10.3847/0004-637X/820/2/89}

\bibitem[{{France} {et~al.}(2019){France}, {Fleming}, {Drake}, {Mason},
  {Youngblood}, {Bourrier}, {Fossati}, {Froning}, {Koskinen}, {Kruczek},
  {Lipscy}, {McEntaffer}, {Romaine}, {Siegmund}, \&
  {Wilkinson}}]{2019SPIE11118E..08F}
{France}, K., {Fleming}, B.~T., {Drake}, J.~J., {et~al.} 2019, in Society of
  Photo-Optical Instrumentation Engineers (SPIE) Conference Series, Vol. 11118,
  UV, X-Ray, and Gamma-Ray Space Instrumentation for Astronomy XXI, ed. O.~H.
  {Siegmund}, 1111808, \dodoi{10.1117/12.2526859}

\bibitem[{{Froning} {et~al.}(2019){Froning}, {Kowalski}, {France}, {Loyd},
  {Schneider}, {Youngblood}, {Wilson}, {Brown}, {Berta-Thompson}, {Pineda},
  {Linsky}, {Rugheimer}, \& {Miguel}}]{2019ApJ...871L..26F}
{Froning}, C.~S., {Kowalski}, A., {France}, K., {et~al.} 2019, \apjl, 871, L26,
  \dodoi{10.3847/2041-8213/aaffcd}

\bibitem[{{Gaia Collaboration} {et~al.}(2021){Gaia Collaboration}, {Brown},
  {Vallenari}, {Prusti}, {de Bruijne}, {Babusiaux}, {Biermann}, {Creevey},
  {Evans}, {Eyer}, {Hutton}, {Jansen}, {Jordi}, {Klioner}, {Lammers},
  {Lindegren}, {Luri}, {Mignard}, {Panem}, {Pourbaix}, {Randich}, {Sartoretti},
  {Soubiran}, {Walton}, {Arenou}, {Bailer-Jones}, {Bastian}, {Cropper},
  {Drimmel}, {Katz}, {Lattanzi}, {van Leeuwen}, {Bakker}, {Cacciari},
  {Casta{\~n}eda}, {De Angeli}, {Ducourant}, {Fabricius}, {Fouesneau},
  {Fr{\'e}mat}, {Guerra}, {Guerrier}, {Guiraud}, {Jean-Antoine Piccolo},
  {Masana}, {Messineo}, {Mowlavi}, {Nicolas}, {Nienartowicz}, {Pailler},
  {Panuzzo}, {Riclet}, {Roux}, {Seabroke}, {Sordo}, {Tanga}, {Th{\'e}venin},
  {Gracia-Abril}, {Portell}, {Teyssier}, {Altmann}, {Andrae}, {Bellas-Velidis},
  {Benson}, {Berthier}, {Blomme}, {Brugaletta}, {Burgess}, {Busso}, {Carry},
  {Cellino}, {Cheek}, {Clementini}, {Damerdji}, {Davidson}, {Delchambre},
  {Dell'Oro}, {Fern{\'a}ndez-Hern{\'a}ndez}, {Galluccio}, {Garc{\'\i}a-Lario},
  {Garcia-Reinaldos}, {Gonz{\'a}lez-N{\'u}{\~n}ez}, {Gosset}, {Haigron},
  {Halbwachs}, {Hambly}, {Harrison}, {Hatzidimitriou}, {Heiter},
  {Hern{\'a}ndez}, {Hestroffer}, {Hodgkin}, {Holl}, {Jan{\ss}en}, {Jevardat de
  Fombelle}, {Jordan}, {Krone-Martins}, {Lanzafame}, {L{\"o}ffler}, {Lorca},
  {Manteiga}, {Marchal}, {Marrese}, {Moitinho}, {Mora}, {Muinonen}, {Osborne},
  {Pancino}, {Pauwels}, {Petit}, {Recio-Blanco}, {Richards}, {Riello},
  {Rimoldini}, {Robin}, {Roegiers}, {Rybizki}, {Sarro}, {Siopis}, {Smith},
  {Sozzetti}, {Ulla}, {Utrilla}, {van Leeuwen}, {van Reeven}, {Abbas}, {Abreu
  Aramburu}, {Accart}, {Aerts}, {Aguado}, {Ajaj}, {Altavilla}, {{\'A}lvarez},
  {{\'A}lvarez Cid-Fuentes}, {Alves}, {Anderson}, {Anglada Varela}, {Antoja},
  {Audard}, {Baines}, {Baker}, {Balaguer-N{\'u}{\~n}ez}, {Balbinot}, {Balog},
  {Barache}, {Barbato}, {Barros}, {Barstow}, {Bartolom{\'e}}, {Bassilana},
  {Bauchet}, {Baudesson-Stella}, {Becciani}, {Bellazzini}, {Bernet}, {Bertone},
  {Bianchi}, {Blanco-Cuaresma}, {Boch}, {Bombrun}, {Bossini}, {Bouquillon},
  {Bragaglia}, {Bramante}, {Breedt}, {Bressan}, {Brouillet}, {Bucciarelli},
  {Burlacu}, {Busonero}, {Butkevich}, {Buzzi}, {Caffau}, {Cancelliere},
  {C{\'a}novas}, {Cantat-Gaudin}, {Carballo}, {Carlucci}, {Carnerero},
  {Carrasco}, {Casamiquela}, {Castellani}, {Castro-Ginard}, {Castro Sampol},
  {Chaoul}, {Charlot}, {Chemin}, {Chiavassa}, {Cioni}, {Comoretto}, {Cooper},
  {Cornez}, {Cowell}, {Crifo}, {Crosta}, {Crowley}, {Dafonte}, {Dapergolas},
  {David}, {David}, {de Laverny}, {De Luise}, {De March}, {De Ridder}, {de
  Souza}, {de Teodoro}, {de Torres}, {del Peloso}, {del Pozo}, {Delbo},
  {Delgado}, {Delgado}, {Delisle}, {Di Matteo}, {Diakite}, {Diener},
  {Distefano}, {Dolding}, {Eappachen}, {Edvardsson}, {Enke}, {Esquej}, {Fabre},
  {Fabrizio}, {Faigler}, {Fedorets}, {Fernique}, {Fienga}, {Figueras},
  {Fouron}, {Fragkoudi}, {Fraile}, {Franke}, {Gai}, {Garabato},
  {Garcia-Gutierrez}, {Garc{\'\i}a-Torres}, {Garofalo}, {Gavras}, {Gerlach},
  {Geyer}, {Giacobbe}, {Gilmore}, {Girona}, {Giuffrida}, {Gomel}, {Gomez},
  {Gonzalez-Santamaria}, {Gonz{\'a}lez-Vidal}, {Granvik},
  {Guti{\'e}rrez-S{\'a}nchez}, {Guy}, {Hauser}, {Haywood}, {Helmi}, {Hidalgo},
  {Hilger}, {H{\l}adczuk}, {Hobbs}, {Holland}, {Huckle}, {Jasniewicz},
  {Jonker}, {Juaristi Campillo}, {Julbe}, {Karbevska}, {Kervella}, {Khanna},
  {Kochoska}, {Kontizas}, {Kordopatis}, {Korn}, {Kostrzewa-Rutkowska},
  {Kruszy{\'n}ska}, {Lambert}, {Lanza}, {Lasne}, {Le Campion}, {Le Fustec},
  {Lebreton}, {Lebzelter}, {Leccia}, {Leclerc}, {Lecoeur-Taibi}, {Liao},
  {Licata}, {Lindstr{\o}m}, {Lister}, {Livanou}, {Lobel}, {Madrero Pardo},
  {Managau}, {Mann}, {Marchant}, {Marconi}, {Marcos Santos}, {Marinoni},
  {Marocco}, {Marshall}, {Martin Polo}, {Mart{\'\i}n-Fleitas}, {Masip},
  {Massari}, {Mastrobuono-Battisti}, {Mazeh}, {McMillan}, {Messina},
  {Michalik}, {Millar}, {Mints}, {Molina}, {Molinaro}, {Moln{\'a}r},
  {Montegriffo}, {Mor}, {Morbidelli}, {Morel}, {Morris}, {Mulone}, {Munoz},
  {Muraveva}, {Murphy}, {Musella}, {Noval}, {Ord{\'e}novic}, {Orr{\`u}},
  {Osinde}, {Pagani}, {Pagano}, {Palaversa}, {Palicio}, {Panahi}, {Pawlak},
  {Pe{\~n}alosa Esteller}, {Penttil{\"a}}, {Piersimoni}, {Pineau}, {Plachy},
  {Plum}, {Poggio}, {Poretti}, {Poujoulet}, {Pr{\v{s}}a}, {Pulone}, {Racero},
  {Ragaini}, {Rainer}, {Raiteri}, {Rambaux}, {Ramos}, {Ramos-Lerate}, {Re
  Fiorentin}, {Regibo}, {Reyl{\'e}}, {Ripepi}, {Riva}, {Rixon}, {Robichon},
  {Robin}, {Roelens}, {Rohrbasser}, {Romero-G{\'o}mez}, {Rowell}, {Royer},
  {Rybicki}, {Sadowski}, {Sagrist{\`a} Sell{\'e}s}, {Sahlmann}, {Salgado},
  {Salguero}, {Samaras}, {Sanchez Gimenez}, {Sanna}, {Santove{\~n}a},
  {Sarasso}, {Schultheis}, {Sciacca}, {Segol}, {Segovia}, {S{\'e}gransan},
  {Semeux}, {Shahaf}, {Siddiqui}, {Siebert}, {Siltala}, {Slezak}, {Smart},
  {Solano}, {Solitro}, {Souami}, {Souchay}, {Spagna}, {Spoto}, {Steele},
  {Steidelm{\"u}ller}, {Stephenson}, {S{\"u}veges}, {Szabados}, {Szegedi-Elek},
  {Taris}, {Tauran}, {Taylor}, {Teixeira}, {Thuillot}, {Tonello}, {Torra},
  {Torra}, {Turon}, {Unger}, {Vaillant}, {van Dillen}, {Vanel}, {Vecchiato},
  {Viala}, {Vicente}, {Voutsinas}, {Weiler}, {Wevers}, {Wyrzykowski}, {Yoldas},
  {Yvard}, {Zhao}, {Zorec}, {Zucker}, {Zurbach}, \&
  {Zwitter}}]{2021A&A...649A...1G}
{Gaia Collaboration}, {Brown}, A.~G.~A., {Vallenari}, A., {et~al.} 2021, \aap,
  649, A1, \dodoi{10.1051/0004-6361/202039657}

\bibitem[{{Gaidos} \& {Gonzalez}(2002)}]{2002NewA....7..211G}
{Gaidos}, E.~J., \& {Gonzalez}, G. 2002, \na, 7, 211,
  \dodoi{10.1016/S1384-1076(02)00108-210.48550/arXiv.astro-ph/0203518}

\bibitem[{{Gingerich} {et~al.}(1971){Gingerich}, {Noyes}, {Kalkofen}, \&
  {Cuny}}]{1971SoPh...18..347G}
{Gingerich}, O., {Noyes}, R.~W., {Kalkofen}, W., \& {Cuny}, Y. 1971, \solphys,
  18, 347, \dodoi{10.1007/BF00149057}

\bibitem[{{Golding} {et~al.}(2017){Golding}, {Leenaarts}, \&
  {Carlsson}}]{2017A&A...597A.102G}
{Golding}, T.~P., {Leenaarts}, J., \& {Carlsson}, M. 2017, \aap, 597, A102,
  \dodoi{10.1051/0004-6361/201629462}

\bibitem[{{Guarcello} {et~al.}(2019){Guarcello}, {Micela}, {Sciortino},
  {L{\'o}pez-Santiago}, {Argiroffi}, {Reale}, {Flaccomio},
  {Alvarado-G{\'o}mez}, {Antoniou}, {Drake}, {Pillitteri}, {Rebull}, \&
  {Stauffer}}]{2019A&A...622A.210G}
{Guarcello}, M.~G., {Micela}, G., {Sciortino}, S., {et~al.} 2019, \aap, 622,
  A210, \dodoi{10.1051/0004-6361/201834370}

\bibitem[{{G{\"u}del}(2007)}]{2007LRSP....4....3G}
{G{\"u}del}, M. 2007, Living Reviews in Solar Physics, 4, 3,
  \dodoi{10.12942/lrsp-2007-3}

\bibitem[{{G{\"u}del} {et~al.}(1997){G{\"u}del}, {Guinan}, \&
  {Skinner}}]{1997ApJ...483..947G}
{G{\"u}del}, M., {Guinan}, E.~F., \& {Skinner}, S.~L. 1997, \apj, 483, 947,
  \dodoi{10.1086/304264}

\bibitem[{{Gudiksen} \& {Nordlund}(2005)}]{2005ApJ...618.1020G}
{Gudiksen}, B.~V., \& {Nordlund}, {\r{A}}. 2005, \apj, 618, 1020,
  \dodoi{10.1086/426063}

\bibitem[{{Hawley} {et~al.}(2014){Hawley}, {Davenport}, {Kowalski},
  {Wisniewski}, {Hebb}, {Deitrick}, \& {Hilton}}]{2014ApJ...797..121H}
{Hawley}, S.~L., {Davenport}, J. R.~A., {Kowalski}, A.~F., {et~al.} 2014, \apj,
  797, 121, \dodoi{10.1088/0004-637X/797/2/121}

\bibitem[{{Hazra} {et~al.}(2022){Hazra}, {Vidotto}, {Carolan}, {Villarreal
  D'Angelo}, \& {Manchester}}]{2022MNRAS.509.5858H}
{Hazra}, G., {Vidotto}, A.~A., {Carolan}, S., {Villarreal D'Angelo}, C., \&
  {Manchester}, W. 2022, \mnras, 509, 5858, \dodoi{10.1093/mnras/stab3271}

\bibitem[{{Hock} {et~al.}(2012){Hock}, {Chamberlin}, {Woods}, {Crotser},
  {Eparvier}, {Woodraska}, \& {Woods}}]{2012SoPh..275..145H}
{Hock}, R.~A., {Chamberlin}, P.~C., {Woods}, T.~N., {et~al.} 2012, \solphys,
  275, 145, \dodoi{10.1007/s11207-010-9520-9}

\bibitem[{{Hotta} \& {Kusano}(2021)}]{2021NatAs...5.1100H}
{Hotta}, H., \& {Kusano}, K. 2021, Nature Astronomy, 5, 1100,
  \dodoi{10.1038/s41550-021-01459-0}

\bibitem[{{Johnstone} {et~al.}(2021){Johnstone}, {Bartel}, \&
  {G{\"u}del}}]{2021A&A...649A..96J}
{Johnstone}, C.~P., {Bartel}, M., \& {G{\"u}del}, M. 2021, \aap, 649, A96,
  \dodoi{10.1051/0004-6361/202038407}

\bibitem[{{Johnstone} {et~al.}(2019){Johnstone}, {Khodachenko},
  {L{\"u}ftinger}, {Kislyakova}, {Lammer}, \&
  {G{\"u}del}}]{2019A&A...624L..10J}
{Johnstone}, C.~P., {Khodachenko}, M.~L., {L{\"u}ftinger}, T., {et~al.} 2019,
  \aap, 624, L10, \dodoi{10.1051/0004-6361/201935279}

\bibitem[{{Kochukhov} {et~al.}(2020){Kochukhov}, {Hackman}, {Lehtinen}, \&
  {Wehrhahn}}]{2020A&A...635A.142K}
{Kochukhov}, O., {Hackman}, T., {Lehtinen}, J.~J., \& {Wehrhahn}, A. 2020,
  \aap, 635, A142, \dodoi{10.1051/0004-6361/201937185}

\bibitem[{{Kowalski} {et~al.}(2013){Kowalski}, {Hawley}, {Wisniewski}, {Osten},
  {Hilton}, {Holtzman}, {Schmidt}, \& {Davenport}}]{2013ApJS..207...15K}
{Kowalski}, A.~F., {Hawley}, S.~L., {Wisniewski}, J.~P., {et~al.} 2013, \apjs,
  207, 15, \dodoi{10.1088/0067-0049/207/1/15}

\bibitem[{{Kretzschmar}(2011)}]{2011A&A...530A..84K}
{Kretzschmar}, M. 2011, \aap, 530, A84, \dodoi{10.1051/0004-6361/201015930}

\bibitem[{{Lammer} {et~al.}(2003){Lammer}, {Selsis}, {Ribas}, {Guinan},
  {Bauer}, \& {Weiss}}]{2003ApJ...598L.121L}
{Lammer}, H., {Selsis}, F., {Ribas}, I., {et~al.} 2003, \apjl, 598, L121,
  \dodoi{10.1086/380815}

\bibitem[{{Lecavelier Des Etangs}(2007)}]{2007A&A...461.1185L}
{Lecavelier Des Etangs}, A. 2007, \aap, 461, 1185,
  \dodoi{10.1051/0004-6361:20065014}

\bibitem[{{Leenaarts} {et~al.}(2016){Leenaarts}, {Golding}, {Carlsson},
  {Libbrecht}, \& {Joshi}}]{2016A&A...594A.104L}
{Leenaarts}, J., {Golding}, T., {Carlsson}, M., {Libbrecht}, T., \& {Joshi}, J.
  2016, \aap, 594, A104, \dodoi{10.1051/0004-6361/201628490}

\bibitem[{{Linsky}(2019)}]{2019LNP...955.....L}
{Linsky}, J. 2019, {Host Stars and their Effects on Exoplanet Atmospheres},
  Vol. 955, \dodoi{10.1007/978-3-030-11452-7}

\bibitem[{{Linsky} {et~al.}(2012){Linsky}, {Bushinsky}, {Ayres}, {Fontenla}, \&
  {France}}]{2012ApJ...745...25L}
{Linsky}, J.~L., {Bushinsky}, R., {Ayres}, T., {Fontenla}, J., \& {France}, K.
  2012, \apj, 745, 25, \dodoi{10.1088/0004-637X/745/1/25}

\bibitem[{{Linsky} {et~al.}(2014){Linsky}, {Fontenla}, \&
  {France}}]{2014ApJ...780...61L}
{Linsky}, J.~L., {Fontenla}, J., \& {France}, K. 2014, \apj, 780, 61,
  \dodoi{10.1088/0004-637X/780/1/61}

\bibitem[{{Linsky} {et~al.}(2020){Linsky}, {Wood}, {Youngblood}, {Brown},
  {Froning}, {France}, {Buccino}, {Cranmer}, {Mauas}, {Miguel}, {Pineda},
  {Rugheimer}, {Vieytes}, {Wheatley}, \& {Wilson}}]{2020ApJ...902....3L}
{Linsky}, J.~L., {Wood}, B.~E., {Youngblood}, A., {et~al.} 2020, \apj, 902, 3,
  \dodoi{10.3847/1538-4357/abb36f}

\bibitem[{{Liu} {et~al.}(2007){Liu}, {Norton}, \&
  {Scherrer}}]{2007SoPh..241..185L}
{Liu}, Y., {Norton}, A.~A., \& {Scherrer}, P.~H. 2007, \solphys, 241, 185,
  \dodoi{10.1007/s11207-007-0296-5}

\bibitem[{{Liu} {et~al.}(2012){Liu}, {Hoeksema}, {Scherrer}, {Schou},
  {Couvidat}, {Bush}, {Duvall}, {Hayashi}, {Sun}, \&
  {Zhao}}]{2012SoPh..279..295L}
{Liu}, Y., {Hoeksema}, J.~T., {Scherrer}, P.~H., {et~al.} 2012, \solphys, 279,
  295, \dodoi{10.1007/s11207-012-9976-x}

\bibitem[{{Louden} {et~al.}(2017){Louden}, {Wheatley}, \&
  {Briggs}}]{2017MNRAS.464.2396L}
{Louden}, T., {Wheatley}, P.~J., \& {Briggs}, K. 2017, \mnras, 464, 2396,
  \dodoi{10.1093/mnras/stw2421}

\bibitem[{{Loyd} {et~al.}(2016){Loyd}, {France}, {Youngblood}, {Schneider},
  {Brown}, {Hu}, {Linsky}, {Froning}, {Redfield}, {Rugheimer}, \&
  {Tian}}]{2016ApJ...824..102L}
{Loyd}, R.~O.~P., {France}, K., {Youngblood}, A., {et~al.} 2016, \apj, 824,
  102, \dodoi{10.3847/0004-637X/824/2/102}

\bibitem[{{Loyd} {et~al.}(2018){Loyd}, {France}, {Youngblood}, {Schneider},
  {Brown}, {Hu}, {Segura}, {Linsky}, {Redfield}, {Tian}, {Rugheimer}, {Miguel},
  \& {Froning}}]{2018ApJ...867...71L}
---. 2018, \apj, 867, 71, \dodoi{10.3847/1538-4357/aae2bd}

\bibitem[{{Loyd} {et~al.}(2022){Loyd}, {Mason}, {Jin}, {Shkolnik}, {France},
  {Youngblood}, {Villadsen}, {Schneider}, {Schneider}, {Llama},
  {Ramiaramanantsoa}, \& {Richey-Yowell}}]{2022ApJ...936..170L}
{Loyd}, R.~O.~P., {Mason}, J.~P., {Jin}, M., {et~al.} 2022, \apj, 936, 170,
  \dodoi{10.3847/1538-4357/ac80c1}

\bibitem[{{MacGregor} {et~al.}(2018){MacGregor}, {Weinberger}, {Wilner},
  {Kowalski}, \& {Cranmer}}]{2018ApJ...855L...2M}
{MacGregor}, M.~A., {Weinberger}, A.~J., {Wilner}, D.~J., {Kowalski}, A.~F., \&
  {Cranmer}, S.~R. 2018, \apjl, 855, L2, \dodoi{10.3847/2041-8213/aaad6b}

\bibitem[{{Maehara} {et~al.}(2015){Maehara}, {Shibayama}, {Notsu}, {Notsu},
  {Honda}, {Nogami}, \& {Shibata}}]{2015EP&S...67...59M}
{Maehara}, H., {Shibayama}, T., {Notsu}, Y., {et~al.} 2015, Earth, Planets, and
  Space, 67, 59, \dodoi{10.1186/s40623-015-0217-z}

\bibitem[{{Maehara} {et~al.}(2012){Maehara}, {Shibayama}, {Notsu}, {Notsu},
  {Nagao}, {Kusaba}, {Honda}, {Nogami}, \& {Shibata}}]{2012Natur.485..478M}
{Maehara}, H., {Shibayama}, T., {Notsu}, S., {et~al.} 2012, Nature, 485, 478,
  \dodoi{10.1038/nature11063}

\bibitem[{{Maehara} {et~al.}(2021){Maehara}, {Notsu}, {Namekata}, {Honda},
  {Kowalski}, {Katoh}, {Ohshima}, {Iida}, {Oeda}, {Murata}, {Yamanaka},
  {Takagi}, {Sasada}, {Akitaya}, {Ikuta}, {Okamoto}, {Nogami}, \&
  {Shibata}}]{2020PASJ..tmp..253M}
{Maehara}, H., {Notsu}, Y., {Namekata}, K., {et~al.} 2021, \pasj,
  \dodoi{10.1093/pasj/psaa098}

\bibitem[{{Mahajan} {et~al.}(2010){Mahajan}, {Lodhi}, \&
  {Upadhayaya}}]{2010JGRA..11512330M}
{Mahajan}, K.~K., {Lodhi}, N.~K., \& {Upadhayaya}, A.~K. 2010, Journal of
  Geophysical Research (Space Physics), 115, A12330,
  \dodoi{10.1029/2010JA015576}

\bibitem[{{Maldonado} {et~al.}(2019){Maldonado}, {Phillips}, {Dumusque},
  {Collier Cameron}, {Haywood}, {Lanza}, {Micela}, {Mortier}, {Saar},
  {Sozzetti}, {Rice}, {Milbourne}, {Cecconi}, {Cegla}, {Cosentino}, {Costes},
  {Ghedina}, {Gonzalez}, {Guerra}, {Hern{\'a}ndez}, {Li}, {Lodi}, {Malavolta},
  {Molinari}, {Pepe}, {Piotto}, {Poretti}, {Sasselov}, {San Juan}, {Thompson},
  {Udry}, \& {Watson}}]{2019A&A...627A.118M}
{Maldonado}, J., {Phillips}, D.~F., {Dumusque}, X., {et~al.} 2019, \aap, 627,
  A118, \dodoi{10.1051/0004-6361/201935233}

\bibitem[{{Milligan} {et~al.}(2014){Milligan}, {Kerr}, {Dennis}, {Hudson},
  {Fletcher}, {Allred}, {Chamberlin}, {Ireland}, {Mathioudakis}, \&
  {Keenan}}]{2014ApJ...793...70M}
{Milligan}, R.~O., {Kerr}, G.~S., {Dennis}, B.~R., {et~al.} 2014, \apj, 793,
  70, \dodoi{10.1088/0004-637X/793/2/70}

\bibitem[{{Namekata} {et~al.}(2022{\natexlab{a}}){Namekata}, {Ichimoto},
  {Ishii}, \& {Shibata}}]{2022ApJ...933..209N}
{Namekata}, K., {Ichimoto}, K., {Ishii}, T.~T., \& {Shibata}, K.
  2022{\natexlab{a}}, \apj, 933, 209, \dodoi{10.3847/1538-4357/ac75cd}

\bibitem[{{Namekata} {et~al.}(2019){Namekata}, {Maehara}, {Notsu}, {Toriumi},
  {Hayakawa}, {Ikuta}, {Notsu}, {Honda}, {Nogami}, \&
  {Shibata}}]{2019ApJ...871..187N}
{Namekata}, K., {Maehara}, H., {Notsu}, Y., {et~al.} 2019, \apj, 871, 187,
  \dodoi{10.3847/1538-4357/aaf471}

\bibitem[{{Namekata} {et~al.}(2020{\natexlab{a}}){Namekata}, {Maehara},
  {Sasaki}, {Kawai}, {Notsu}, {Kowalski}, {Allred}, {Iwakiri}, {Tsuboi},
  {Murata}, {Niwano}, {Shiraishi}, {Adachi}, {Iida}, {Oeda}, {Honda}, {Tozuka},
  {Katoh}, {Onozato}, {Okamoto}, {Isogai}, {Kimura}, {Kojiguchi}, {Wakamatsu},
  {Tampo}, {Nogami}, \& {Shibata}}]{2020PASJ...72...68N}
{Namekata}, K., {Maehara}, H., {Sasaki}, R., {et~al.} 2020{\natexlab{a}},
  \pasj, 72, 68, \dodoi{10.1093/pasj/psaa051}

\bibitem[{{Namekata} {et~al.}(2020{\natexlab{b}}){Namekata}, {Davenport},
  {Morris}, {Hawley}, {Maehara}, {Notsu}, {Toriumi}, {Ikuta}, {Notsu}, {Honda},
  {Nogami}, \& {Shibata}}]{2020ApJ...891..103N}
{Namekata}, K., {Davenport}, J. R.~A., {Morris}, B.~M., {et~al.}
  2020{\natexlab{b}}, \apj, 891, 103, \dodoi{10.3847/1538-4357/ab7384}

\bibitem[{{Namekata} {et~al.}(2022{\natexlab{b}}){Namekata}, {Maehara},
  {Honda}, {Notsu}, {Okamoto}, {Takahashi}, {Takayama}, {Ohshima}, {Saito},
  {Katoh}, {Tozuka}, {Murata}, {Ogawa}, {Niwano}, {Adachi}, {Oeda},
  {Shiraishi}, {Isogai}, {Seki}, {Ishii}, {Ichimoto}, {Nogami}, \&
  {Shibata}}]{2022NatAs...6..241N}
{Namekata}, K., {Maehara}, H., {Honda}, S., {et~al.} 2022{\natexlab{b}}, Nature
  Astronomy, 6, 241, \dodoi{10.1038/s41550-021-01532-8}

\bibitem[{{Namekata} {et~al.}(2022{\natexlab{c}}){Namekata}, {Maehara},
  {Honda}, {Notsu}, {Okamoto}, {Takahashi}, {Takayama}, {Ohshima}, {Saito},
  {Katoh}, {Tozuka}, {Murata}, {Ogawa}, {Niwano}, {Adachi}, {Oeda},
  {Shiraishi}, {Isogai}, {Nogami}, \& {Shibata}}]{2022ApJ...926L...5N}
---. 2022{\natexlab{c}}, \apjl, 926, L5, \dodoi{10.3847/2041-8213/ac4df0}

\bibitem[{{Notsu} {et~al.}(2019){Notsu}, {Maehara}, {Honda}, {Hawley},
  {Davenport}, {Namekata}, {Notsu}, {Ikuta}, {Nogami}, \&
  {Shibata}}]{2019ApJ...876...58N}
{Notsu}, Y., {Maehara}, H., {Honda}, S., {et~al.} 2019, \apj, 876, 58,
  \dodoi{10.3847/1538-4357/ab14e6}

\bibitem[{{Otsu} {et~al.}(2022){Otsu}, {Asai}, {Ichimoto}, {Ishii}, \&
  {Namekata}}]{2022ApJ...939...98O}
{Otsu}, T., {Asai}, A., {Ichimoto}, K., {Ishii}, T.~T., \& {Namekata}, K. 2022,
  \apj, 939, 98, \dodoi{10.3847/1538-4357/ac9730}

\bibitem[{{Parker}(1972)}]{1972ApJ...174..499P}
{Parker}, E.~N. 1972, \apj, 174, 499, \dodoi{10.1086/151512}

\bibitem[{{Parker}(1988)}]{1988ApJ...330..474P}
---. 1988, \apj, 330, 474, \dodoi{10.1086/166485}

\bibitem[{{Pesnell} {et~al.}(2012){Pesnell}, {Thompson}, \&
  {Chamberlin}}]{2012SoPh..275....3P}
{Pesnell}, W.~D., {Thompson}, B.~J., \& {Chamberlin}, P.~C. 2012, \solphys,
  275, 3, \dodoi{10.1007/s11207-011-9841-3}

\bibitem[{{Pevtsov} {et~al.}(2003){Pevtsov}, {Fisher}, {Acton}, {Longcope},
  {Johns-Krull}, {Kankelborg}, \& {Metcalf}}]{2003ApJ...598.1387P}
{Pevtsov}, A.~A., {Fisher}, G.~H., {Acton}, L.~W., {et~al.} 2003, \apj, 598,
  1387, \dodoi{10.1086/378944}

\bibitem[{{Reiners} {et~al.}(2009){Reiners}, {Basri}, \&
  {Browning}}]{2009ApJ...692..538R}
{Reiners}, A., {Basri}, G., \& {Browning}, M. 2009, \apj, 692, 538,
  \dodoi{10.1088/0004-637X/692/1/538}

\bibitem[{{Reiners} {et~al.}(2022){Reiners}, {Shulyak}, {K{\"a}pyl{\"a}},
  {Ribas}, {Nagel}, {Zechmeister}, {Caballero}, {Shan}, {Fuhrmeister},
  {Quirrenbach}, {Amado}, {Montes}, {Jeffers}, {Azzaro}, {B{\'e}jar},
  {Chaturvedi}, {Henning}, {K{\"u}rster}, \& {Pall{\'e}}}]{2022A&A...662A..41R}
{Reiners}, A., {Shulyak}, D., {K{\"a}pyl{\"a}}, P.~J., {et~al.} 2022, \aap,
  662, A41, \dodoi{10.1051/0004-6361/202243251}

\bibitem[{{Ribas} {et~al.}(2005){Ribas}, {Guinan}, {G{\"u}del}, \&
  {Audard}}]{2005ApJ...622..680R}
{Ribas}, I., {Guinan}, E.~F., {G{\"u}del}, M., \& {Audard}, M. 2005, \apj, 622,
  680, \dodoi{10.1086/427977}

\bibitem[{{Rottman}(2005)}]{2005SoPh..230....7R}
{Rottman}, G. 2005, \solphys, 230, 7, \dodoi{10.1007/s11207-005-8112-6}

\bibitem[{{Saar}(2001)}]{2001ASPC..223..292S}
{Saar}, S.~H. 2001, in Astronomical Society of the Pacific Conference Series,
  Vol. 223, 11th Cambridge Workshop on Cool Stars, Stellar Systems and the Sun,
  ed. R.~J. {Garcia Lopez}, R.~{Rebolo}, \& M.~R. {Zapaterio Osorio}, 292

\bibitem[{{Sakaue} \& {Shibata}(2021)}]{2021ApJ...919...29S}
{Sakaue}, T., \& {Shibata}, K. 2021, \apj, 919, 29,
  \dodoi{10.3847/1538-4357/ac0e34}

\bibitem[{{Sanz-Forcada} {et~al.}(2011){Sanz-Forcada}, {Micela}, {Ribas},
  {Pollock}, {Eiroa}, {Velasco}, {Solano}, \&
  {Garc{\'\i}a-{\'A}lvarez}}]{2011A&A...532A...6S}
{Sanz-Forcada}, J., {Micela}, G., {Ribas}, I., {et~al.} 2011, \aap, 532, A6,
  \dodoi{10.1051/0004-6361/201116594}

\bibitem[{{Sanz-Forcada} {et~al.}(2010){Sanz-Forcada}, {Ribas}, {Micela},
  {Pollock}, {Garc{\'\i}a-{\'A}lvarez}, {Solano}, \&
  {Eiroa}}]{2010A&A...511L...8S}
{Sanz-Forcada}, J., {Ribas}, I., {Micela}, G., {et~al.} 2010, \aap, 511, L8,
  \dodoi{10.1051/0004-6361/200913670}

\bibitem[{{Scelsi} {et~al.}(2005){Scelsi}, {Maggio}, {Peres}, \&
  {Pallavicini}}]{2005A&A...432..671S}
{Scelsi}, L., {Maggio}, A., {Peres}, G., \& {Pallavicini}, R. 2005, \aap, 432,
  671, \dodoi{10.1051/0004-6361:20041739}

\bibitem[{{Scherrer} {et~al.}(1995){Scherrer}, {Bogart}, {Bush}, {Hoeksema},
  {Kosovichev}, {Schou}, {Rosenberg}, {Springer}, {Tarbell}, {Title},
  {Wolfson}, {Zayer}, \& {MDI Engineering Team}}]{1995SoPh..162..129S}
{Scherrer}, P.~H., {Bogart}, R.~S., {Bush}, R.~I., {et~al.} 1995, \solphys,
  162, 129, \dodoi{10.1007/BF00733429}

\bibitem[{{Scherrer} {et~al.}(2012){Scherrer}, {Schou}, {Bush}, {Kosovichev},
  {Bogart}, {Hoeksema}, {Liu}, {Duvall}, {Zhao}, {Title}, {Schrijver},
  {Tarbell}, \& {Tomczyk}}]{2012SoPh..275..207S}
{Scherrer}, P.~H., {Schou}, J., {Bush}, R.~I., {et~al.} 2012, \solphys, 275,
  207, \dodoi{10.1007/s11207-011-9834-2}

\bibitem[{{Segura} {et~al.}(2010){Segura}, {Walkowicz}, {Meadows}, {Kasting},
  \& {Hawley}}]{2010AsBio..10..751S}
{Segura}, A., {Walkowicz}, L.~M., {Meadows}, V., {Kasting}, J., \& {Hawley}, S.
  2010, Astrobiology, 10, 751, \dodoi{10.1089/ast.2009.0376}

\bibitem[{{Shibata} \& {Magara}(2011)}]{2011LRSP....8....6S}
{Shibata}, K., \& {Magara}, T. 2011, Living Reviews in Solar Physics, 8, 6,
  \dodoi{10.12942/lrsp-2011-6}

\bibitem[{{Shoda} \& {Takasao}(2021)}]{2021A&A...656A.111S}
{Shoda}, M., \& {Takasao}, S. 2021, \aap, 656, A111,
  \dodoi{10.1051/0004-6361/202141563}

\bibitem[{{Shoda} {et~al.}(2020){Shoda}, {Suzuki}, {Matt}, {Cranmer},
  {Vidotto}, {Strugarek}, {See}, {R{\'e}ville}, {Finley}, \&
  {Brun}}]{2020ApJ...896..123S}
{Shoda}, M., {Suzuki}, T.~K., {Matt}, S.~P., {et~al.} 2020, \apj, 896, 123,
  \dodoi{10.3847/1538-4357/ab94bf}

\bibitem[{{Snow} {et~al.}(2022){Snow}, {McClintock}, {Woods}, \&
  {Elliott}}]{2022SoPh..297...55S}
{Snow}, M., {McClintock}, W.~E., {Woods}, T.~N., \& {Elliott}, J.~P. 2022,
  \solphys, 297, 55, \dodoi{10.1007/s11207-022-01984-9}

\bibitem[{{Sreejith} {et~al.}(2020){Sreejith}, {Fossati}, {Youngblood},
  {France}, \& {Ambily}}]{2020A&A...644A..67S}
{Sreejith}, A.~G., {Fossati}, L., {Youngblood}, A., {France}, K., \& {Ambily},
  S. 2020, \aap, 644, A67, \dodoi{10.1051/0004-6361/202039167}

\bibitem[{{Sutherland} \& {Dopita}(1993)}]{1993ApJS...88..253S}
{Sutherland}, R.~S., \& {Dopita}, M.~A. 1993, \apjs, 88, 253,
  \dodoi{10.1086/191823}

\bibitem[{{Suzuki} {et~al.}(2013){Suzuki}, {Imada}, {Kataoka}, {Kato},
  {Matsumoto}, {Miyahara}, \& {Tsuneta}}]{2013PASJ...65...98S}
{Suzuki}, T.~K., {Imada}, S., {Kataoka}, R., {et~al.} 2013, \pasj, 65, 98,
  \dodoi{10.1093/pasj/65.5.98}

\bibitem[{{Takasao} {et~al.}(2020){Takasao}, {Mitsuishi}, {Shimura}, {Yoshida},
  {Kunitomo}, {Tanaka}, \& {Ishihara}}]{2020ApJ...901...70T}
{Takasao}, S., {Mitsuishi}, I., {Shimura}, T., {et~al.} 2020, \apj, 901, 70,
  \dodoi{10.3847/1538-4357/abad34}

\bibitem[{{Takeda} {et~al.}(2007){Takeda}, {Ford}, {Sills}, {Rasio}, {Fischer},
  \& {Valenti}}]{2007ApJS..168..297T}
{Takeda}, G., {Ford}, E.~B., {Sills}, A., {et~al.} 2007, \apjs, 168, 297,
  \dodoi{10.1086/50976310.48550/arXiv.astro-ph/0607235}

\bibitem[{{Testa} {et~al.}(2015){Testa}, {Saar}, \&
  {Drake}}]{2015RSPTA.37340259T}
{Testa}, P., {Saar}, S.~H., \& {Drake}, J.~J. 2015, Philosophical Transactions
  of the Royal Society of London Series A, 373, 20140259,
  \dodoi{10.1098/rsta.2014.0259}

\bibitem[{{Toriumi} \& {Airapetian}(2022)}]{2022ApJ...927..179T}
{Toriumi}, S., \& {Airapetian}, V.~S. 2022, \apj, 927, 179,
  \dodoi{10.3847/1538-4357/ac5179}

\bibitem[{{Toriumi} {et~al.}(2020){Toriumi}, {Airapetian}, {Hudson},
  {Schrijver}, {Cheung}, \& {DeRosa}}]{2020ApJ...902...36T}
{Toriumi}, S., {Airapetian}, V.~S., {Hudson}, H.~S., {et~al.} 2020, \apj, 902,
  36, \dodoi{10.3847/1538-4357/abadf9}

\bibitem[{{Toriumi} {et~al.}(2022){Toriumi}, {Airapetian}, {Namekata}, \&
  {Notsu}}]{2022ApJS..262...46T}
{Toriumi}, S., {Airapetian}, V.~S., {Namekata}, K., \& {Notsu}, Y. 2022, \apjs,
  262, 46, \dodoi{10.3847/1538-4365/ac8b15}

\bibitem[{{Toriumi} \& {Hotta}(2019)}]{2019ApJ...886L..21T}
{Toriumi}, S., \& {Hotta}, H. 2019, \apjl, 886, L21,
  \dodoi{10.3847/2041-8213/ab55e7}

\bibitem[{{Toriumi} \& {Wang}(2019)}]{2019LRSP...16....3T}
{Toriumi}, S., \& {Wang}, H. 2019, Living Reviews in Solar Physics, 16, 3,
  \dodoi{10.1007/s41116-019-0019-7}

\bibitem[{{Vernazza} {et~al.}(1981){Vernazza}, {Avrett}, \&
  {Loeser}}]{1981ApJS...45..635V}
{Vernazza}, J.~E., {Avrett}, E.~H., \& {Loeser}, R. 1981, \apjs, 45, 635,
  \dodoi{10.1086/190731}

\bibitem[{{Veronig} {et~al.}(2021){Veronig}, {Odert}, {Leitzinger}, {Dissauer},
  {Fleck}, \& {Hudson}}]{2021NatAs...5..697V}
{Veronig}, A.~M., {Odert}, P., {Leitzinger}, M., {et~al.} 2021, Nature
  Astronomy, 5, 697, \dodoi{10.1038/s41550-021-01345-9}

\bibitem[{{Vida} {et~al.}(2016){Vida}, {Kriskovics}, {Ol{\'a}h}, {Leitzinger},
  {Odert}, {K{\H{o}}v{\'a}ri}, {Korhonen}, {Greimel}, {Robb}, {Cs{\'a}k}, \&
  {Kov{\'a}cs}}]{2016A&A...590A..11V}
{Vida}, K., {Kriskovics}, L., {Ol{\'a}h}, K., {et~al.} 2016, \aap, 590, A11,
  \dodoi{10.1051/0004-6361/201527925}

\bibitem[{{Vidotto} {et~al.}(2014){Vidotto}, {Gregory}, {Jardine}, {Donati},
  {Petit}, {Morin}, {Folsom}, {Bouvier}, {Cameron}, {Hussain}, {Marsden},
  {Waite}, {Fares}, {Jeffers}, \& {do Nascimento}}]{2014MNRAS.441.2361V}
{Vidotto}, A.~A., {Gregory}, S.~G., {Jardine}, M., {et~al.} 2014, \mnras, 441,
  2361, \dodoi{10.1093/mnras/stu728}

\bibitem[{{Wilson} {et~al.}(2021){Wilson}, {Froning}, {Duvvuri}, {France},
  {Youngblood}, {Schneider}, {Berta-Thompson}, {Brown}, {Buccino}, {Hawley},
  {Irwin}, {Kaltenegger}, {Kowalski}, {Linsky}, {Parke Loyd}, {Miguel},
  {Pineda}, {Redfield}, {Roberge}, {Rugheimer}, {Tian}, \&
  {Vieytes}}]{2021ApJ...911...18W}
{Wilson}, D.~J., {Froning}, C.~S., {Duvvuri}, G.~M., {et~al.} 2021, \apj, 911,
  18, \dodoi{10.3847/1538-4357/abe771}

\bibitem[{{Wood} {et~al.}(2021){Wood}, {M{\"u}ller}, {Redfield}, {Konow},
  {Vannier}, {Linsky}, {Youngblood}, {Vidotto}, {Jardine},
  {Alvarado-G{\'o}mez}, \& {Drake}}]{2021ApJ...915...37W}
{Wood}, B.~E., {M{\"u}ller}, H.-R., {Redfield}, S., {et~al.} 2021, \apj, 915,
  37, \dodoi{10.3847/1538-4357/abfda5}

\bibitem[{{Woods} \& {Elliott}(2022)}]{2022SoPh..297...64W}
{Woods}, T.~N., \& {Elliott}, J. 2022, \solphys, 297, 64,
  \dodoi{10.1007/s11207-022-01997-4}

\bibitem[{{Woods} {et~al.}(2018){Woods}, {Eparvier}, {Harder}, \&
  {Snow}}]{2018SoPh..293...76W}
{Woods}, T.~N., {Eparvier}, F.~G., {Harder}, J., \& {Snow}, M. 2018, \solphys,
  293, 76, \dodoi{10.1007/s11207-018-1294-5}

\bibitem[{{Woods} \& {Rottman}(2005)}]{2005SoPh..230..375W}
{Woods}, T.~N., \& {Rottman}, G. 2005, \solphys, 230, 375,
  \dodoi{10.1007/s11207-005-2555-7}

\bibitem[{{Woods} {et~al.}(2005){Woods}, {Eparvier}, {Bailey}, {Chamberlin},
  {Lean}, {Rottman}, {Solomon}, {Tobiska}, \&
  {Woodraska}}]{2005JGRA..110.1312W}
{Woods}, T.~N., {Eparvier}, F.~G., {Bailey}, S.~M., {et~al.} 2005, Journal of
  Geophysical Research (Space Physics), 110, A01312,
  \dodoi{10.1029/2004JA010765}

\bibitem[{{Woods} {et~al.}(2008){Woods}, {Chamberlin}, {Peterson}, {Meier},
  {Richards}, {Strickland}, {Lu}, {Qian}, {Solomon}, {Iijima}, {Mannucci}, \&
  {Tsurutani}}]{2008SoPh..250..235W}
{Woods}, T.~N., {Chamberlin}, P.~C., {Peterson}, W.~K., {et~al.} 2008,
  \solphys, 250, 235, \dodoi{10.1007/s11207-008-9196-6}

\bibitem[{{Woods} {et~al.}(2012){Woods}, {Eparvier}, {Hock}, {Jones},
  {Woodraska}, {Judge}, {Didkovsky}, {Lean}, {Mariska}, {Warren}, {McMullin},
  {Chamberlin}, {Berthiaume}, {Bailey}, {Fuller-Rowell}, {Sojka}, {Tobiska}, \&
  {Viereck}}]{2012SoPh..275..115W}
{Woods}, T.~N., {Eparvier}, F.~G., {Hock}, R., {et~al.} 2012, \solphys, 275,
  115, \dodoi{10.1007/s11207-009-9487-6}

\bibitem[{{Wright} {et~al.}(2011){Wright}, {Drake}, {Mamajek}, \&
  {Henry}}]{2011ApJ...743...48W}
{Wright}, N.~J., {Drake}, J.~J., {Mamajek}, E.~E., \& {Henry}, G.~W. 2011,
  \apj, 743, 48, \dodoi{10.1088/0004-637X/743/1/48}

\bibitem[{{Youngblood} {et~al.}(2016){Youngblood}, {France}, {Loyd}, {Linsky},
  {Redfield}, {Schneider}, {Wood}, {Brown}, {Froning}, {Miguel}, {Rugheimer},
  \& {Walkowicz}}]{2016ApJ...824..101Y}
{Youngblood}, A., {France}, K., {Loyd}, R.~O.~P., {et~al.} 2016, \apj, 824,
  101, \dodoi{10.3847/0004-637X/824/2/101}

\bibitem[{{Youngblood} {et~al.}(2017){Youngblood}, {France}, {Loyd}, {Brown},
  {Mason}, {Schneider}, {Tilley}, {Berta-Thompson}, {Buccino}, {Froning},
  {Hawley}, {Linsky}, {Mauas}, {Redfield}, {Kowalski}, {Miguel}, {Newton},
  {Rugheimer}, {Segura}, {Roberge}, \& {Vieytes}}]{2017ApJ...843...31Y}
---. 2017, \apj, 843, 31, \dodoi{10.3847/1538-4357/aa76dd}

\end{thebibliography}
\bibliographystyle{aasjournal}



\end{document}